\documentclass{aa}

\newcommand\dosingle[1]{#1}  \newcommand\dodouble[1]{ } 

\newcommand\postrefereechanges[1]{#1}    
\newcommand\posteditorchanges[1]{{#1}}

\newcommand\nice[1]{#1}    \newcommand\subm[1]{}   
\subm{ \renewcommand\dosingle[1]{ }   \renewcommand\dodouble[1]{#1} }

\dodouble{ \documentclass[referee]{aa} }
\usepackage{graphicx}

\usepackage{amsfonts}
\usepackage{mathpi}

\usepackage{url} 

 \usepackage{hyperref}
\nice{ 
\providecommand{\eprint}[1]{\href{http://arxiv.org/abs/#1}{[arXiv:#1]}}

\providecommand{\url}[1]{\href{#1}{#1}}

}

\providecommand{\adsurl}[1]{} 
\newcommand\SSS{Sect.~}

\nice{ \newcommand\mycaptionfont{\protect\footnotesize} }

\newcommand\kms{\,km\,s$^{-1}$}

\newcommand\gtapprox{\,\lower.6ex\hbox{$\buildrel >\over \sim$} \, }
\newcommand\ltapprox{\,\lower.6ex\hbox{$\buildrel <\over \sim$} \, }
\newcommand\propapprox{\,\lower.6ex\hbox{$\buildrel \propto\over \sim$} \, }

\newcommand\e{ {\scriptstyle \times} 10^}

\newcommand\arcs{\ifmmode {'' }\else $'' $\fi}     
\newcommand\arcm{\ifmmode {' }\else $' $\fi}       
\newcommand\ddeg{\ifmmode^\circ\else$^\circ$\fi}    

\newcommand\frtoday{Le\space\number\day\space\ifcase\month\or
  janvier\or f\'evrier\or mars\or avril\or mai\or juin\or
  juillet\or ao\^ut\or septembre\or octobre\or novembre\or 
d\'ecembre\fi\space \number\year}

\newcommand\cqg{ClassQuantGra}   %
\newcommand\BASI{Bull. Astr. Soc. India}

\newcommand\hGpc{\mbox{$h^{-1}$ Gpc}}

\newcommand\rSLS{r_{\mbox{\rm \small SLS}}}  

\newcommand\Omm{\Omega_{\mbox{\rm \small m}}}
\newcommand\Omtot{\Omega_{\mbox{\rm \small tot}}}

\newcommand\ximc{\xi_{\mbox{\rm \small C}}} 
\newcommand\xisc{\xi_{\mbox{\rm \small A}}} 

\newcommand\ximctiny{\xi_{C}} 
\newcommand\xisctiny{\xi_{A}} 

\newcommand\phiWMAP{\phi_{\mbox{\rm \small WMAP}}}
\newcommand\Npoint{N_{\mbox{\rm \small p}}}
\newcommand\Npair{N_{\mbox{\rm \small pair}}}
\newcommand\Nchain{N_{\mbox{\rm \small chain}}}
\newcommand\sigmaMCMC{\sigma_{\mbox{\rm \small MCMC}}}
\newcommand\Nburnin{N_{\mbox{\rm \small burn-in}}}
\newcommand\Pmin{P_{\mbox{\rm \small min}}}
\newcommand\ILC{\mbox{\rm ILC}}
\newcommand\TOH{\mbox{\rm TOH}}
\newcommand\kpzero{\mbox{\rm kp0}}
\newcommand\kptwo{\mbox{\rm kp2}}

\newcommand\lII{l}
\newcommand\bII{b}

\newcommand\sigmalbth{\sigma_{\left<(l,b)\right>}} 
\newcommand\sigmaalpha{\sigma_{\left<\alpha\right>}} 
\newcommand\sigmaphi{\sigma_{\left<\phi\right>}} 
\newcommand\phii{\phi_{\mbox{i}}}
\newcommand\phio{\phi_{\mbox{o}}}

\newcommand\langed[1]{}

\subm{ \usepackage{color} \renewcommand\langed[1]{{\bf \color{red} [{\em lang.ed.:} #1] }} }

\title{The optimal phase of the generalised Poincar\'e 
dodecahedral space {\langed{(lower case ``d'', ``s'')}} 
hypothesis implied by the
spatial cross-correlation function of the WMAP sky maps}

\author{Boudewijn F. Roukema\inst{1},
Zbigniew Buli\'nski\inst{1},
Agnieszka Szaniewska\inst{1},
Nicolas E. Gaudin\inst{2,1}  
}

\institute{Toru\'n Centre for Astronomy, N. Copernicus University,
ul. Gagarina 11, PL-87-100 Toru\'n, Poland 
\and
{\'Ecole nationale sup\'erieure de physique de
Strasbourg, Universit\'e Louis Pasteur, 
Bd. S\'ebastien Brant, BP 10413, F-67412 Illkirch Cedex,
France}
}

\date{\frtoday}

\titlerunning{Optimal phase of PDS for WMAP sky map}
\authorrunning{Roukema et al.}

\begin{document}

\newcommand\probPDS{{92-95\%}}   
\newcommand\probifnotPDS{{6-9\%}}   

\abstract
{
Small universe models predicted a cutoff in large-scale power
in the cosmic microwave background (CMB). This was detected by
the Wilkinson Microwave Anisotropy Probe (WMAP). 
Several studies have since proposed that the preferred
model of the comoving spatial 3-hypersurface of the Universe 
may be a Poincar\'e
dodecahedral space (PDS) rather than
a simply connected, flat space. Both
models assume an FLRW metric and are close to flat
with about 30\% matter density. 
}
{
We study two predictions of the PDS model. 
(i) For the correct astronomical positioning of the fundamental
domain, the 
spatial two-point cross-correlation function $\ximc$ of temperature
fluctuations in the covering space (where the two points
in any pair are on different copies of the surface of last
scattering (SLS)) should have a similar order of magnitude to 
the auto-correlation function $\xisc$ on a single copy of the SLS. 
(ii) Consider a ``generalised'' PDS model for an
{{\em arbitrary}} ``twist'' phase $\phi \in \left[0,2\pi\right]$. 
The optimal orientation and 
identified circle radius for a generalised PDS model found by maximising 
$\ximc$ relative to $\xisc$ in the WMAP
maps should yield one of the two twist angles $ \pm 36\ddeg$.
}
{
Comparison of $\ximc$ to $\xisc$ 
extends the identified circles method, 
using a much larger number of data points.
We optimise the ratio of these functions at scales $\ltapprox 4.0${\hGpc}
using a Markov chain Monte Carlo (MCMC) method 
over orientation ($l$, $b$, $\theta$), 
circle size $\alpha$, and twist $\phi$.
}
{
Both predictions were satisfied: 
(i) An optimal generalised PDS solution
was found for 
two different foreground-reduced versions of the WMAP 3-year 
all-sky map, both with and without the kp2 galactic contamination mask.
This solution yields a strong
cross-correlation between points which
would be distant and only weakly correlated 
according to the simply connected hypothesis. 
The face centres are
$\{(l,b)\}_{i=1,6}\approx \{
(184\ddeg, 62\ddeg),
(305\ddeg, 44\ddeg),
(46\ddeg, 49\ddeg),
(117\ddeg, 20\ddeg),
(176\ddeg, -4\ddeg),
(240\ddeg, 13\ddeg) \}$ 
(and their antipodes) to within $\approx 2\ddeg$ ;
(ii) This solution has twist $\phi= (+39 \pm 2.5)\ddeg$,   
in agreement with the PDS model.
The chance of this occurring in the simply connected model,
assuming a uniform distribution 
$\phi \in [0,2\pi]$, 
is about {\probifnotPDS}.
}
{
  The PDS model now satisfies several different observational constraints.
}


\maketitle

\dodouble{ \clearpage } 


\newcommand\tdodec{
\begin{table}
\caption{Sky positions of the best estimate of 
the six face centres for the fundamental dodecahedron 
for the ILC map with the kp2 
mask, as shown in Fig.~\protect\ref{f-ilc_lbth_N}, for three values
of the minimum probability $\Pmin$
(see \SSS\protect\ref{s-res-lbtheta}).
\label{t-dodec}}
$$\begin{array}{c c r  r r r} \hline 
\rule[-1.5ex]{0ex}{4.5ex}
\Pmin &
i & 
n & 
{\lII} \ddeg & {\bII} \ddeg 
& \sigmalbth \ddeg\\ \hline 
   0.4 &  1 & 12117 &  \ \ 182.7 &   61.5 &    0.9  \\
   0.4 &  2 & 11370 &  304.2 &   44.6 &    1.2  \\
   0.4 &  3 & 13353 &   46.9 &   49.0 &    0.4  \\
   0.4 &  4 & 13785 &  117.6 &   20.2 &    1.0  \\
   0.4 &  5 & 13663 &  175.6 &   -4.9 &    2.3  \\
   0.4 &  6 & 11246 &  239.7 &   13.0 &    0.8  \\
\hline
   0.5 &  1 &  6342 &  183.5 &   62.5 &    1.1  \\
   0.5 &  2 &  6013 &  304.7 &   44.1 &    1.4  \\
   0.5 &  3 &  6903 &   46.5 &   48.7 &    0.4  \\
   0.5 &  4 &  7187 &  117.4 &   20.3 &    1.0  \\
   0.5 &  5 &  6886 &  175.9 &   -3.5 &    2.2  \\
   0.5 &  6 &  5894 &  239.8 &   13.2 &    1.2  \\
\hline
   0.6 &  1 &  2889 &  184.6 &   62.3 &    1.0  \\
   0.6 &  2 &  2862 &  303.9 &   43.0 &    1.8  \\
   0.6 &  3 &  3099 &   44.9 &   49.0 &    0.8  \\
   0.6 &  4 &  3201 &  116.8 &   20.7 &    0.8  \\
   0.6 &  5 &  3035 &  176.4 &   -1.8 &    1.6  \\
   0.6 &  6 &  2680 &  240.2 &   12.9 &    0.9  \\
\hline
\end{array}$$
\end{table}
}  

\newcommand\talphaphi{
\begin{table}
\caption{Estimates of matched circle radius $\alpha$ and twist
phase $\phi$ using 
the same points in the MCMC chains used
for the optimal dodecahedral face solution indicated in 
Table~\protect\ref{t-dodec} (see \SSS\protect\ref{s-res-alpha} regarding
systematic uncertainties in estimating $\alpha$). 
\label{t-alpha-phi}}
$$\begin{array}{c r r c c c} \hline 
\Pmin \rule[-1.5ex]{0ex}{4.5ex}
&
n & 
\alpha \ddeg & \sigmaalpha \ddeg  &
 \phi \ddeg & \sigmaphi \ddeg 
\\ \hline 
0.4& 12589.0 & \ \ 20.6 & 0.6 & 39.0 & 2.4  \\
0.5& 6537.5 & 20.8 & 0.7 & 38.7 & 2.2 \\
0.6&  2961.0 & 22.1 & 0.5 & 37.4 & 2.1 \\
\hline
\end{array}$$
\end{table}
}  

\newcommand\talphaphinomask{
\begin{table}
\caption{Estimates of matched circle radius $\alpha$ and twist
phase $\phi$ as per
Table~\protect\ref{t-alpha-phi}, but for the ILC map without any
mask, and the TOH map with and without the kp2 mask, each
based on only $N=2$ ``independent experiments'',
i.e., two independent concatenations of five MCMC chains.
\label{t-alpha-phi-nomask}}
$$\begin{array}{c r r c c c} \hline 
\Pmin \rule[-1.5ex]{0ex}{4.5ex}
&
n & 
\alpha \ddeg & \sigmaalpha \ddeg  &
 \phi \ddeg & \sigmaphi \ddeg 
\\ \hline 
\multicolumn{6}{c}{\mbox{ILC no mask}} \\
0.4 & 6103.5 & \ \ 18.7 & 0.9 & 35.1 & 0.7 \\
 0.5 & 3513.2 & 17.6 & 0.5 & 37.4 & 0.2 \\
 0.6 & 1735.0 & 17.4 & 0.4 & 38.3 & 0.6 \\
\hline
\multicolumn{6}{c}{\mbox{TOH kp2}} \\
 0.4 & 7144.3 & 22.1 & 0.8 & 30.8 & 4.3 \\
 0.5 & 4139.3 & 21.9 & 0.4 & 30.9 & 3.6 \\
 0.6 & 2173.8 & 22.9 & 0.3 & 28.7 & 5.3 \\
\hline
\multicolumn{6}{c}{\mbox{TOH no mask}} \\
 0.4 & 7909.5 & 19.6 & 0.4 & 27.5 & 4.1 \\
 0.5 & 4549.8 & 19.3 & 0.3 & 30.8 & 2.4 \\
 0.6 & 2203.2 & 19.1 & 0.2 & 31.2 & 1.8 \\
\hline
\end{array}$$
\end{table}
}  

\newcommand\falphaphi{
\begin{figure}
\centering 
\includegraphics[width=8cm,bb=65 173 489 570]{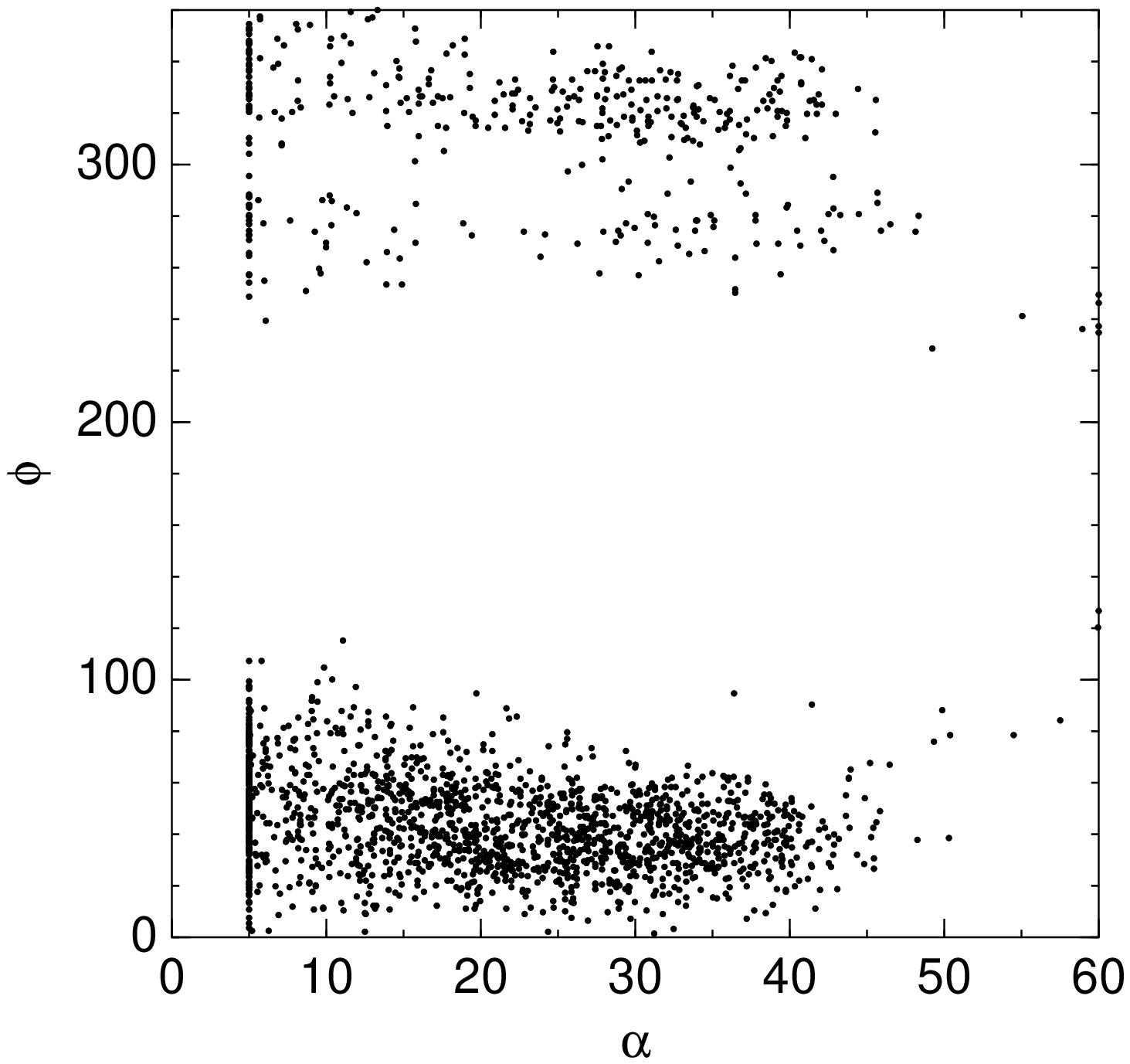} 
\caption[]{ \mycaptionfont
Distribution of $\alpha$ and $\phi$ states in the MCMC
chains of the dodecahedral solution used 
in Table~\protect\ref{t-dodec}. 
A single point is shown independently 
\protect\langed{``shown independently''}
of whether the chain spends a long
time at that point --- implying that it is a highly probable point --- 
or a shorter time at that point, so the density of points does not
fully represent the information in the distribution. In addition to the main 
cluster of points with $\phi \sim +36\ddeg$ 
(see Table~\protect\ref{t-alpha-phi} for a precise estimate),
a secondary weak cluster of points exists for $\phi \sim -36\ddeg = 324\ddeg$,
and an even weaker, tertiary cluster exists for $\phi \sim 280\ddeg$.
}
\label{f-alpha-phi}
\end{figure} 
} 

\newcommand\tsecondary{
\begin{table}
\caption{Estimates of matched circle radius $\alpha$ and twist
phase $\phi$ for the secondary and tertiary weak features
visible in Fig.~\protect\ref{f-alpha-phi}.$^{\mathrm{a}}$ 
\label{t-secondary}}
$$\begin{array}{c r r c c c} \hline 
\rule[-1.5ex]{0ex}{4.5ex}
\Pmin &
n & 
\alpha $\ddeg$ & \sigmaalpha $\ddeg$  &
 \phi $\ddeg$ & \sigmaphi $\ddeg$ 
\\ \hline  
\multicolumn{6}{c}{\mbox{secondary: \rule{0ex}{2.5ex}} 
                   8\pi /5 \le \phi \le 2\pi } 
\\
0.4&   970.2 & \ \ 22.8 & 1.3 & -32.3 & 2.7 \\
0.5&   474.0 & 22.1 & 1.7 & -30.6 & 2.3 \\
0.6&   193.2 & 22.4 & 3.0 & -30.5 & 3.4 \\
\hline  
\multicolumn{6}{c}{\mbox{tertiary: \rule{0ex}{2.5ex}} 
                   6\pi /5 \le \phi \le 8\pi/5 }  
 \\
0.4&   341.2 & 31.2 & 2.8 & -94.3 & 3.2 \\
0.5&   151.5 & 27.9 & 3.6 & -91.9 & 3.5 \\
0.6&    54.8 & 27.9 & 5.8 & -94.2 & 4.9 \\
\hline
\end{array}$$
\protect\langed{A note using the style of the first example in the example 
file {\protect\url{ftp://ftp.edpsciences.org/pub/aa/aa.dem}}.}
\begin{list}{}{}
\item[$^{\mathrm{a}}$] Apart from
restricting $\phi$ to the ranges 
$8\pi /5 \le \phi \le 2\pi$ and
$6\pi /5 \le \phi \le 8\pi/5$ respectively, the method is
the same as for Table~\protect\ref{f-alpha-phi}.
\end{list}
\end{table}
}  

\newcommand\tilckptwo{
\begin{table}
\caption{Estimates of optimal parameters $(l,b,\alpha,\phi)$ 
from WMAP 3yr ILC map,
above a ``probability'' threshold $\Pmin$, estimated 
from the whole parameter space ($i=0$) and then iterated ($i=1,2,
\ldots,12$) over smaller regions around the previous optimal point.
The kp2 mask of the Galaxy is used, covering about 15\% of the sky.
In a given iteration, the total number of MCMC steps with higher
``probability'' than $\Pmin$ and lying in the region is shown as $N$,
and the number of different MCMC chains (out of the full 24 chains 
for this map plus mask combination)
which have steps in this region is $j$.
See \SSS\protect\ref{s-res-optimal} for details. 
\label{t-ilc_kptwo}}
$$\begin{array}{c c c c c c c c } \hline 
\rule[-1.5ex]{0ex}{4.5ex}
\Pmin & i & N & j &
\lII \mbox{ in $\ddeg$} & \bII \mbox{ in $\ddeg$} 
& \alpha \mbox{ in $\ddeg$}
& \phi \mbox{ in $\ddeg$}
\\ \hline 
   0.6 &  0 & 24 &   5668 &  186.1 &   77.7 & 22.0 &   25.8 \\
   0.6 &  1 & 23 &   4275 &  182.9 &   70.8 & 22.4 &   39.4 \\
   0.6 &  2 & 23 &   3319 &  183.8 &   65.2 & 24.4 &   38.5 \\
   0.6 &  3 & 23 &   3258 &  184.5 &   64.9 & 24.8 &   38.0 \\
   0.6 &  4 & 23 &   3243 &  184.6 &   64.9 & 24.9 &   38.0 \\
 \ldots \\
   0.6 & 12 & 23 &   3240 &  184.6 &   64.9 & 24.9 &   37.9 \\
\hline
\end{array}$$
\end{table}
}  

\newcommand\tilckpzero{
\begin{table}
\caption{Estimates of optimal parameters $(l,b,\alpha,\phi)$ 
from WMAP 3yr ILC map, as for Table~\protect\ref{t-ilc_kptwo}, but
using the kp0 mask which covers 25\% of the sky, invalidating about
44\% of random pairs of points on the SLS.
Since multiple local maxima exist, an additional criterion constraining
the twist parameter $\phi_1 \le \phi \le \phi_2$  is used here.
\label{t-ilc_kpzero}}
$$\begin{array}{c c c c  c c c c } \hline 
\rule[-1.5ex]{0ex}{4.5ex}
\Pmin & i & N & j &
\lII \mbox{ in $\ddeg$} & \bII \mbox{ in $\ddeg$} 
& \alpha \mbox{ in $\ddeg$}
& \phi \mbox{ in $\ddeg$}
\\ \hline 
\multicolumn{8}{l}{0 \le \phi \le \pi } \\
   0.6 &  0 &  7 &   1005 &  185.3 &   71.2 & 24.6 &   39.7 \\
   0.6 &  1 &  7 &    851 &  187.1 &   67.7 & 25.4 &   38.5 \\
   0.6 &  2 &  7 &    830 &  187.5 &   66.4 & 25.8 &   38.5 \\
   0.6 &  3 &  7 &    807 &  187.7 &   66.0 & 25.8 &   37.8 \\
   0.6 &  4 &  7 &    807 &  187.7 &   66.0 & 25.8 &   37.8 \\
\ldots \\
   0.6 & 12 &  7 &    807 &  187.7 &   66.0 & 25.8 &   37.8 \\
\hline
\multicolumn{8}{l}{6\pi/5 \le \phi \le 8\pi/5 } \\
   0.6 &  0 &  8 &    628 &  339.3 &   83.3 & 19.5 &  -98.2 \\
   0.6 &  1 &  8 &    628 &  339.3 &   83.3 & 19.5 &  -98.2 \\
   0.6 &  2 &  1 &     15 &    1.2 &   66.7 & 15.1 & -115.3 \\
   0.6 &  3 &  7 &    236 &   18.6 &   59.0 & 17.2 &  -98.5 \\
   0.6 &  4 &  7 &    236 &   18.6 &   59.0 & 17.2 &  -98.5 \\
\ldots \\
   0.6 & 12 &  7 &    236 &   18.6 &   59.0 & 17.2 &  -98.5 \\
\hline
\multicolumn{8}{l}{8\pi/5 \le \phi \le 2\pi } \\
   0.6 &  0 &  9 &    567 &   14.1 &   80.1 & 30.4 &  -35.2 \\
   0.6 &  1 &  9 &    567 &   14.1 &   80.1 & 30.4 &  -35.2 \\
   0.6 &  2 &  8 &    550 &   10.6 &   78.0 & 33.0 &  -38.8 \\
   0.6 &  3 &  8 &    537 &   14.6 &   77.2 & 32.9 &  -37.8 \\
   0.6 &  4 &  8 &    537 &   14.6 &   77.2 & 32.9 &  -37.8 \\
\ldots \\
   0.6 & 12 &  8 &    537 &   14.6 &   77.2 & 32.9 &  -37.8 \\
\hline
\end{array}$$
\end{table}
}  

\newcommand\tresultsall{
\begin{table*}
\caption{Summary of estimates of optimal parameters $(l,b,\alpha,\phi)$ 
from WMAP 3yr ILC and TOH maps, as for Table~\protect\ref{t-ilc_kptwo}, 
for no mask, the kp2 mask and the kp0 mask.
Since multiple local maxima exist for the kp0 case, 
constraints on 
the twist parameter $\phi_1 \le \phi \le \phi_2$  are used in that case
(but not in other cases). Circle sizes $\alpha < 10.0\ddeg$ are excluded.
\label{t-results_all}}
$$\begin{array}{c c c  c c c  c c c c } \hline 
\rule[-1.5ex]{0ex}{5.0ex} 
\mbox{map} & \mbox{mask} & \mbox{$\phi$ constraint} &
\Pmin & N & j &
\lII \mbox{ in $\ddeg$} & \bII \mbox{ in $\ddeg$} & 
 \alpha \mbox{ in $\ddeg$} &
 \phi \mbox{ in $\ddeg$}
\\ \hline 
\rule{0ex}{4.0ex} 
\ILC & \mbox{(no mask)} & - &
0.4  & 10 &   4650 &  176.9 &   70.2 & 23.6 &   43.5 \\
&&&
 0.5  & 10 &   2795 &  178.5 &   68.6 & 23.1 &   43.5 \\
&&&
  0.6 & 10 &   1483 &  178.0 &   66.1 & 22.7 &   42.5 \\
&&&
   0.7  & 10 &    665 &  179.0 &   63.8 & 22.6 &   41.1 \\
\hline 
\rule{0ex}{4.0ex} 
\ILC & \kptwo &  
 - &
   0.4  & 23 &  11950 &  180.5 &   70.2 & 27.2 &   39.3 \\
&&&
   0.5  & 23 &   6632 &  181.9 &   69.1 & 27.1 &   39.1 \\
&&&
   0.6  & 23 &   3279 &  183.8 &   68.2 & 27.5 &   39.0 \\
&&&
   0.7  & 23 &   1282 &  184.8 &   68.9 & 27.1 &   38.9 \\
\hline 
\rule{0ex}{4.0ex} 
\ILC & \kpzero & 
0 \le \phi \le \pi &
   0.4  &   8 &   3079 &  172.4 &   75.2 & 27.4 &   46.8 \\
&&&
   0.5  &   8 &   1758 &  177.4 &   73.2 & 26.9 &   42.6 \\
&&&
   0.6  &   7 &    650 &  189.6 &   63.7 & 26.1 &   32.5 \\
&&&
   0.7  &    8 &    318 &  187.7 &   67.1 & 27.2 &   31.7 \\
\cline{4-10} 
\ILC & \kpzero & 
6\pi/5 \le \phi \le 8\pi/5 &
   0.4  &   7 &    265 &   23.0 &   79.4 & 20.1 &  -96.2 \\
&&&
   0.5  &  8 &   1039 &   21.5 &   59.6 & 14.5 &  -98.7 \\
&&&
   0.6  &   9 &   1675 &  294.3 &   84.0 & 22.7 &  -95.9 \\
&&&
   0.7  & 10 &   3290 &  279.1 &   83.2 & 23.2 &  -92.6 \\
\ILC & \kpzero & 
8\pi/5 \le \phi \le 2\pi &
   0.4  &  10 &   2469 &  349.4 &   79.4 & 31.9 &  -38.1 \\
&&&
   0.5  &   9 &   1176 &   10.6 &   79.1 & 34.8 &  -36.2 \\
&&&
   0.6  &  8 &    496 &   10.5 &   77.5 & 34.7 &  -36.6 \\
&&&
   0.7  &   8 &    232 &    7.7 &   78.0 & 36.6 &  -35.3 \\
\hline 
\rule{0ex}{4.0ex} 
\TOH &  \mbox{(no mask)} & 
 - &
{\bf NOT YET} \\
\hline 
\TOH & \kptwo  & 
 - &
   0.4  &  9 &   7731 &  183.7 &   70.8 & 29.7 &   34.5 \\
&&&
   0.5  &  9 &   4365 &  187.5 &   68.2 & 29.2 &   35.0 \\
&&&
   0.6  &   9 &   2406 &  190.7 &   68.4 & 29.7 &   35.4 \\
&&&
   0.7  &  9 &   1039 &  189.6 &   67.2 & 28.9 &   33.4 \\
\hline 
\rule{0ex}{4.0ex} 
\TOH & \kpzero & 
 - &
 0.4 &  9 &   4057 &  351.5 &   79.7 & 32.1 &  -40.2 \\
&&&  0.5 &  9 &   3038 &  183.0 &   77.8 & 29.3 &   35.4 \\
&&&  0.6 &   8 &   1673 &  173.7 &   77.2 & 30.6 &   40.8 \\
&&&  0.7 &   8 &    873 &  183.5 &   69.0 & 29.4 &   41.8 \\
\TOH & \kpzero & 
0 \le \phi \le \pi &
   0.4  &  8 &   4979 &  171.2 &   78.7 & 30.8 &   41.6 \\
&&&   0.5  &  8 &   3009 &  173.7 &   78.2 & 31.0 &   41.9 \\
&&&   0.6  &  8 &   1673 &  173.7 &   77.2 & 30.6 &   40.8 \\
&&&   0.7  &  8 &    873 &  183.5 &   69.0 & 29.4 &   41.8 \\
\TOH & \kpzero & 
6\pi/5 \le \phi \le 8\pi/5 &
   0.4  &  8 &   2312 &   37.1 &   82.6 & 33.2 &  -90.3 \\
&&&   0.5  &  7 &    829 &  336.9 &   83.1 & 29.6 &  -96.7 \\
&&&   0.6  &  7 &    454 &  346.6 &   84.8 & 28.6 &  -95.8 \\
&&&   0.7  &  7 &    155 &   58.9 &   78.6 & 28.1 &  -96.8 \\
\TOH & \kpzero & 
8\pi/5 \le \phi \le 2\pi &
   0.4  &  9 &   4057 &  351.5 &   79.7 & 32.1 &  -40.2 \\
&&&   0.5  &  9 &   2318 &  355.3 &   79.0 & 32.8 &  -39.7 \\
&&&   0.6  &  9 &   1227 &    3.0 &   79.0 & 33.0 &  -40.1 \\
&&&   0.7  &  9 &    563 &   10.5 &   78.4 & 36.0 &  -39.4 \\
\hline
\end{array}$$
\end{table*}
}  

\newcommand\trlcmb{
\begin{table}
\caption{Estimates of optimal parameters $(l,b,\alpha,\phi)$ 
from WMAP 3yr ILC maps, as for Table~\protect\ref{t-ilc_kptwo}, 
with the kp2 mask, but starting each time at the point in parameter
space of the \protect\nocite{RLCMB04}{Roukema} {et~al.} (2004) suggested PDS orientation and
size.  The minimum circle size is $\alpha = 5.0\ddeg$.
\label{t-rlcmb}}
$$\begin{array}{  c c c  c c c c } \hline 
\rule[-1.5ex]{0ex}{4.5ex}
\Pmin & N & j &
\lII \mbox{ in $\ddeg$} & \bII \mbox{ in $\ddeg$} 
& \alpha \mbox{ in $\ddeg$}
& \phi \mbox{ in $\ddeg$}
\\ \hline 
 0.6 & 6 &   1183 &  179.7 &   65.6 & 27.3 &   37.0  {\bf NOT FINAL} \\ 
 0.5 &  6 &   2181 &  179.7 &   65.0 & 26.8 &   37.3 \\
\hline
\end{array}$$
\end{table}
}  

\newcommand\fomm{
\begin{figure}
\centering 
\includegraphics[width=8cm,bb=0 0 493 383]{fifj.eps} 
\caption[]{ \mycaptionfont
Dependence of $\Omm$ on $\alpha$, similarly to 
The slope in the range $5\ddeg < \alpha < 15\ddeg$ is 
$\mbox{d} \Omm /\mbox{d}\alpha  \approx 2.2\e{-3}~\mbox{deg}^{-1}$,
with only weak dependence on $\Omtot$ over this interval.
}
\label{f-omm}
\end{figure} 
} 

\newcommand\fthreedcorr{
\begin{figure}
\centering 
\includegraphics[width=8cm]{corr3d.eps} 
\caption[]{ \mycaptionfont
Pairs of comoving spatial points used for auto- and cross-correlation function 
calculations in the 
comoving covering space (apparent space) diagram of three overlapping copies of the
surface of last scattering (SLS, a sphere of radius $R$) showing the intersection 
of the three copies. A single copy of the SLS is shown by a continuous arc, 
and two copies of the SLS, mapped by isometries $g_1$ and $g_2$ which generate
the 3-manifold from the covering space, are shown as dashed arcs. The 
auto-correlation function for the {\em simply} connected hypothesis, 
$\xisc$ (Eq.~\protect\ref{e-xisc-defn}) includes pairs
such as $(x_1, x_2)$,
separated by $ d\left( x_{i_1}, x_{i_2} \right) = r$,
shown here by a continous line segment joining the two
points, in which case both points are on a single
copy of the SLS.
The cross-correlation 
function for the {\em multiply} connected hypothesis includes pairs
$(x_{i_1}, g_j(x_{i_2}))$,
separated by $d\left( x_{i_1}, \left[g_j({x_{i_2}})\right] \right) = r$,
where $g_i$ is a holonomy transformation which is not the 
identity. For illustrative
purposes, here we show the pairs
$(x_1, g_2(x_3)),$
$(x_1, g_1(x_4)),$
$(x_1, g_2(x_5)),$
$(x_1, g_1(x_6))$ by joining them with dashed line segments.
Each such pair contributes to the mean product of temperature fluctuations
at separations $r$, either $\xisc(r)$ or $\ximc(r)$.
}
\label{f-threedcorr}
\end{figure} 
} 

\newcommand\fcauto{
\begin{figure}
\centering 
\includegraphics[width=6cm]{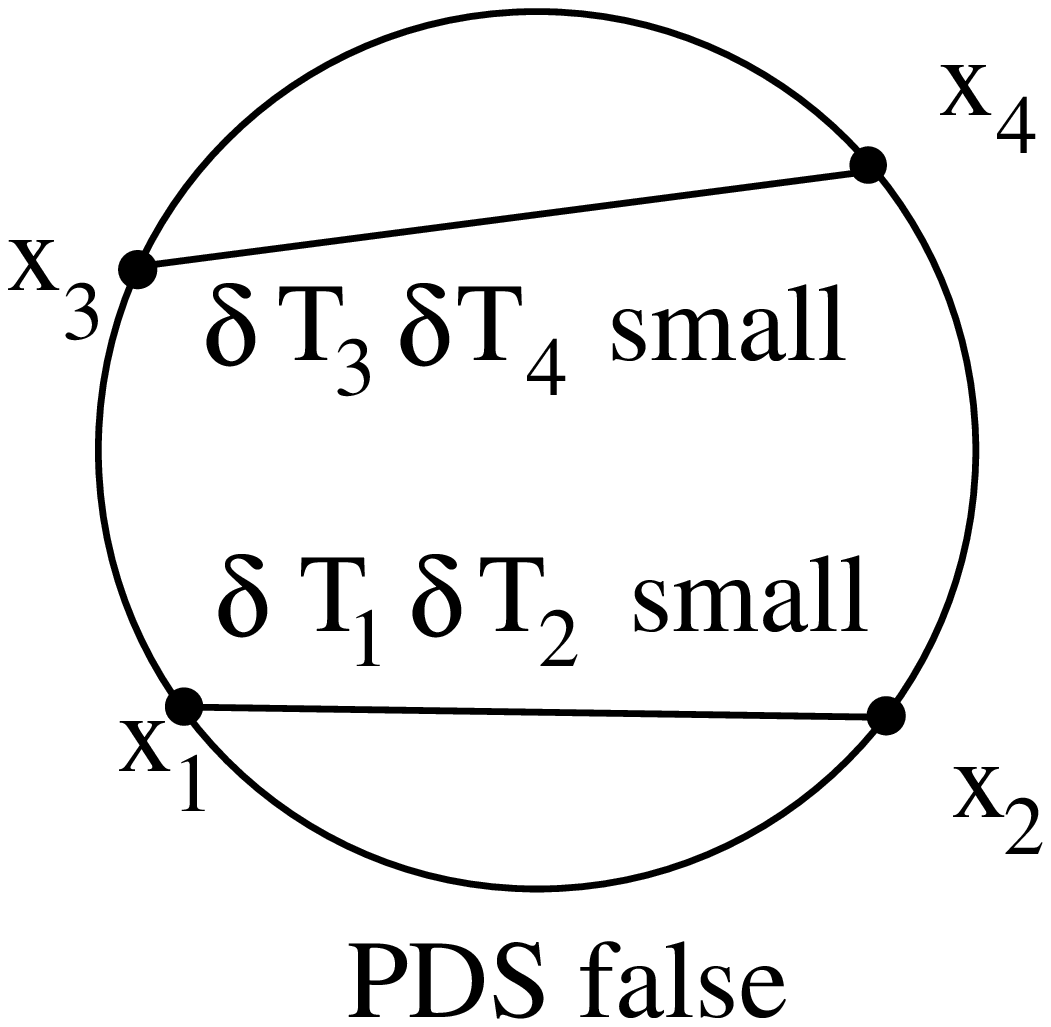}
\caption[]{ \mycaptionfont
Schematic diagram of temperature {\em auto}-correlations
in a simply connected universe, showing
``typical'' temperature correlations $\delta T_i \delta T_j$ for 
two pairs of points $(i,j)= (1,2), (3,4)$ that are widely separated
on a single copy of the surface of last scattering (SLS).
Since the correlation is in general high at low separations and low at
high separations, the product  $\delta T_i \delta T_j$ is low in both
cases.
}
\label{f-c_auto}
\end{figure} 
} 

\newcommand\fccross{
\begin{figure}
\centering 
\includegraphics[width=8cm]{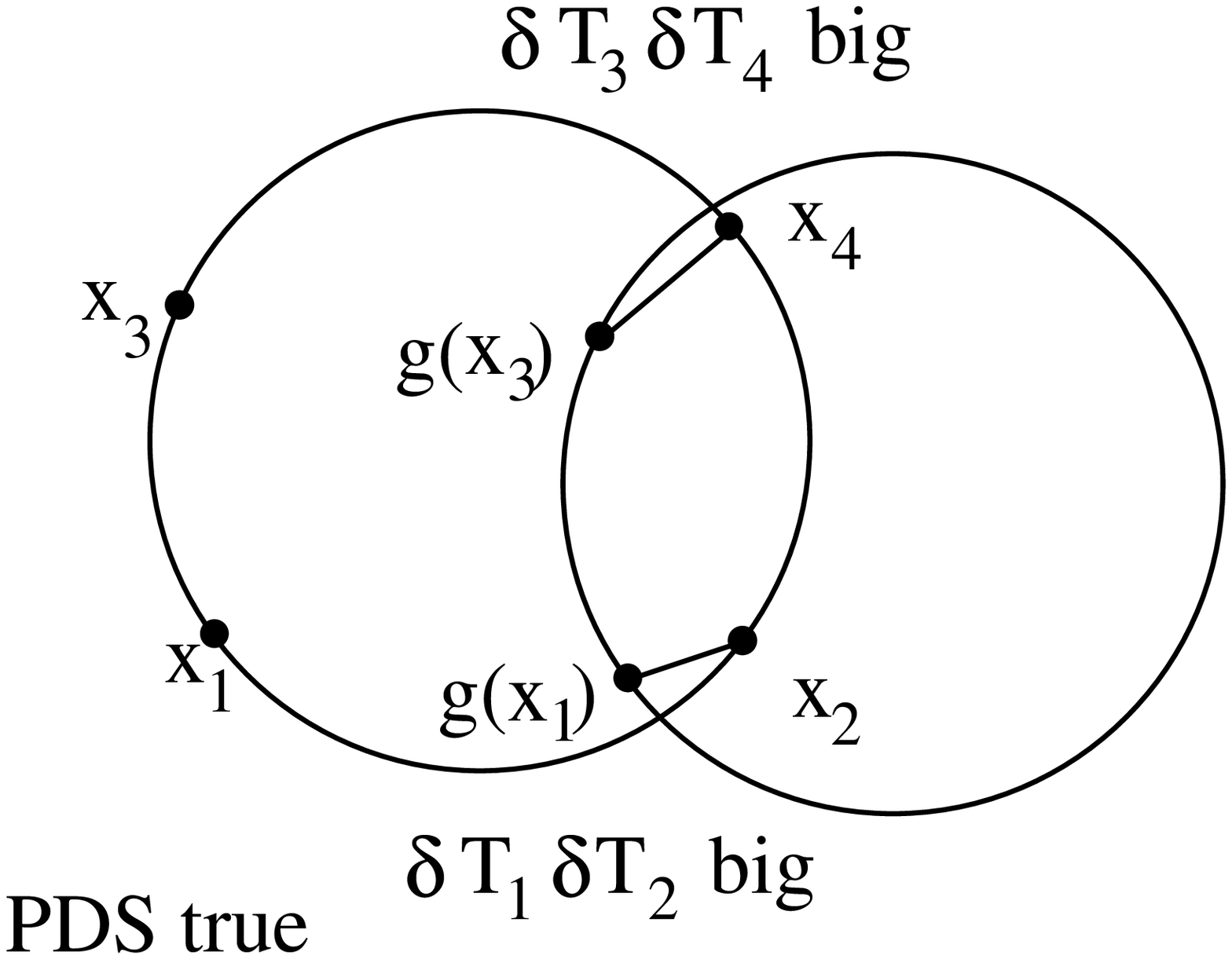}
\caption[]{ \mycaptionfont
Schematic diagram of temperature {\em cross}-correlations
in a multiply connected universe (e.g. a Poincar\'e dodecahedral space, PDS,
model), showing
``typical'' temperature correlations $\delta T_i \; \delta T_j$ for 
two pairs of points $(i,j)= (1,2), (3,4)$ that are widely separated
on a single copy of the SLS, as in 
Fig.~\protect\ref{f-c_auto}. The holonomy transformation $g$ maps points in space
to copies of themselves in the covering space. The points $x_1$ and $g(x_1)$ 
are the same physical point,
and similarly, $x_3$ and $g(x_3)$ are physically identical.
Since the comoving separations 
$d[g(x_1),x_2]$ and $d[g(x_3),x_4]$ are both small, 
and the correlation is, in general, high at low separations,
the product  $\delta T_i \delta T_j$ is high in both
cases. This should only occur if the holonomy transformation $g$ 
is physically correct.
This is described statistically in Eq.~(\protect\ref{e-ximc-pdsgood}).
}
\label{f-c_cross}
\end{figure} 
} 

\newcommand\fkeyangles{
\begin{figure}
\centering 
\includegraphics[width=8cm]{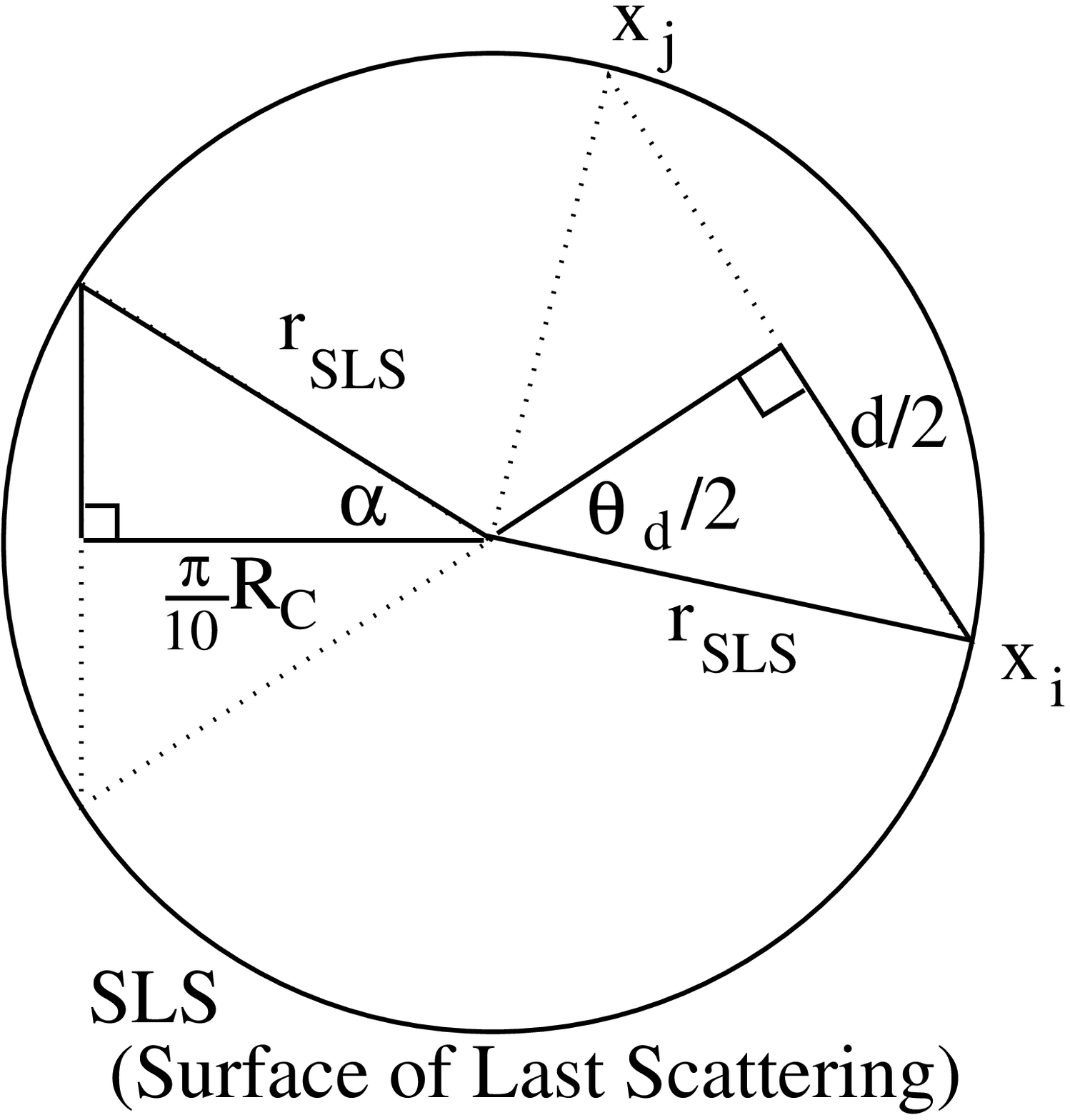}
\caption[]{ \mycaptionfont
\posteditorchanges{Surface of last scattering and its interior, i.e. the 
comoving 3-volume of the observable Universe as a subset of $S^3$, 
embedded in $\mathbb{R}^4$
for convenience.}
All lines appearing
straight in this figure correspond to spatial geodesics in $S^3$,
i.e. arcs in $\mathbb{R}^4$. 
The triangle to the left illustrates the geometry of a matched
circle. In the $\mathbb{R}^4$ embedding, 
the PDS mapping from one matched circle to another rotates
by $2\pi/10 = 36\ddeg$ between circle centres, so an arclength of half
this (from the observer to one matched circle centre), i.e. 
$\pi/10$ multiplied by the 3-sphere radius, $R_C$, is shown. 
The circle radius as measured on the SLS is shown as $\alpha$.
The relevant equation describing this triangle 
is Eq.~(\protect\ref{e-alpha-tri}).
The right-hand 
part of the figure shows half the geodesic distance $d/2$
between two points $x_i, x_j$ lying on the SLS and the angle $\theta_d/2$
subtending this. 
The relevant equation describing this triangle 
is Eq.~(\protect\ref{e-d-theta-tri}).
\posteditorchanges{The 
relations between angles and arclengths of the sides of these triangles
are determined by spherical
trigonometry.}
}
\label{f-keyangles}
\end{figure} 
} 

\newcommand\fdiscs{
\begin{figure}
\centering 
\includegraphics[width=8cm]{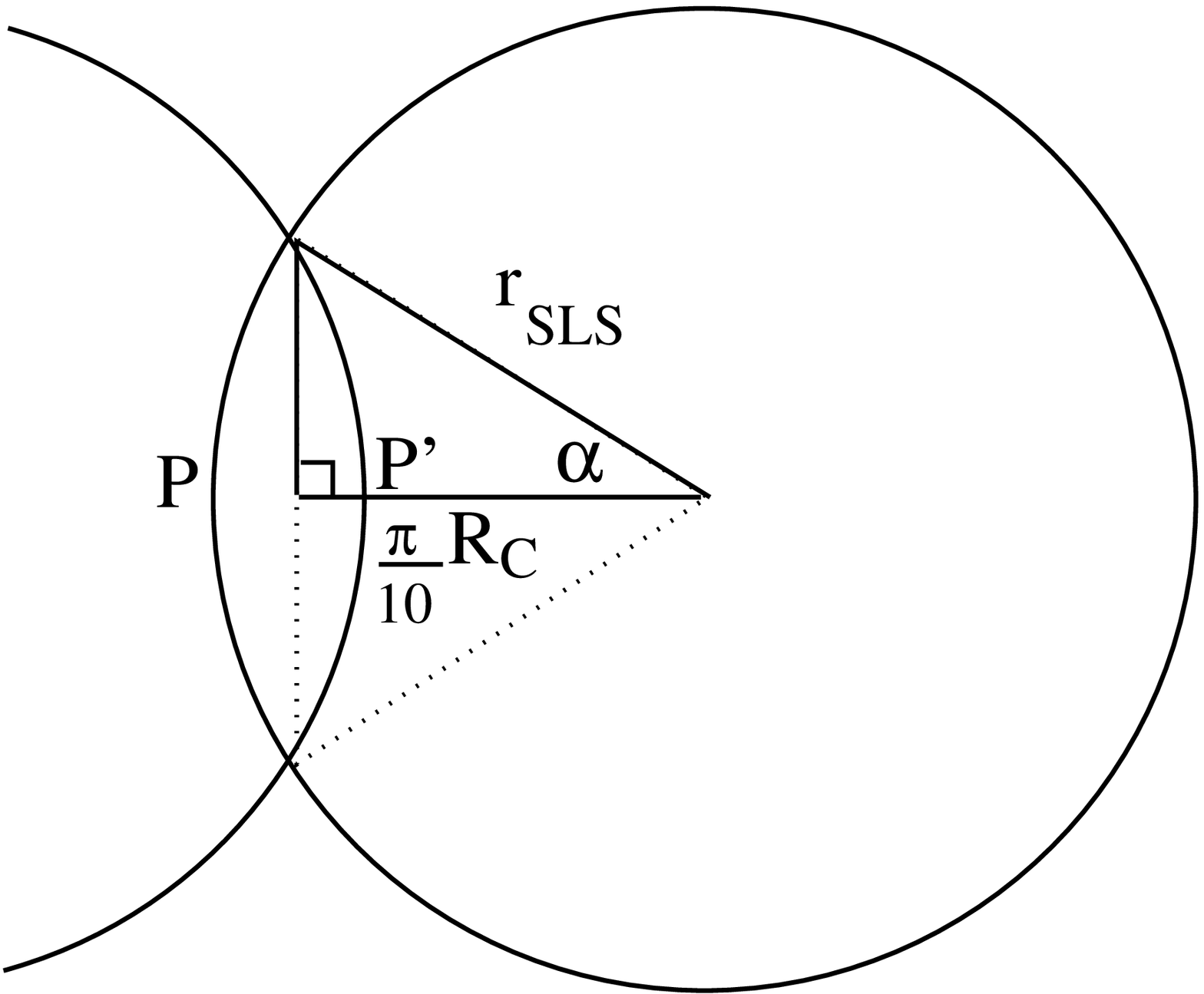}
\caption[]{ \mycaptionfont
Schematic diagram showing the approximately parallel
positioning of two ``approximately matched annuli'' or ``approximately 
matched discs'' in two copies of the SLS,
as per Fig.~\protect\ref{f-keyangles}, 
when 
the matched circle angular radius $\alpha$ is relatively small.
The centres of the two matched circles projected onto their respective
spheres (copies of the SLS), i.e. P and P', are separated by 
$2 [ \rSLS - (\pi/10)\; R_C ]$.
}
\label{f-discs}
\end{figure} 
} 

\newcommand\filclbthN{
\begin{figure}
\centering 
\includegraphics[width=6cm]{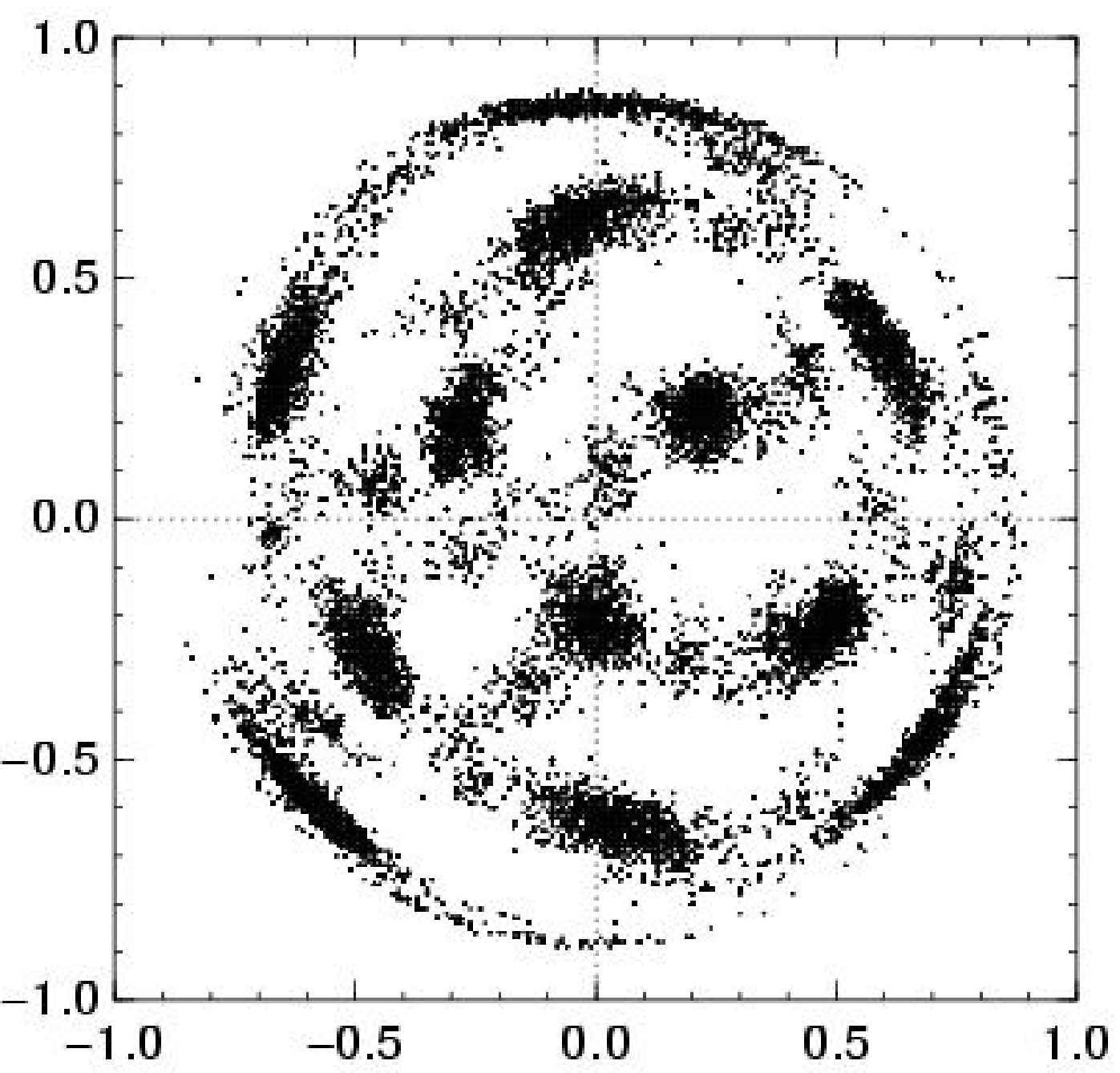}
\includegraphics[width=6cm]{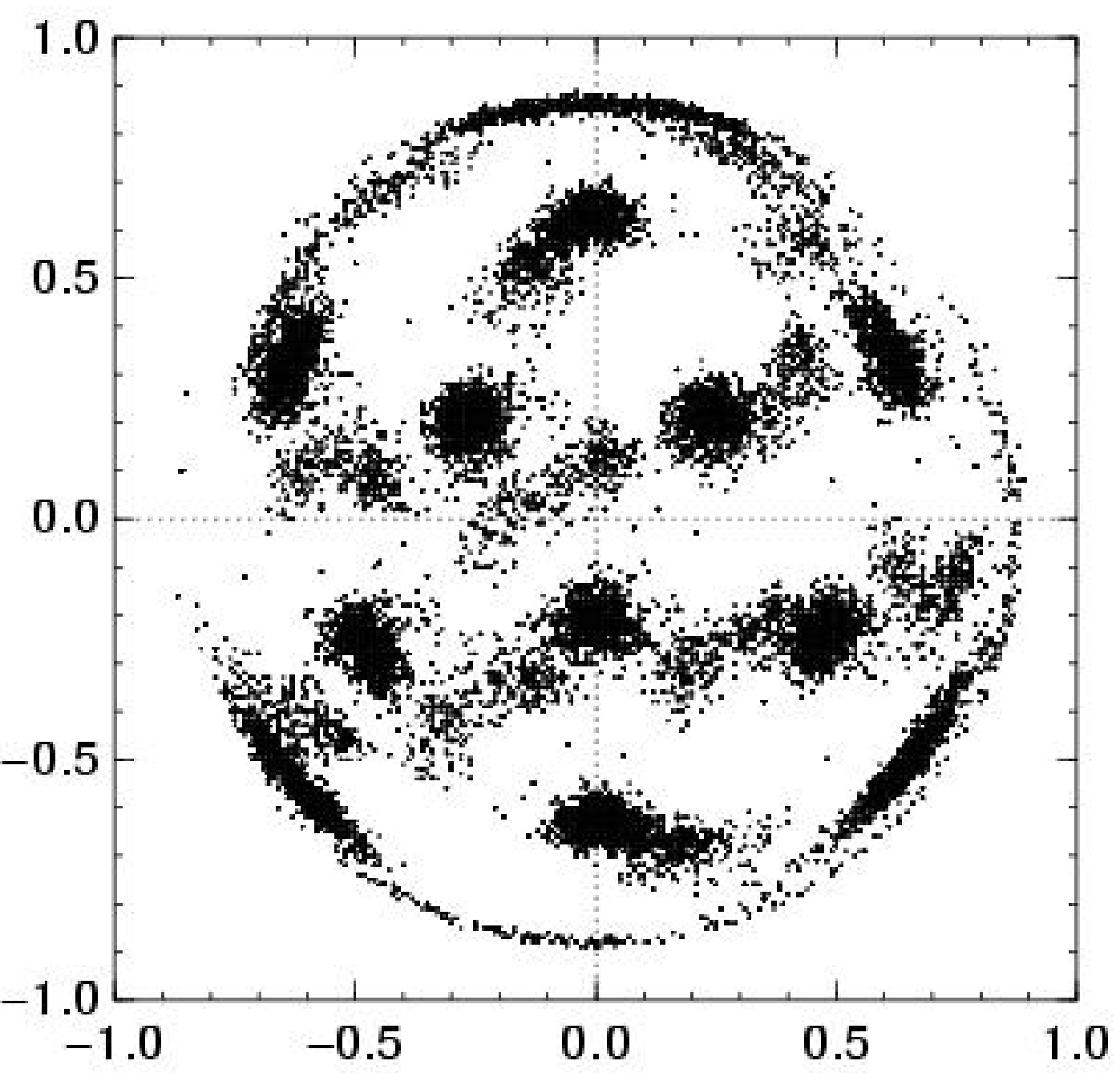}
\caption[]{ \mycaptionfont Full sky map showing the optimal
  orientation of dodecahedral face centres based on 100,000 steps in
  10 MCMC chains, using the ILC map and either the kp2 mask (upper panel)
  or no mask (lower panel), showing face centres
  for which $P > 0.5$ (see Eq.~(\protect\ref{e-prob-defn})). The
  projection is a Lambert azimuthal equal area projection
  \protect\nocite{Lambert1772}({Lambert} 1772) of the full sky, centred on the North
  Galactic Pole (NGP). The $0\ddeg$ meridian is the positive vertical axis
  and galactic longitude increases clockwise.   
  \posteditorchanges{These face centres are derived from the MCMC chains
  {\em without any constraint on the twist phase $\phi$}.}
}
\label{f-ilc_lbth_N}
\end{figure} 
} 

\newcommand\filclbthS{
\begin{figure}
\centering 
\includegraphics[width=6cm]{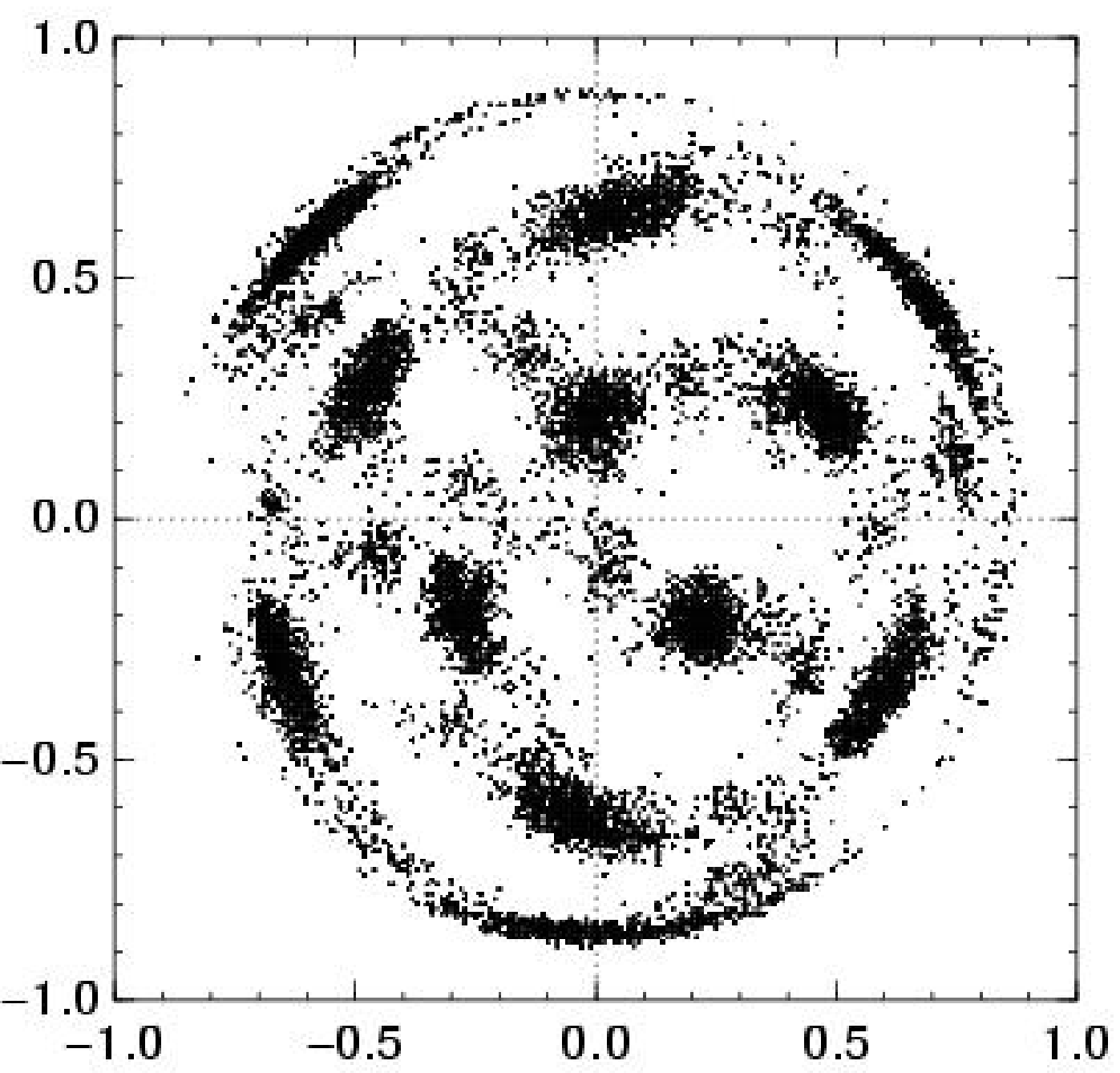}
\includegraphics[width=6cm]{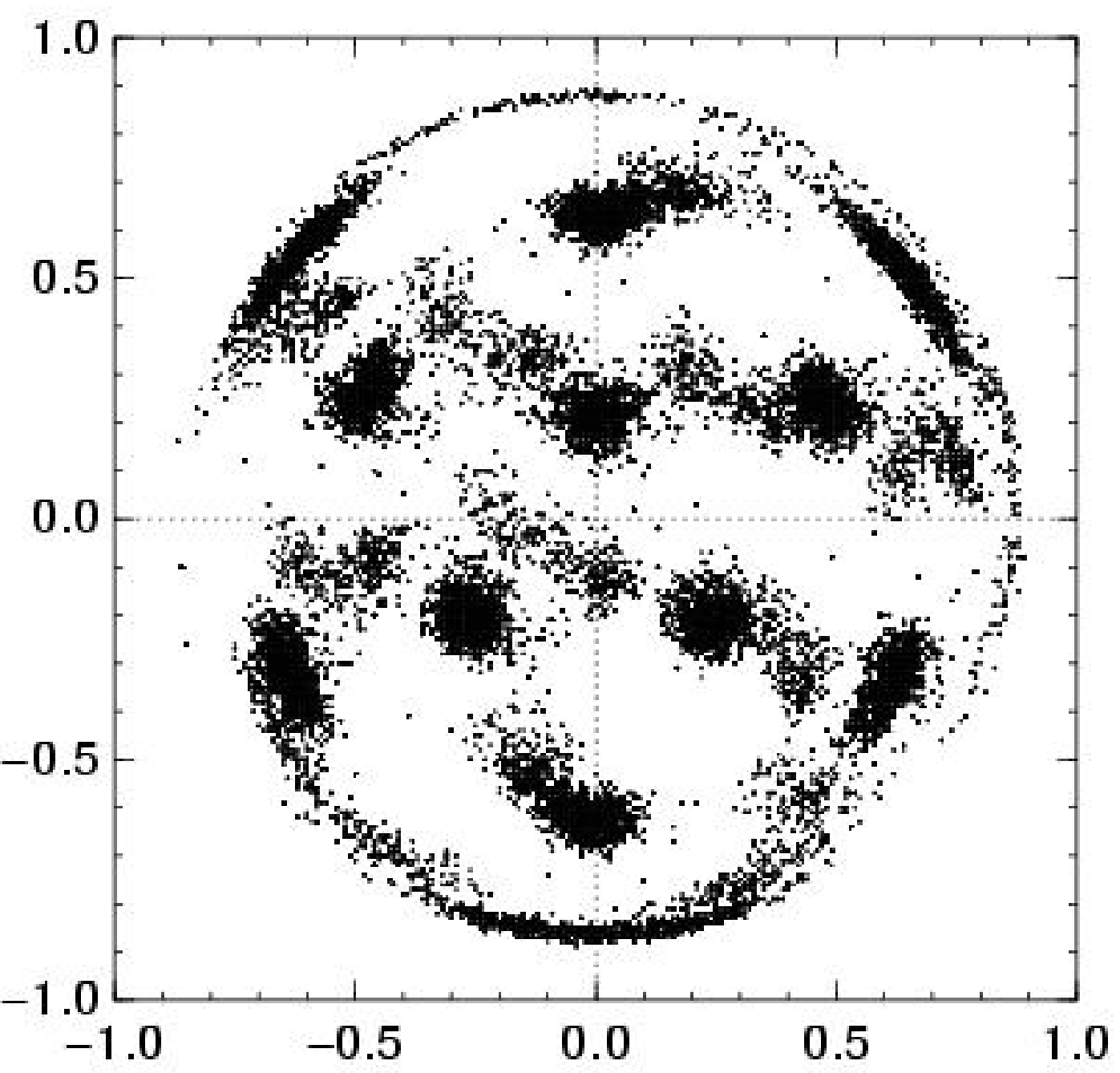}
\caption[]{ \mycaptionfont Full sky map showing the optimal
  dodecahedral face centres for the ILC map with the kp2 mask (upper panel)
  and no mask (lower panel), as for Fig.~\protect\ref{f-ilc_lbth_N},
  but centred on the South Galactic Pole.  The $0\ddeg$ meridian is
  the negative vertical axis and galactic longitude increases
  anticlockwise. 
 }
\label{f-ilc_lbth_S}
\end{figure} 
} 

\newcommand\ftohlbthN{
\begin{figure}
\centering 
\includegraphics[width=6cm]{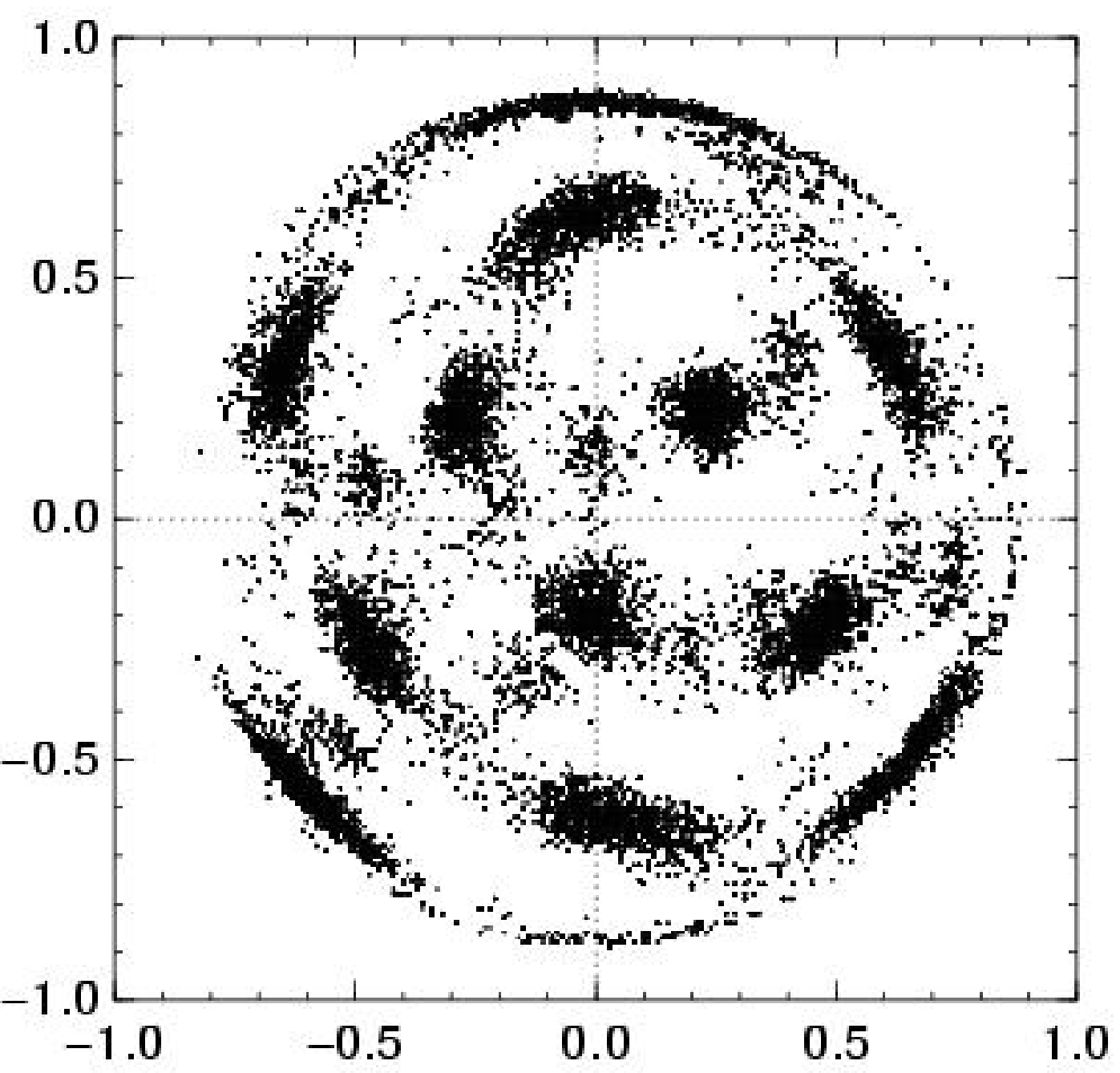}
\includegraphics[width=6cm]{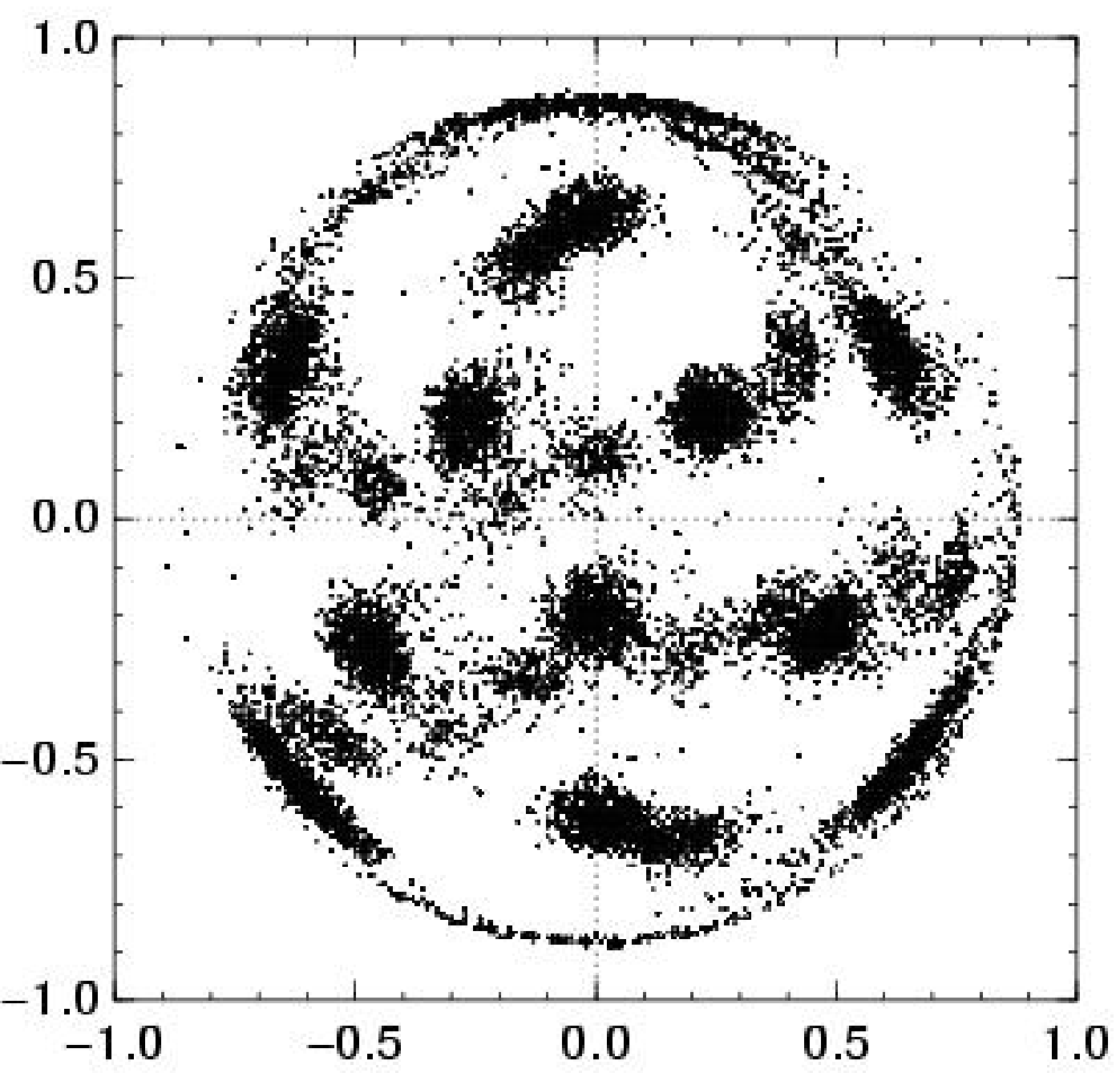}
\caption[]{ \mycaptionfont Full sky map showing the optimal dodecahedral
  orientation for the TOH map with the kp2 mask (upper panel)
  and no mask (lower panel), as for Fig.~\protect\ref{f-ilc_lbth_N},
  centred on the North Galactic Pole.
}
\label{f-toh_lbth_N}
\end{figure} 
} 

\newcommand\ftohlbthS{
\begin{figure}
\centering 
\includegraphics[width=6cm]{toh_kp2_SGP.ps}
\includegraphics[width=6cm]{toh_nomask_SGP.ps}
\caption[]{ \mycaptionfont Full sky map showing the optimal dodecahedral
  orientation for the TOH map with the kp2 mask (upper panel)
  and no mask (lower panel), as for Fig.~\protect\ref{f-ilc_lbth_S},
  centred on the South Galactic Pole.
}
\label{f-toh_lbth_S}
\end{figure} 
} 

\newcommand\frlcmb{
\begin{figure}
\centering 
\includegraphics[width=6cm]{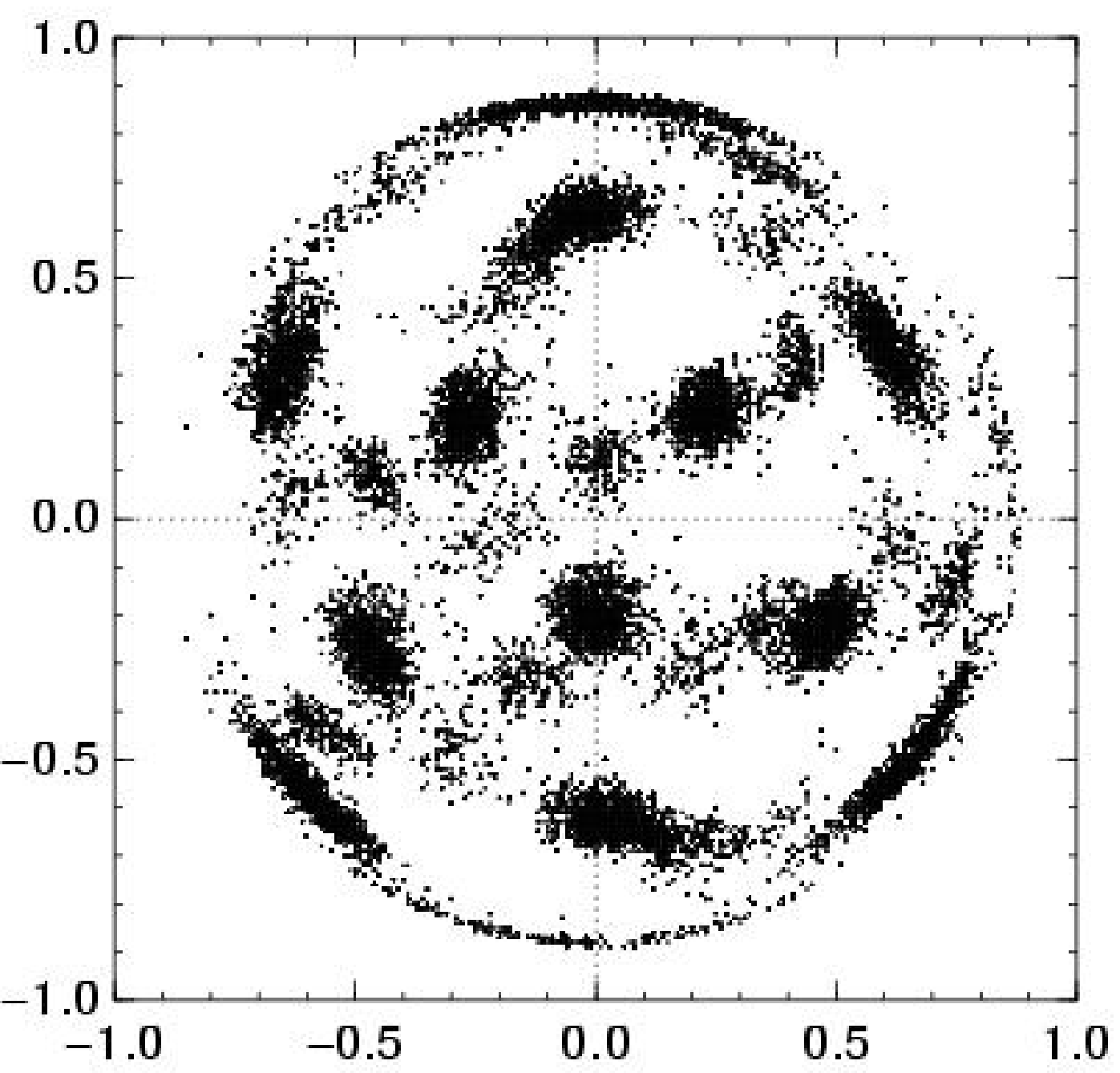}
\caption[]{ \mycaptionfont Full sky map showing the optimal
  dodecahedral orientation for the ILC map with the kp2 mask, as for
  Fig.~\protect\ref{f-ilc_lbth_N}, centred on the North Galactic Pole,
  from an MCMC chain starting at
  the PDS orientation and circle size suggested in
  \protect\nocite{RLCMB04}{Roukema} {et~al.} (2004), for an initial twist of $\phi=-\pi/5$. The
  optimal orientation is clearly very close to what is 
  found from arbitrary initial
  positions, shown in the previous figures.
}
\label{f-rlcmb}
\end{figure} 
} 

\newcommand\filclbthkpzero{
\begin{figure}
\centering 
\includegraphics[width=6cm]{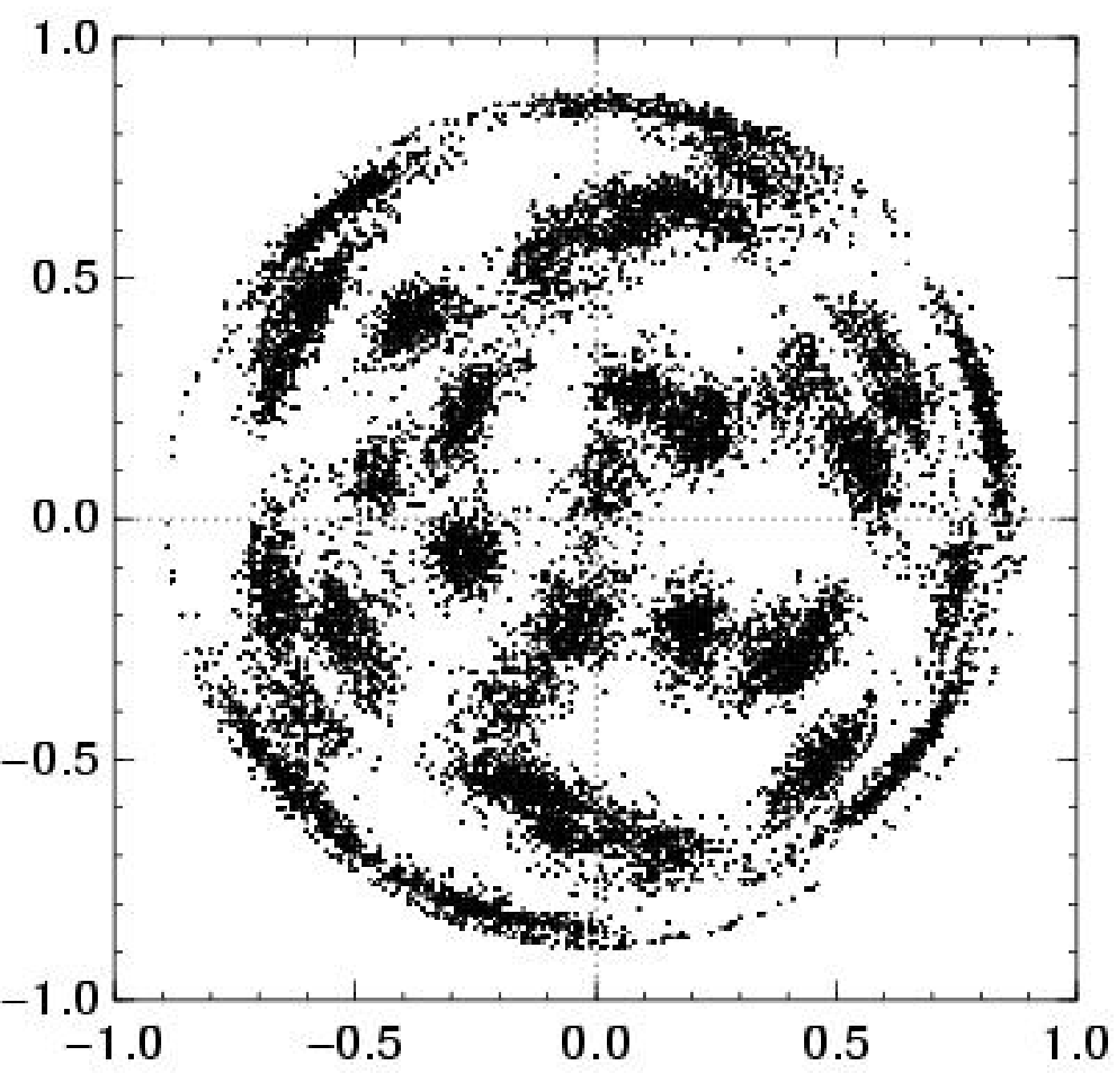}
\caption[]{ \mycaptionfont Full sky map showing the optimal
  dodecahedral orientation for the ILC map with the kp0 galactic contamination mask, as for
  Fig.~\protect\ref{f-ilc_lbth_N}, centred on the North Galactic Pole.
}
\label{f-ilc_lbth_kpzero}
\end{figure} 
} 

\newcommand\ftohlbthkpzero{
\begin{figure}
\centering 
\includegraphics[width=6cm]{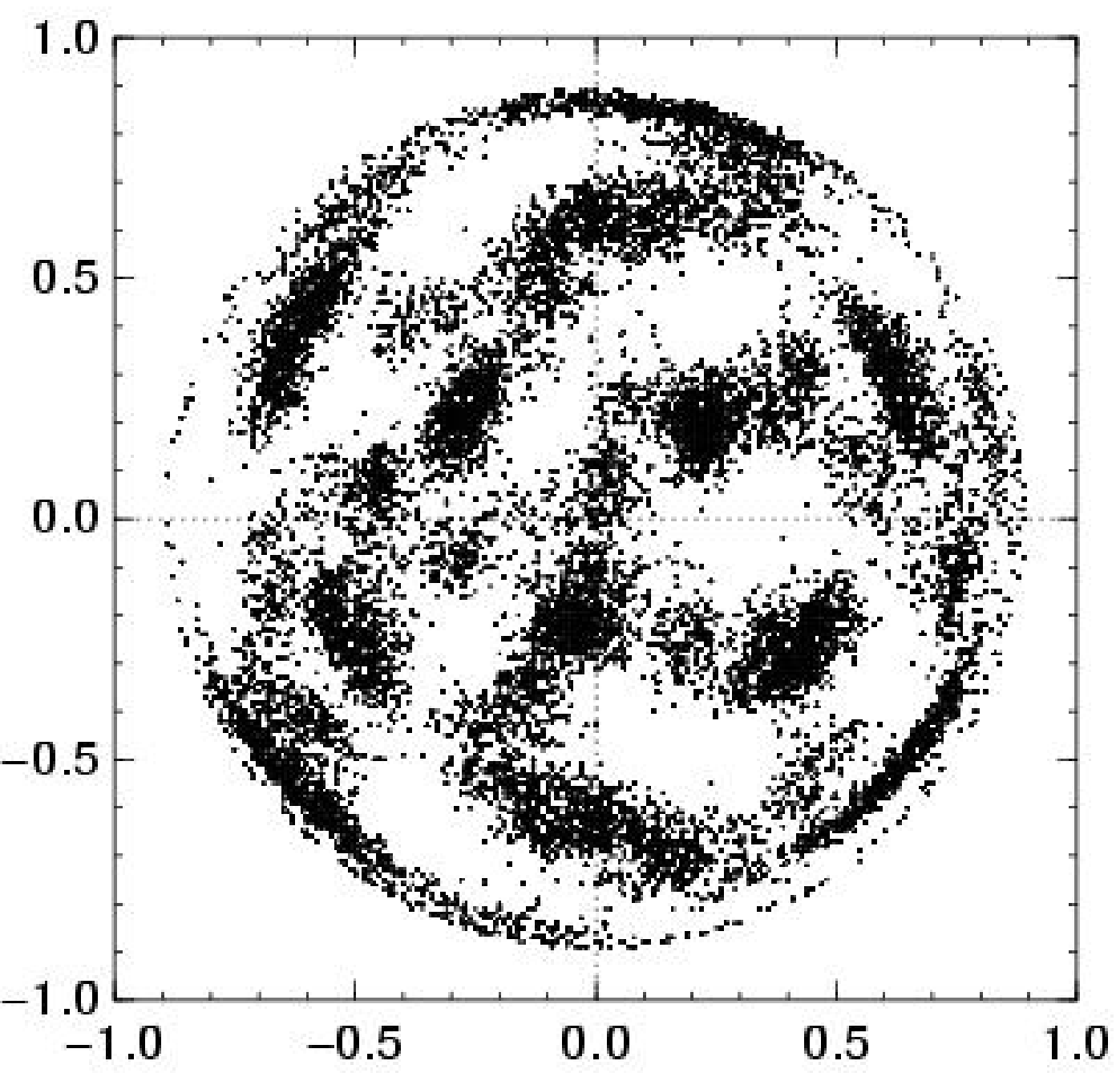}
\caption[]{ \mycaptionfont Full sky map showing the optimal
  dodecahedral orientation for the TOH map with the kp0 galactic contamination mask, as for
  Fig.~\protect\ref{f-ilc_lbth_N}, centred on the North Galactic Pole.
}
\label{f-toh_lbth_kpzero}
\end{figure} 
} 

\newcommand\fwmapQ{
\begin{figure}
\centering 
\includegraphics[width=6cm]{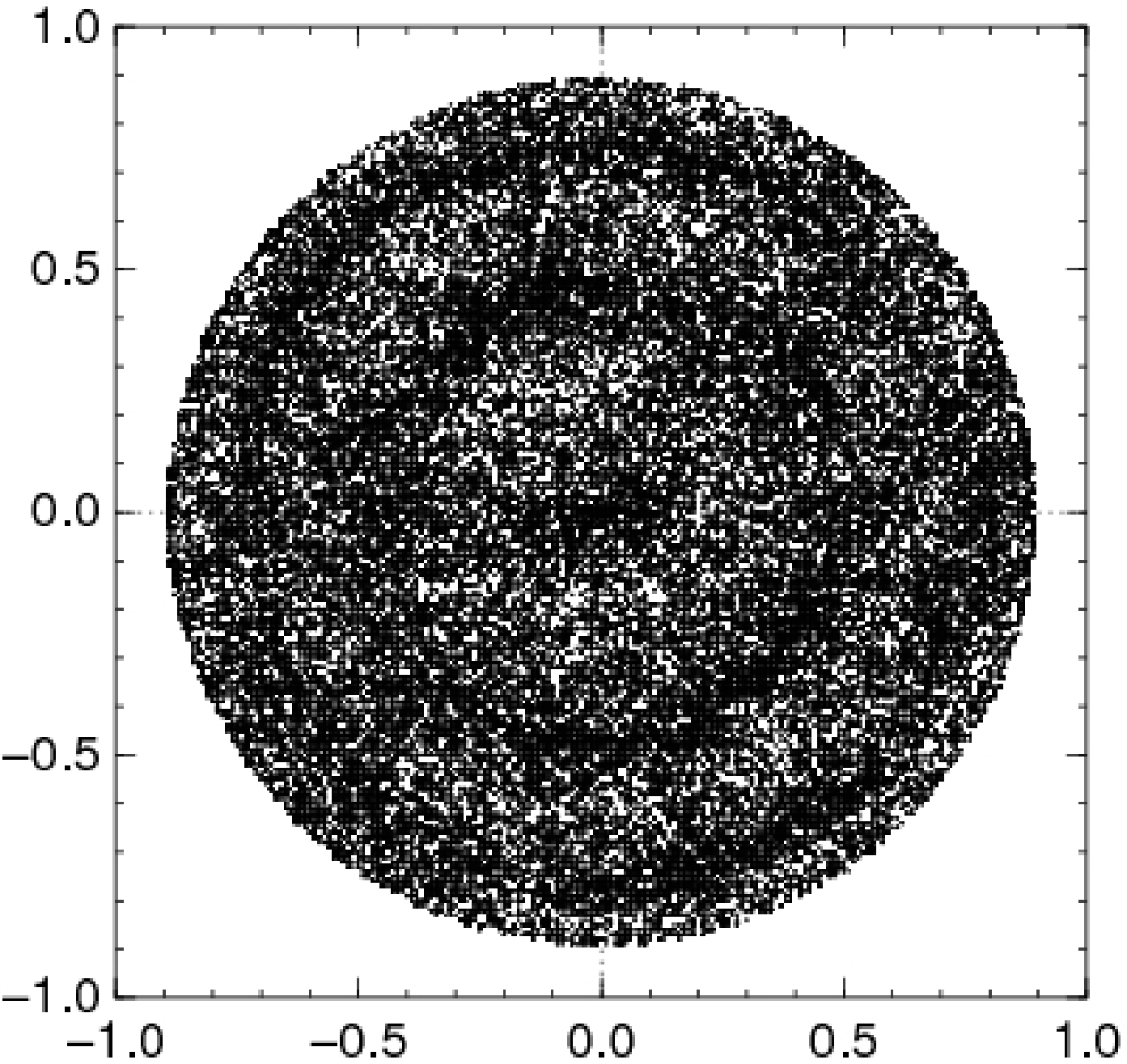}
\includegraphics[width=6cm]{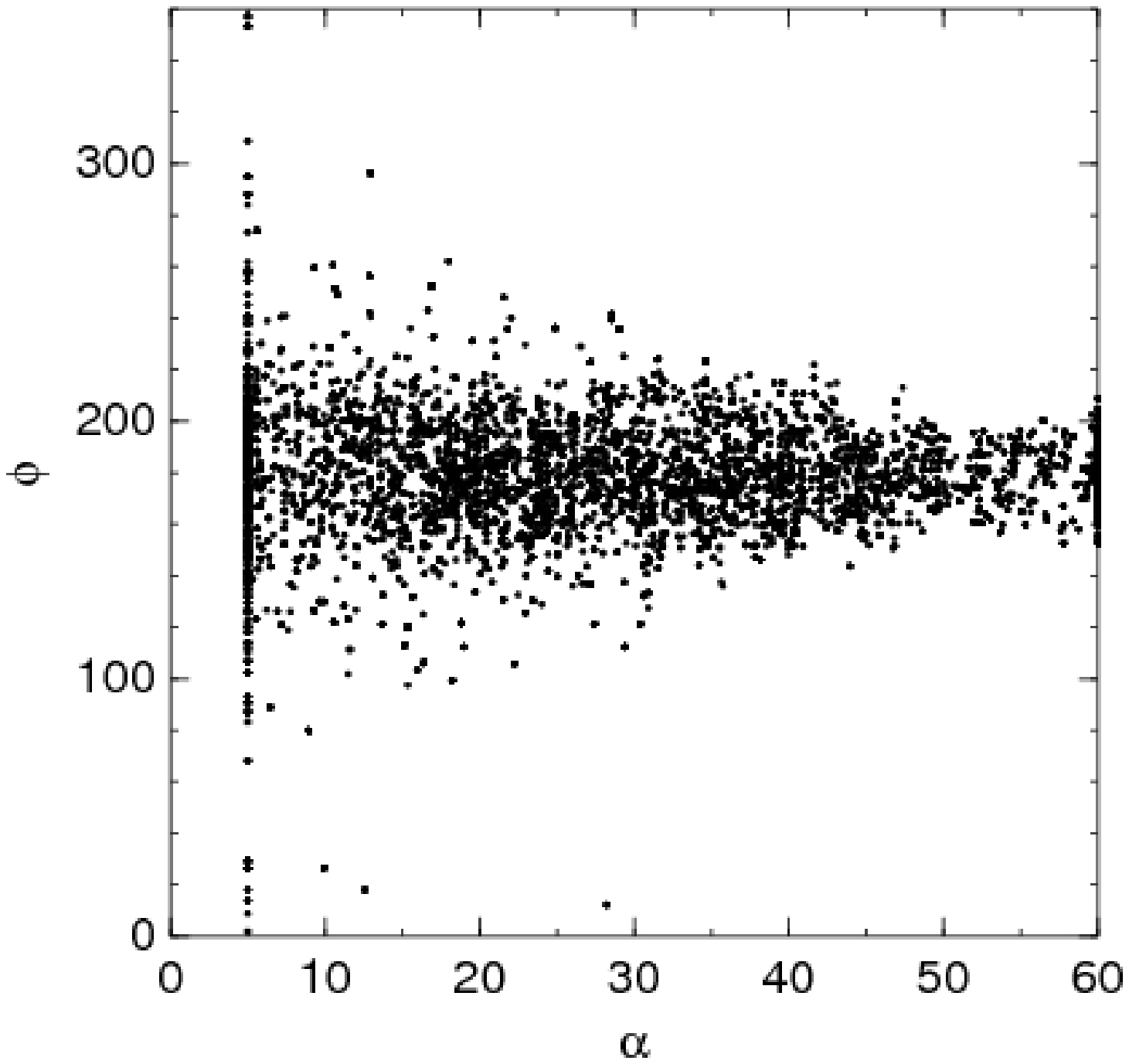}
\caption[]{ \mycaptionfont Full sky map showing the optimal
  dodecahedral orientation for $\Pmin > 0.5$,  as in 
  Fig.~\protect\ref{f-ilc_lbth_N}, centred on the North Galactic Pole
  (upper panel), 
  and the corresponding $\alpha, \phi$
  values in degrees (lower panel) for the WMAP Q frequency band map, 
  in which the Galaxy very strongly dominates, with no galactic mask. 
  Clearly, the signal is very different from what is found in 
  the foreground corrected maps.
}
\label{f-wmapQ}
\end{figure} 
} 

\newcommand\fgplanesig{
\begin{figure}
\centering 
\includegraphics[width=6cm]{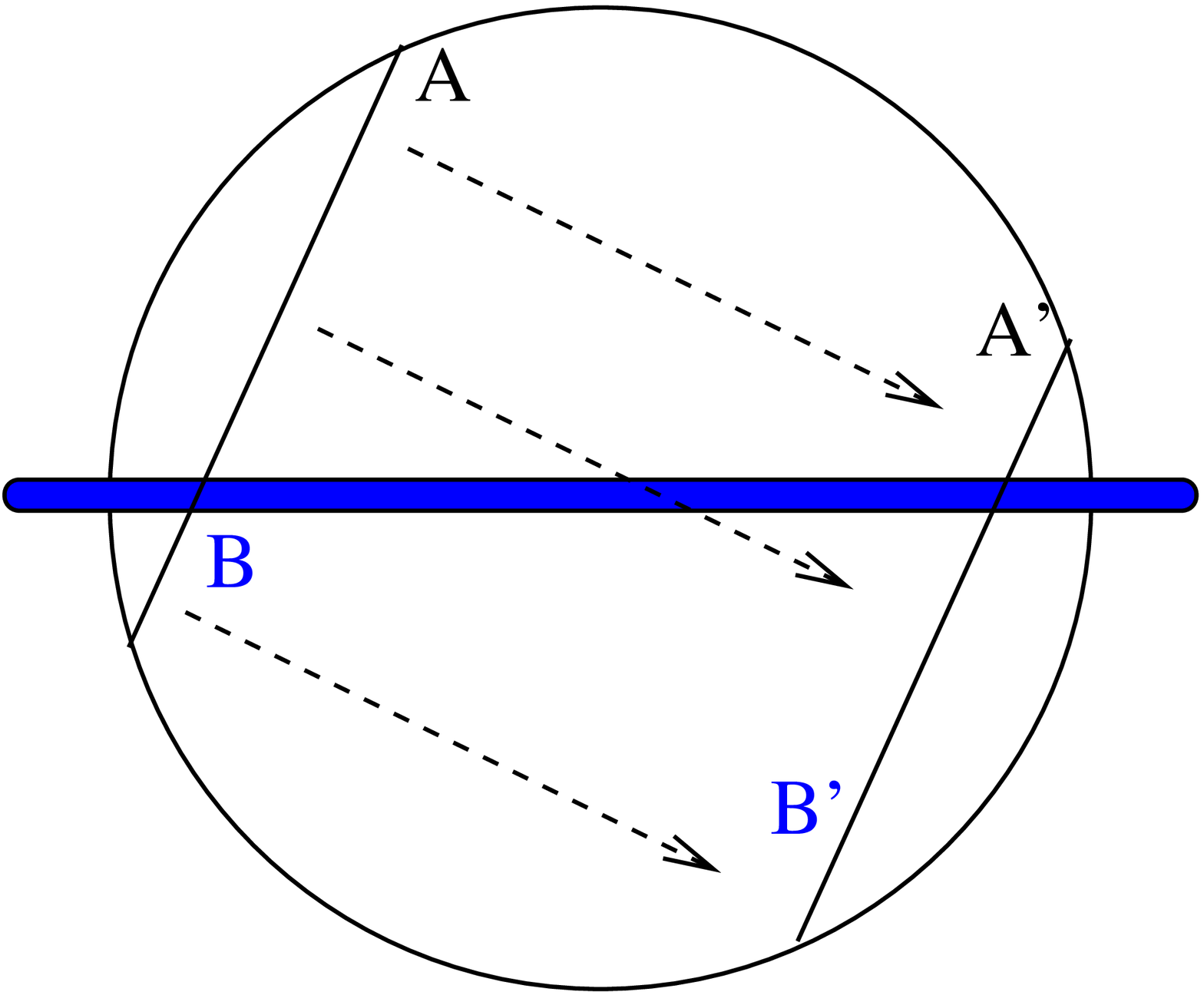}
\caption[]{ \mycaptionfont Schematic derivation of the 
preferred twist of a matched circles
pair implied by a map dominated by the Galactic Plane (GP). 
The SLS is shown by a circle. The GP is shown as
a horizontal edge-on, solid disc. The sky map can be thought of
as a binary map, zero everywhere except at the GP where it is positively
valued. One copy, AB, of a matched circle
is shown to the left, at an arbitrary angle with respect to the GP, in 
projection, containing point A, far from the GP, and B,
intersecting with the GP. The circle intersects
with the GP at only two points, B and another point behind B
in this projection.
Dashed arrows indicate the translation (with no twist) 
from one side of the sky to the other, so that A maps to A', 
B maps to B'. 
\posteditorchanges{See \SSS\protect\ref{s-gal-expected}.}
}
\label{f-gplane_sig}
\end{figure} 
} 

\newcommand\fcalib{
\begin{figure}
\centering 
\includegraphics[width=6cm]{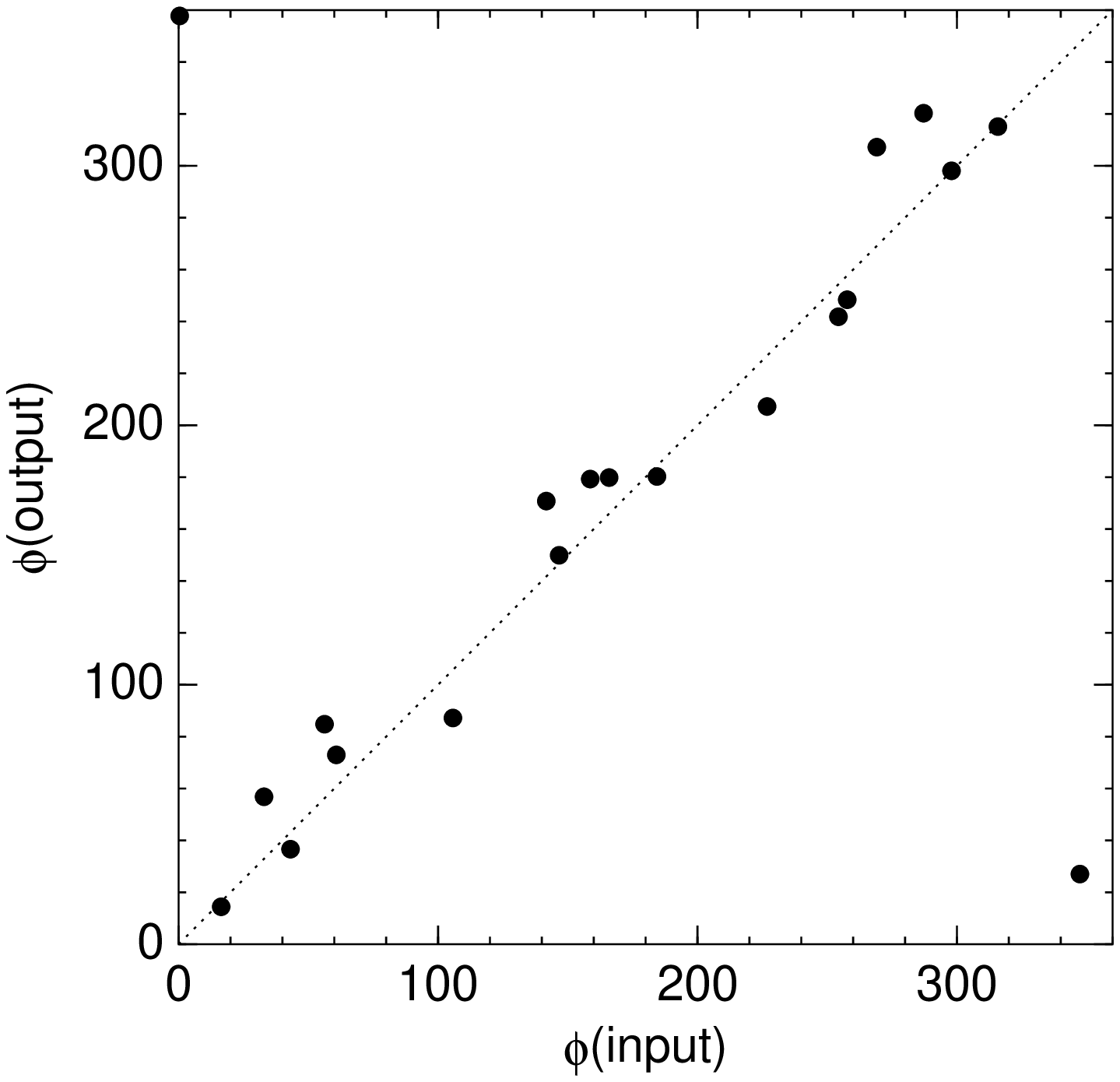}
\caption[]{ \mycaptionfont Comparison of simulated ``generalised''
PDS maps with random input twists $\phii$, to  
the estimated (output) twists $\phio$,
both in degrees. The two correlate well, with no sign of any 
tendency to favour $\pm 36\ddeg$.  The r.m.s. difference between 
$\phii$ and $\phio$ is $20.2\ddeg$.
See \SSS\protect\ref{s-toys} for details.
}
\label{f-calib}
\end{figure} 
} 

\newcommand\fauto{
\begin{figure}
\centering 
\includegraphics[width=6cm]{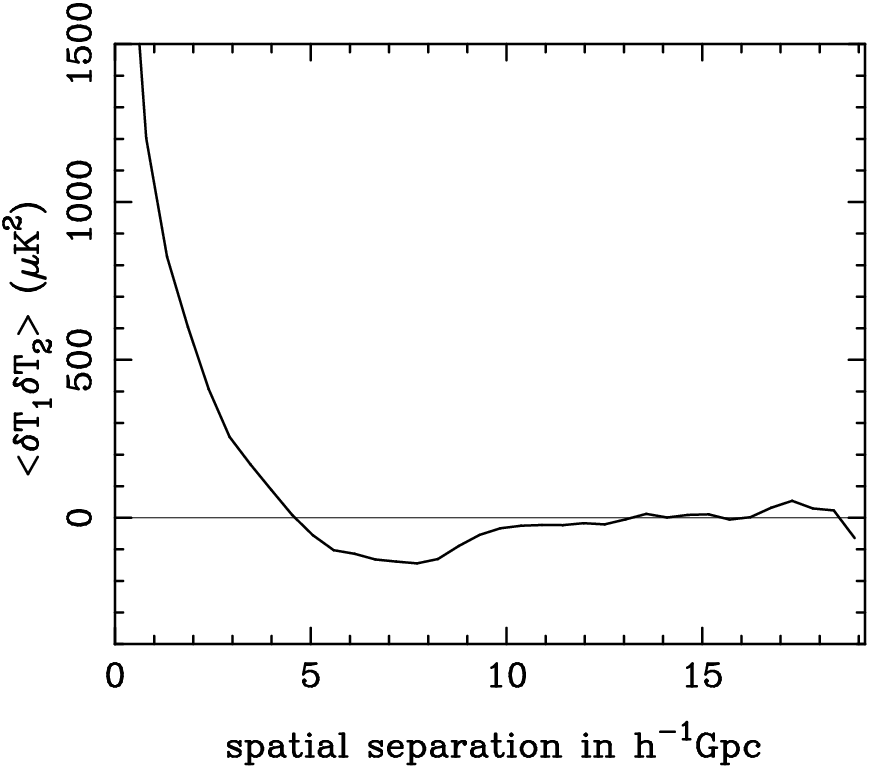}
\caption[]{ \mycaptionfont Estimate of the spatial 
auto-correlation function, $\xisc$, of temperature fluctuations
in the ILC WMAP map.
The estimate assumes a simply connected space with
$\Omtot \approx 1.0$, $\Omm \approx 0.3$, using 30,000 points
randomly chosen from a uniform distribution on the 2-sphere, excluding 
points falling within the kp2 Galaxy mask.
Correlations are shown in $\mu K^2$ against comoving separation
in {\hGpc} ranging from zero separation up to the diameter
of the SLS. 
An approximate conversion from spatial to angular separations
on small scales ($ d \ltapprox 4${\hGpc}) can be obtained by setting 
$19.2\hGpc = \mbox{2 rad} = 360\ddeg/\pi$, i.e. 
$1\hGpc \approx 6.0\ddeg$.  
}
\label{f-auto}
\end{figure} 
} 

\newcommand\fcross{
\begin{figure}
\centering 
\includegraphics[width=6cm]{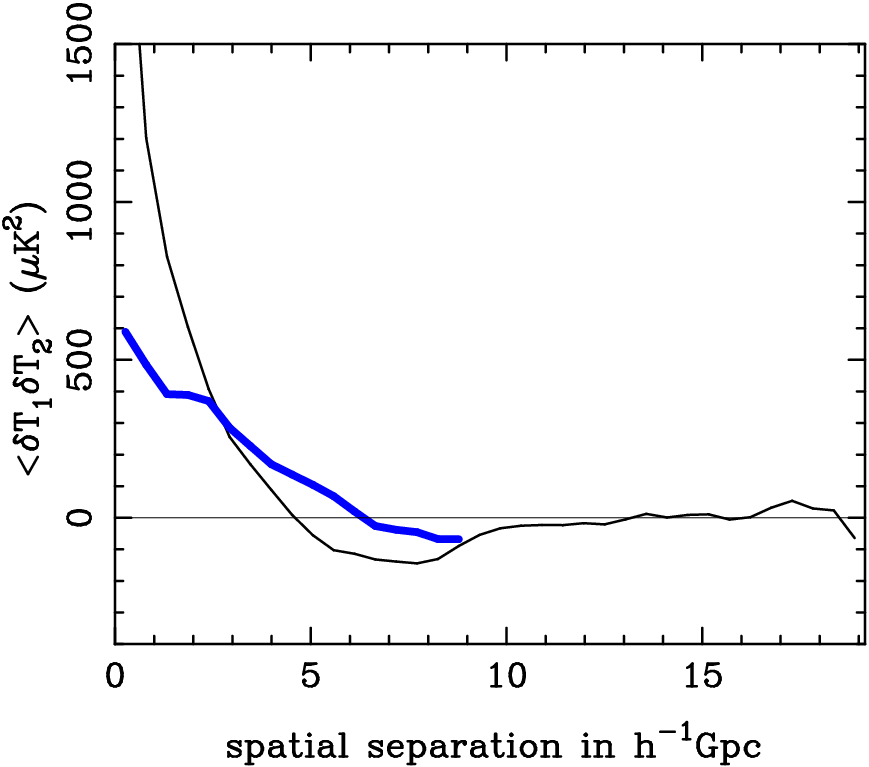}
\caption[]{ \mycaptionfont As for Fig.~\protect\ref{f-auto}, 
together with an estimate of the spatial 
cross-correlation function, $\ximc$ (thick curve), made assuming 
the Poincar\'e dodecahedral space with 
the best estimate parameters listed in Tables~\protect\ref{t-dodec} and
\protect\ref{t-alpha-phi}, corrected to an exact dodecahedral 
solution: $(l,b,\theta,\alpha, \phi)= (184.0,62.0,34.0,21.0,36.0)$ 
in degrees.  Values of $\ximc$ are meaningful
up to $d \ltapprox 9.0 \hGpc$, half the distance from the centre of
one face of the fundamental domain to the matching face. 
}
\label{f-cross}
\end{figure} 
} 

\newcommand\fcircles{
\begin{figure}
\centering 
\includegraphics[width=55mm,angle=270,bb=180 330 620 380]{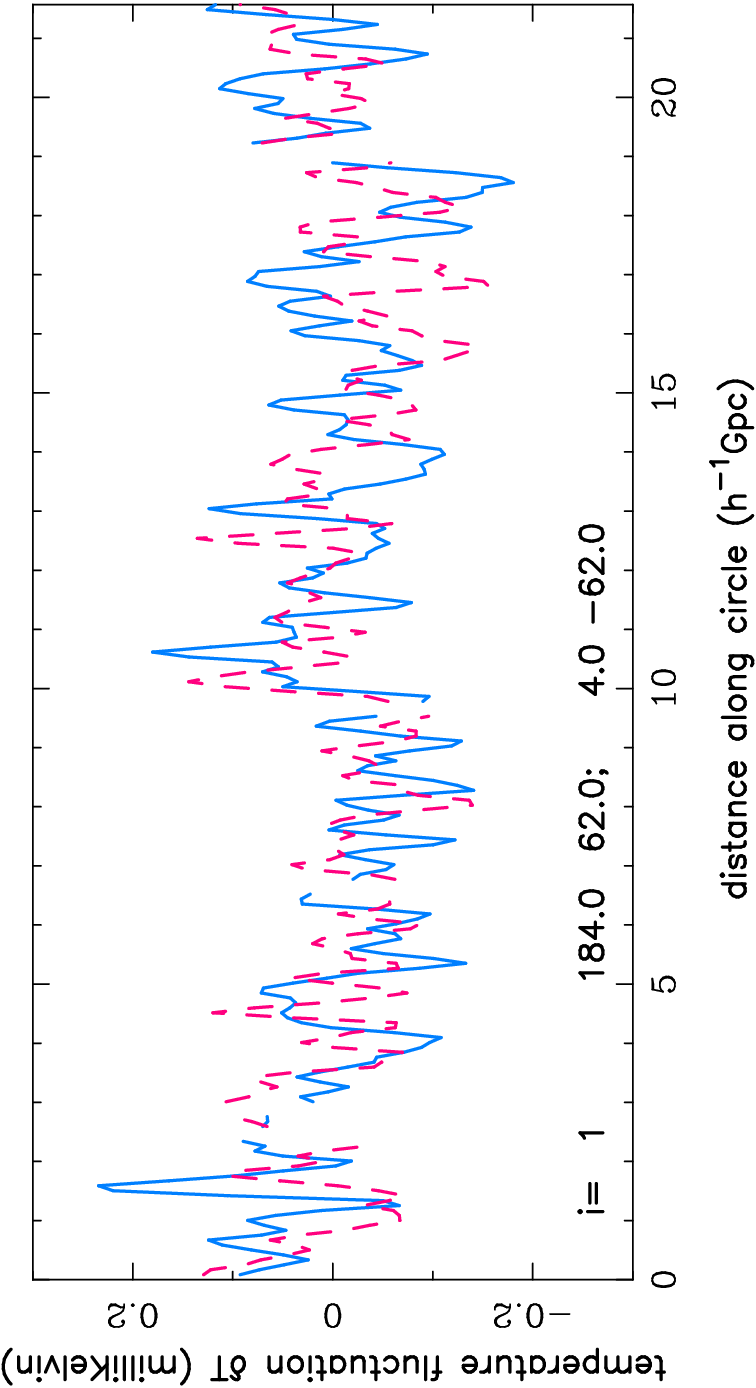}
\caption[]{ \mycaptionfont Matched circle pair 1: this and the
  following figures show matched circles for the best PDS model found
  here, corrected to an exact PDS solution as indicated in
  Fig.~\protect\ref{f-cross}.  Temperature fluctuations are from the
  ILC 3-year WMAP map using the kp2 galactic contamination mask, shown
  in mK against comoving distance along a circle in {\hGpc}. The
  coordinates of the two dodecahedral face centres $(l,b)_i$ are
  indicated. Solid lines show the northern galactic member of a pair;
  dashed lines show the southern member.  The pixels used for these
  plots contributed almost nothing to the method used for {\em
    finding} the optimal cross-correlation, since pairs of points at
  nearly zero implied spatial separation are rare.  (See
  \SSS\ref{s-matched-circles}.)  }
\label{f-circle-1}
\end{figure} 

\begin{figure}
\centering 
\includegraphics[width=55mm,angle=270,bb=180 330 620 380]{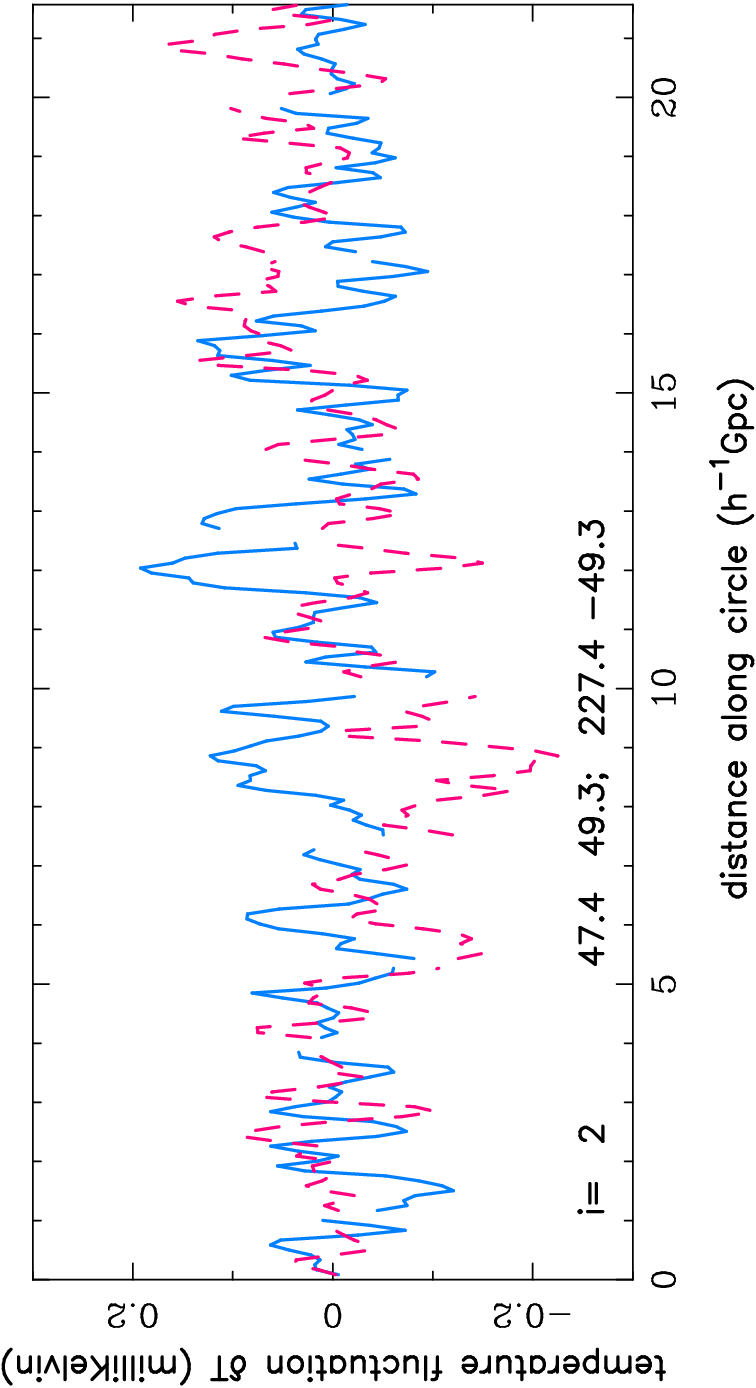}
\caption[]{ \mycaptionfont Matched circle pair 2, as in 
  Fig.~\protect\ref{f-circle-1}.
}
\label{f-circle-2}
\end{figure} 

\begin{figure}
\centering 
\includegraphics[width=55mm,angle=270,bb=180 330 620 380]{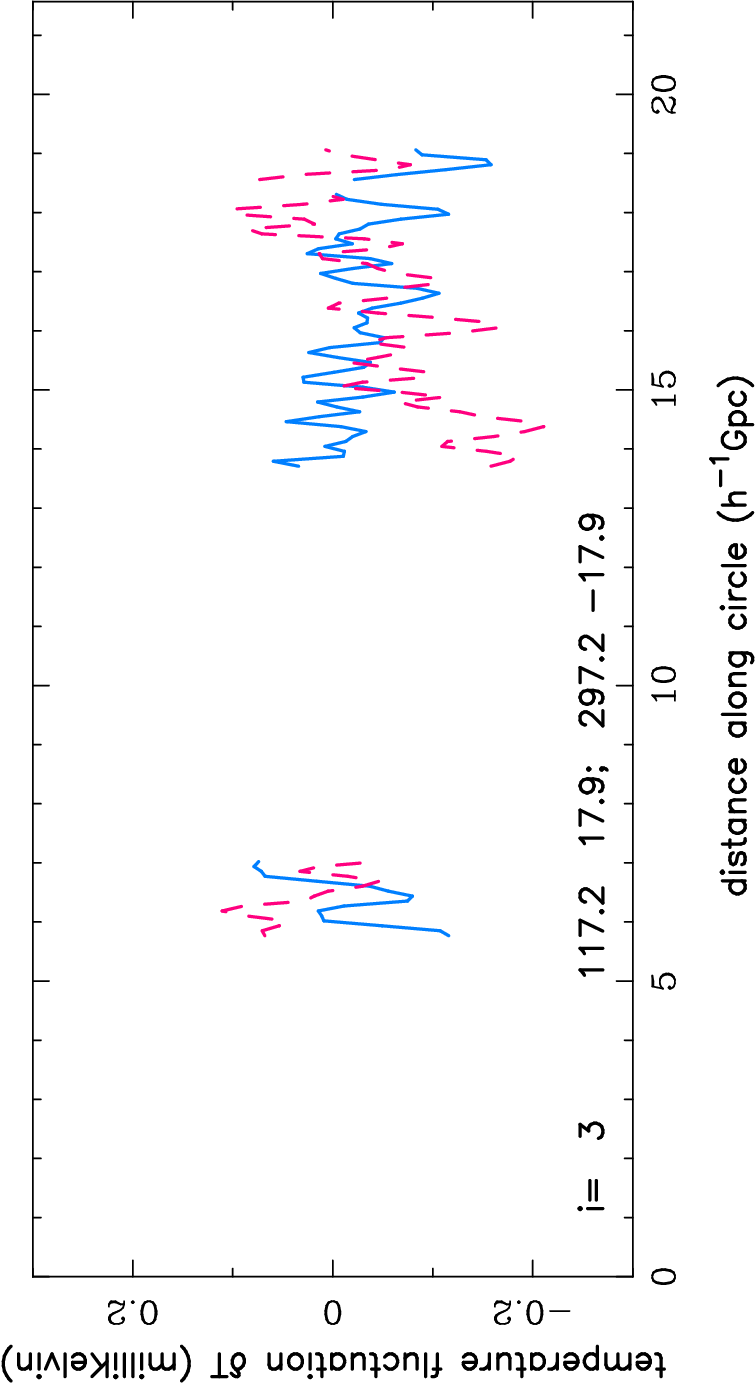}
\caption[]{ \mycaptionfont Matched circle pair 3, as in 
  Fig.~\protect\ref{f-circle-1}. The kp2 galactic contamination
  mask cuts severely into these two circles. 
}
\label{f-circle-3}
\end{figure} 

\begin{figure}
\centering 
\includegraphics[width=55mm,angle=270,bb=180 330 620 380]{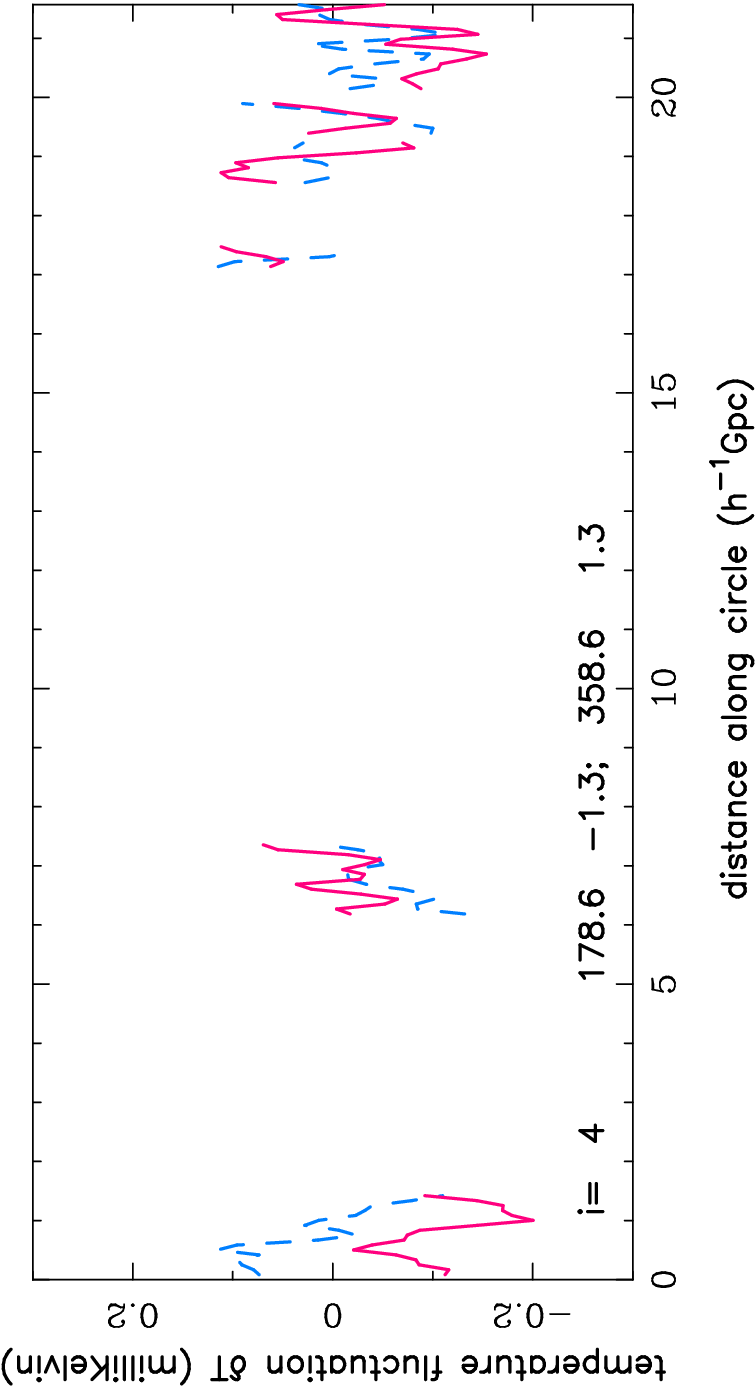}
\caption[]{ \mycaptionfont Matched circle pair 4, as in 
  Fig.~\protect\ref{f-circle-1}.  As in Fig.~\protect\ref{f-circle-3},
  the Galaxy cut is severe.
}
\label{f-circle-4}
\end{figure} 

\begin{figure}
\centering 
\includegraphics[width=55mm,angle=270,bb=180 330 620 380]{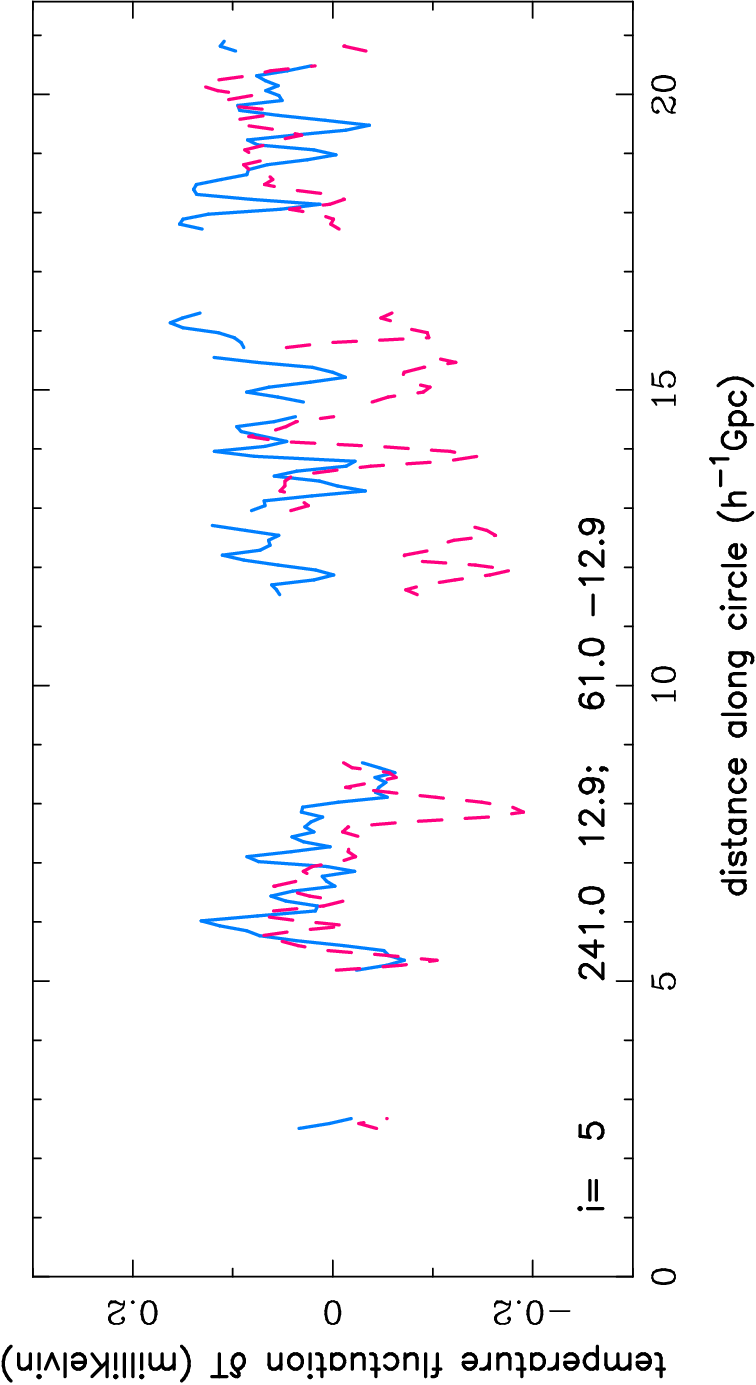}
\caption[]{ \mycaptionfont Matched circle pair 5, as in 
  Fig.~\protect\ref{f-circle-1}.
}
\label{f-circle-5}
\end{figure} 

\begin{figure}
\centering 
\includegraphics[width=55mm,angle=270,bb=180 330 620 380]{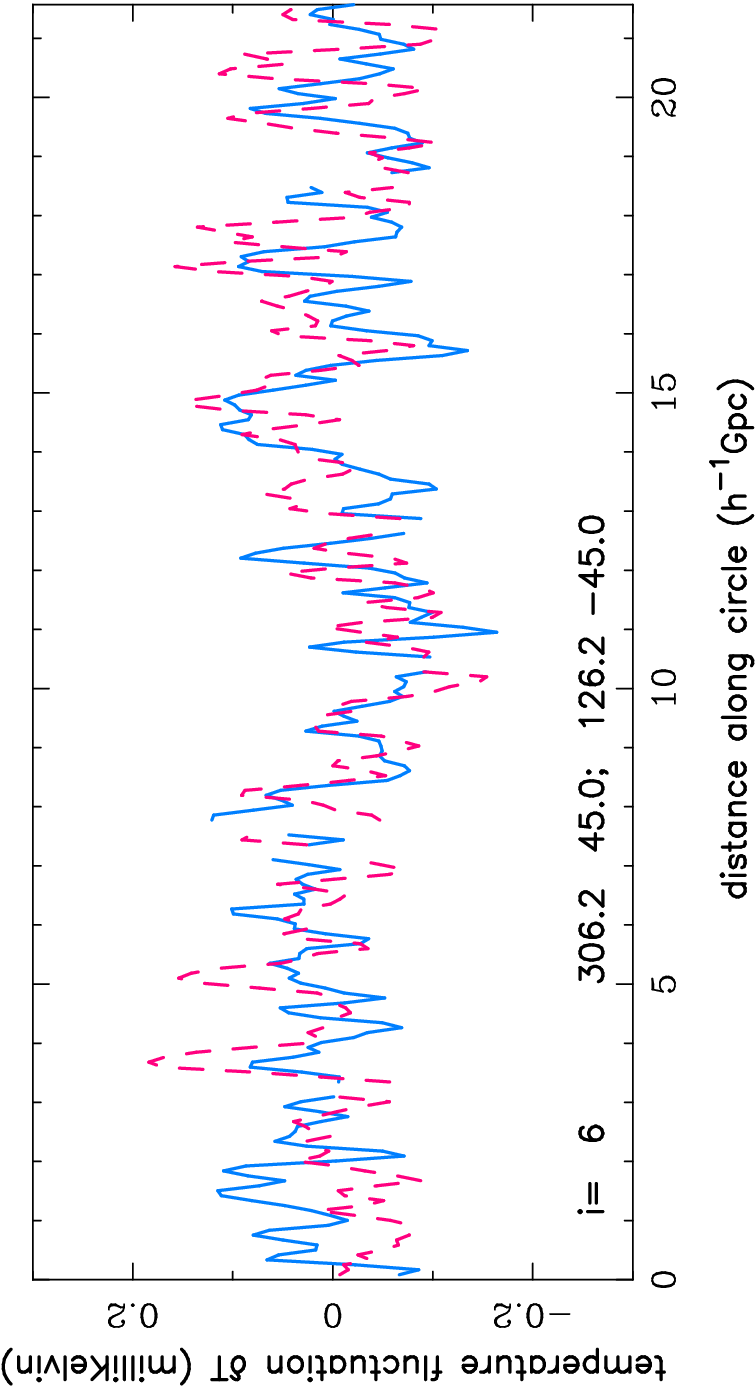}
\caption[]{ \mycaptionfont Matched circle pair 6, as in 
  Fig.~\protect\ref{f-circle-1}.
}
\label{f-circle-6}
\end{figure} 

} 


\section{Introduction}  \label{s-intro}

The past decade and a half have shown considerable growth in attempts
to determine the global shape, i.e. not only the curvature, but also
the topology, of
the spatial comoving section of the Universe, i.e. of the 3-manifold
to which a 3-hypersurface corresponds, informally known as ``space''.  
As noted by
several authors, in particular \nocite{Star93}{Starobinsky} (1993) and \nocite{Stevens93}{Stevens} {et~al.} (1993), a
space that is ``small'' compared to the surface of last scattering
(SLS) cannot contain eigenmodes, which are used for expressing density
perturbations, which are themselves 
larger than the space itself. This should lead 
to a cutoff of power in statistics representing these fluctuations,
above which power should drop to zero. This prediction was made after
COBE data were available, but {\em before} the WMAP satellite was 
launched.
For practical, observational reasons, spherical harmonic analyses of
temperature fluctuations on the 2-sphere are frequently made. 
However, a physically more natural statistic to use is
one in three-dimensional space, e.g. the two-point auto-correlation function.

\fauto

The predicted cutoff in large-scale power appears to have been confirmed by the first-year
observations of the Wilkinson Microwave Anisotropy Probe (WMAP)
experiment. With this data, \nocite{WMAPSpergel}{Spergel} {et~al.} (2003) published a figure
approximately equivalent to such a function, i.e. the black ``WMAP data''
curve of Fig.~16 in their paper.  Their figure shows the auto-correlation
calculated as a function of angular separation, shown against 
projected spatial separation for a (first year) template-cleaned V map with the
kp0 galactic contamination mask.  The authors note the surprisingly
flat correlation on large scales and suggest a multiply connected
universe model to match this function.

In Fig.~\ref{f-auto}, we calculated the auto-correlation function
directly as a function of three-dimensional spatial separation, not of
angular separation, using the 3-year 
integrated linear combination (ILC)
map with the kp2 cut. 
\posteditorchanges{This figure can be approximately compared
to the black ``WMAP data'' curve of Fig.~16 of \protect\nocite{WMAPSpergel}{Spergel} {et~al.} (2003),
except that the present figure shows the 
auto-correlation calculated as a function
of three-dimensional spatial separation, not of angular separation; 
\langed{``however'' would be wrong here; 3-d space vs angle already does
imply non-linearity, but this needs to be emphasised.}
the
relation between the two is {\em not} linear. Also,
our figure uses the 3-year ILC map with the kp2 cut, not the 1-year 
template-cleaned V map with the kp0 cut.}

In spatial
comoving units, we confirm that the auto-correlation is very close to
flat for separations larger than $\approx 10 {\hGpc}$. The relation 
between angular and spatial scales is, of course, not linear. 
Equation~(\ref{e-alpha-tri}) below (for either a spherical covering space
or for a flat covering space using the limit $R_C \gg \rSLS \ge d/2$) can be
used to calculate this.

If the size of the Universe\footnote{See 
e.g. Fig~10 of \protect\nocite{LR99}{Luminet} \& {Roukema} (1999) for a schematic
diagram of various definitions of the ``size'' of the fundamental domain, 
including the injectivity radius, the in-radius, and the out-radius.} 
is about 10{\hGpc}, as this figure seems to indicate, then which of the
various 3-manifolds correctly describes comoving space?
Motivated by indications that the Universe may have positive curvature,
and using eigenmode-based simulations to study the 
spherical harmonic ($C_l$) spectrum of the WMAP data,
\nocite{LumNat03}{Luminet} {et~al.} \nocite{LumNat03,Caillerie07}({Luminet} {et~al.} 2003; {Caillerie} {et~al.} 2007) 
argue that the Poincar\'e dodecahedral space (PDS) is favoured by the WMAP 
data over an infinite, simply connected flat space.
\nocite{Caillerie07}{Caillerie} {et~al.} (2007) state that, by requiring
maximal repression of the quadrupole signal, an optimal total density of 
$\Omtot = 1.018$ is favoured (for a non-relativistic matter density 
$\Omm \equiv 0.27$ and Hubble constant $H_0 = 70$\kms/Mpc) for the PDS model.
Several other authors \nocite{Aurich2005a,Aurich2005b,Gundermann2005}({Aurich} {et~al.} 2005a, 2005b; {Gundermann} 2005)
have also compared simulations for PDS models to the observed first-year and
three-year Wilkinson Microwave Anisotropy Probe (WMAP) maps of the
cosmic microwave background (CMB).

In all these studies, both the infinite flat models and the PDS models
are used in the context of a standard, hot big bang model, i.e. where the
Universe has a Friedmann-Lema\^{\i}tre-Robertson-Walker (FLRW) metric,
perturbed by fluctuations that collapse gravitationally to form
structures such as filaments and clusters of galaxies, and where values of
the metric parameters consistent with the consensus obtained during
the last decade of observations are adopted: the Universe is close to flat on
length scales up to the SLS and has about 30\% non-relativistic matter
density.

In other work, \nocite{RLCMB04}{Roukema} {et~al.} (2004) used the identified circles principle
\nocite{Corn96,Corn98b}({Cornish} {et~al.} 1996, 1998) to find a specific optimal orientation of the
PDS model based on the WMAP first-year 
ILC map, and published a tentative set of coordinates.
\nocite{KeyCSS06}{Key} {et~al.} (2007) confirmed the presence of a signal at the celestial
coordinates, circle radius, and $-36\ddeg$ twist published in
\nocite{RLCMB04}{Roukema} {et~al.} (2004), but argue that it should be considered a false
positive.  
Using software independent of that used in \langed{We want to avoid
ambiguity between independence of the article Roukema et al. 2004 
(true) and independence of the people Roukema et al. (false).} \nocite{RLCMB04}{Roukema} {et~al.} (2004), 
updating to the 3-year WMAP data, and using Gaussian simulations, 
\nocite{LewRouk2008}{Lew} \& {Roukema} (2008) find similar conclusions. They find that 
a local maximum in the statistic 
\langed{``'Statistics' is usually plural, unless this is a very
specific mathematical or physical sort. Do you mean this as a single
piece of data, which is what the singular form means?'' {\em ``statistic''
is used here as:
``a function of a sample where the function itself is independent of
the sample's distribution.''
\url{http://en.wikipedia.org/wiki/Statistic}}}
used for finding matched circles exists
for a circle radius $\sim 11\ddeg$ and a $-36\ddeg$ twist, but it is not
statistically significant.

\nocite{Aurich2005circ}{Aurich} {et~al.} (2006) also made a circles analysis of the
WMAP first-year data, using their own estimator and weight function,
and, in contrast with \nocite{RLCMB04}{Roukema} {et~al.} (2004), \nocite{KeyCSS06}{Key} {et~al.} (2007), and \nocite{LewRouk2008}{Lew} \& {Roukema} (2008), 
did not find any signal at $11\ddeg$. On the other hand, they did find
a tentative PDS signal at $\Omtot \approx 1.015$, or equivalently, a circle
radius $\alpha \approx 40\ddeg$. The signal is weaker than they
expected, but the authors note that uncertainties due to foregrounds and
noise structures in the data make it premature to draw firm conclusions.

A disadvantage of the identified circles approach is that it is based
on the information in a relatively small number of points on the sky
map of temperature fluctuations, making it sensitive to small errors in
the data or analysis and requiring prohibitively long computations. 
Is it 
\langed{No, the word ``even'' would be wrong here.}
possible to generalise from the identified circles principle?

Moreover, leaving aside the debate about matched circles statistics,
observably multiply connected models could reasonably be said to have 
satisfied only
one prediction so far, that of a cutoff in the density fluctuation spectrum
on a large scale. 
Can predictions of the PDS model itself be tested,
for example, using the identified circles principle or an extension of it?

In searches for matched circles, the statistic used is usually some
variation on what can be considered to be the value of the two-point
cross-correlation function of observed temperature fluctuations at pairs
of points in the covering space, 
$\ximc(r)$,
where the two points lie on different
copies of the SLS, and the separation of the two points in
the pair (on the different copies of the SLS) is zero in the covering space.


\postrefereechanges{

We can write this function as if 
it were possible to sample temperature fluctuations
as point objects at arbitrary spatial points throughout the covering space, 
i.e. including arbitrarily low redshifts, as well as points beyond the SLS: 
\begin{eqnarray}
  \ximc(r) &\equiv& \left<
  \delta T 
  (x_{i_1}) 
  \delta T 
  \left[g_j({x_{i_2}})\right]
  \right>_{i_1,i_2,j}  
  \label{e-ximc-defn}
\end{eqnarray}
averaging over triples $(i_1,i_2,j)$
satisfying 
$d\left( x_{i_1}, \left[g_j({x_{i_2}})\right] \right) = r$
and $g_j\not= I$ 
(the identity $I$ is removed because we only want the
cross-correlation).
Here, $x_{i_k}$ for $k=1,2$ are arbitrary points on the SLS
considered to be located at their comoving spatial
positions, $g_j$ for  $j=1, \ldots, 12$ are the 12 
holonomy transformations in the binary icosahedral group $\Gamma = I^*$ operating
on the covering space $S^3$ that match opposite
faces of the fundamental polyhedron of the PDS, $d(x,y)$ is the
comoving spatial geodesic distance between two points $x,y$ in the comoving
covering space (i.e. an arc-length on the covering space $S^3$
embedded in $\mathbb{R}^4$, e.g. \nocite{Rouk01-4D}{Roukema} 2001).
The temperature fluctuations on one copy of the SLS, $\delta T(x)$, 
are extended to the whole covering space via the holonomy transformations, 
i.e.,
\begin{equation}
\forall g_j \in \Gamma, \;\;  \delta T(x) = \delta T \left[g_j(x)\right] .
\label{e-temp-covspace}
\end{equation}

The latter assumption, i.e. Eq.~(\ref{e-temp-covspace}), enables
rewriting Eq.~(\ref{e-ximc-defn}) in a way that becomes observationally
realistic,
\begin{eqnarray}
  \ximc(r) &\equiv& \left<
  \delta T 
  (x_{i_1}) 
  \delta T 
  ({x_{i_2}})
  \right>_{i_1,i_2,j},  
  \label{e-ximc-two}
\end{eqnarray}
again subject to the conditions that the average is taken over pairs 
satisfying $d\left( x_{i_1}, \left[g_j({x_{i_2}})\right] \right) = r$
and $g_j\not= I$.
The auto-correlation function can then be written 
\begin{eqnarray}
  \xisc(r) \equiv \left<
  \delta T
  (x_{i_1}) 
  \delta T
  ({x_{i_2}})
  \right>_{i_1,i_2} ,
  \label{e-xisc-defn}
\end{eqnarray}
averaging over pairs 
satisfying
$ d\left( x_{i_1}, x_{i_2} \right) = r$.
In words, $\xisc$ is the 3-spatial auto-correlation function for pairs
of points on a single copy of the SLS, while $\ximc$ is the 3-spatial
cross-correlation function for pairs of points where the two
members lie on different copies of the SLS in the covering space.


Now rewrite \langed{$S$ was ``written'' in Roukema et al. 2004, so now
it is {\bf re}written.} the statistic $S$ used in matched circles searches 
as follows [see e.g. Eq.~(9) of \nocite{RLCMB04}{Roukema} {et~al.} (2004), 
with the normalisations (denominator and monopole $T$) 
ignored for simplicity], 
\begin{eqnarray}
  S &\equiv& { \left<
      {\delta T }_i \;
      {\delta T }_j  \right>_{(i,j)} }
  \label{e-s-rlcmb04}  \nonumber \\
  & = & \ximc (0)
\label{e-ximc-zero}
\end{eqnarray}
for the case where
pairs of points $(i,j)$ are (hypothetically) multiply-imaged locations 
located on a pair of matching circles in the covering space.
We use $r=0$ here because that is the
defining characteristic of matched circles --- a pair of matching 
points is a match because the two points are the same space-time
points, i.e. they are separated by $r=0$ when their topological
images located on adjacent copies of the SLS in 
the covering space are considered.

}

\newcommand\rmax{2}
Here, generalise from the zero separation cross-correlation $\ximc(0)$ to 
cross-correlations at larger separations $r>0$. We can expect
that 
\begin{equation}
\ximc(r > 0) < \ximc(0),
\end{equation}
i.e. that correlations weaken with separation and that, in general,
$\ximc(r)$ should be approximately equal to the auto-correlation 
function $\xisc(r)$ on scales much smaller than \langed{$\ll$ and $<$ 
differ.} the ``size'' of the
fundamental domain, e.g. the in-diameter $2r_{-}$ \nocite{LR99}(e.g. fig~10,  {Luminet} \& {Roukema} 1999)
\begin{eqnarray}
\ximc(r > 0) \sim  \xisc(r > 0) & &, \mbox{ if } r \ll 2r_{-},
\end{eqnarray}
since there is no statistical, physical distinction between a pair of points
on different copies of the SLS and a pair of points on a single copy of
the SLS, apart from effects that are not locally isotropic, 
such as the Doppler effect.

Since $\xisc$ is generally large at small $r$ and small at large $r$,
obtaining a large 
\langed{Using ``strong'' would be confusing, since we use ``large'' 
a few words earlier in the same sentence. Moreover, ``strong'' has
connotations either of physical force or human force, which is not
a useful connotation here. ``Close'' would be utterly confusing to
the reader.}
correlation requires using
a range of length scales $r$ that are relatively small, e.g.,
\begin{eqnarray}
  \ximc(r \ltapprox \rmax \hGpc) &\sim& \xisc(r \ltapprox \rmax \hGpc) 
  \label{e-ximc-pdsgood}
\end{eqnarray}
should be a relatively high positive value 
if the multiply connected model being studied is the correct model. 

Another way of saying this is that this test compares the
spatial two-point cross-correlation function of mapped 
(in the sense of the holonomy transformation $g$,
which maps one copy of a point in the
covering space to one of its images) and unmapped temperature
fluctuations in the covering space, where the two points
in any pair are on different copies of the SLS, 
against the auto-correlation function on a single copy
of the SLS. 

To see yet another way of thinking about this, suppose that the PDS
model at a given orientation and implied circle size is correct and
that temperature fluctuations occur as point objects, each emitting
isotropically. The covering space {\em cross}-correlation function, based on
pairs of points that are {\em close to one another due to the
  application of one or more holonomy transformations $g$, but are not close to one
  another on a single copy of the SLS}, should then be statistically
equivalent (apart from the Doppler effect and foreground/projection
effects) to a sampling from the spatial {\rm auto}-correlation function
$\xisc$, which can be estimated using points that lie on a single
copy of the SLS.

{\em This gives us our first prediction to test the PDS hypothesis:
  Eq.~(\ref{e-ximc-pdsgood}) should only be expected to hold if
  the physically correct PDS (or other) 3-manifold is assumed and is
  modelled at its correct astronomical orientation.}
If the PDS model is correct but 
a wrong orientation is chosen, then 
the cross-correlation $\ximc$ inferred from it
on small length scales --- in the covering space since it is the 
cross-correlation --- should represent correlations of pairs of points
that in reality are {\em not} close together, due to the erroneous
orientation that causes mappings from one copy of the SLS
to another to be incorrect. Since the correlation of distant points
is, in general, small, we have for this incorrect PDS model:
\begin{eqnarray}
  \ximc(r \ltapprox \rmax \hGpc) &\sim& \xisc(r \gg \rmax \hGpc) \nonumber \\
  & \ll &  \xisc(r \ltapprox \rmax\hGpc) 
\label{e-ximc-pdsbad-first}
\end{eqnarray}
i.e.
\begin{eqnarray}
  \ximc(r \ltapprox \rmax \hGpc) & \ll &  \xisc(r \ltapprox \rmax\hGpc),
\label{e-ximc-pdsbad}
\end{eqnarray}
in contrast to Eq.~(\ref{e-ximc-pdsgood}).
Similarly, if the PDS model is incorrect, then arbitrary orientations
should also 
yield small cross-correlations on small length scales, as 
in Eq.~(\ref{e-ximc-pdsbad}). 

In this paper, the initial aim is to maximise $\ximc(r \ltapprox \rmax \hGpc)$
relative to  $\xisc(r \ltapprox \rmax \hGpc)$, varying
PDS models over the 
parameter space of different orientations and circle sizes.
(In the actual calculations, we used a range of scales 
below and a little above $\sim \rmax \hGpc$: see 
\SSS\ref{s-scales} and Eq.~(\ref{e-range-Gpc}) for details.)
This should lead to an estimate of the best orientation and circle size
for the PDS model, given the observational data.

However, to further test the PDS model,
consider a ``generalisation'' 
from the mathematically correct PDS
model to a class of pseudo-models for which the cross-correlation function
can be calculated, but for which most\footnote{The mathematical term
``almost all'' could be used here, 
but when taking observational 
and numerical calculation uncertainties into account, the word
``most'' is more realistic.} of the members of the class are
physically invalid.
Maximising $\ximc(r \ltapprox \rmax \hGpc)$ (relative to $\xisc(r
\ltapprox \rmax \hGpc)$) over the extended parameter space defined by this
class should then have little chance of yielding an optimal model 
that is valid,
unless the PDS model is astronomically correct. 

The ``generalisation'' that we define here is to allow the ``twist'' angle
$\phi$ to be arbitrary in $\left[0,2\pi\right]$. 
The twist angle can be described as follows. In a 
single-action, spherical 3-manifold \nocite{GausSph01}(e.g., Sect. 4.1,  {Gausmann} {et~al.} 2001)
thought of as embedded in 4-dimensional Euclidean space,
$\mathbb{R}^4$, any holonomy transformation is a Clifford translation 
that rotates
in $\mathbb{R}^4$ about the centre of the hypersphere in one 2-plane
by an angle of $\pi/5$, and also by $\pi/5$ in an orthogonal 2-plane.

From a close-up perspective of the SLS rather than looking at the
whole hypersphere $S^3$, or in other words, from a projection into
$\mathbb{R}^3$, one 4-rotation can be thought of as a {\em
  translation} in $\mathbb{R}^3$ from one circle in a matched circle
pair to the other member of the pair on the opposite side of the SLS,
and the second 4-rotation is then thought of as a {\em twist} around a
vector in the translation direction joining the centres of the two
identified circles. This full motion is frequently termed a ``screw
motion''.  The ``twist'' is the second of these two rotation angles.

Physically,
the only two possible twist angles are $\pm \pi/5 .$ 
If the cross-correlation were to be calculated by filling one copy 
of the fundamental domain with a uniform distribution of points and then
mapping this set of points to copies of the fundamental domain in the
covering space, then for an invalid twist angle, some regions of space
near the first copy would have either zero, two, three, or more times
the density of points in the first copy of the fundamental domain.
In other words, sharp discontinuities would occur in calculating the
cross-correlation.

However, 
a practical way of estimating the cross-correlation is the method represented 
in Eq.~(\ref{e-s-rlcmb04}),
i.e. multiplying temperature
fluctuations and then averaging them. 
Using this method rather than multiplying densities of points, 
the effect mentioned in the previous paragraph should only lead to slightly more
or less frequent sampling in some regions of 
the cross product of the covering space with itself, not to a modification
of the correlations themselves.

The continuous nature of this method, as opposed to the discrete nature 
of starting with a uniform
distribution of points in the fundamental domain, can be seen as follows.
For a given pair of observed locations 
$(x_i, x_j)$ on one copy of the SLS, their comoving spatial separation 
$d[x_i,g(x_j,\phi)]$ for a given ``generalised'' 
\postrefereechanges{ isometry } $g$, which 
depends on the 
generalised twist $\phi$, changes continuously through the values
$\phi = \pm \pi/5,$ not discretely. 
\langed{Either ``and'' or ``but'' would be wrong. ``i.e.'' would be
correct but would sound pedantic.}
Moreover, the product of the observed
temperature fluctuations remains constant. Intuitively, we could say 
that as $\phi$ varies,
the geodesics formed by pairs $[x_i,g_1(x_j,\phi)]$ do not ``know'' 
when they meet other pairs  $[x_i,g_2(x_j,\phi)]$, where $g_1$ and
$g_2$ are two holonomy transformations mapping the fundamental domain
to two copies of the fundamental domain 
adjacent to one another in the covering space. 
\langed{This should be clearer.}

If the PDS model is correct, then calculating $\ximc(r)$ using
temperature fluctuations and applying
isometries using an arbitrary twist angle should give
$\ximc (r \ltapprox \rmax \hGpc)\ll \xisc (r \ltapprox \rmax \hGpc)$ as in 
Eq.~(\ref{e-ximc-pdsbad}), but give
$\ximc (r \ltapprox \rmax \hGpc)\sim \xisc (r \ltapprox \rmax \hGpc)$ as in 
Eq.~(\ref{e-ximc-pdsgood}) when the twist angle is correct.

{\em This gives the second prediction of the PDS model.
The maximal cross-correlation for this generalised PDS
model estimated using correlations of temperature fluctuations
should not only exist as a robust maximum, 
but it should also give a twist angle of either $\pm \pi/5$.}

If the simply connected, perfectly flat model is correct,
then the PDS model is incorrect, 
and it should either be difficult to find
a robust maximal correlation $\ximc (r \ltapprox \rmax \hGpc)$ in this extended
5-parameter space, or else an arbitrary twist angle $\phi$ should result,
with only a small chance of $\phi$ being close to either of the two
values expected for the PDS.

An estimate of how close an ``arbitrary'' twist angle should lie
to one of the two PDS values can be made as follows.
Given the assumption that
\begin{eqnarray}
  \xisc(r \gg \rmax \hGpc) & \ll &  \xisc(r \ltapprox \rmax\hGpc),
\label{e-xisc-zeroatbigr}
\end{eqnarray}
we can expect that 
this arbitrary angle should be selected from 
a uniform probability density distribution on $\left[0,2\pi\right]$. 
In principle, depending on the assumptions made about 
the complex statistical properties of the WMAP temperature fluctuation
maps, it could be possible for this distribution to be non-uniform,
even for a non-PDS space model. However, the estimated spatial
auto-correlation function $\xisc$ for $r \gtapprox 10${\hGpc} 
is close to zero, as shown in fig~16 of \nocite{WMAPSpergel}{Spergel} {et~al.} (2003) as a function
of angular separation on the SLS ($S^2$), and
explicitly as a function of spatial separation 
in Fig.~\ref{f-auto} here, so the assumption does appear to 
be supported by the 
\langed{The only serious direct empirical evidence is that of WMAP,
hence ``the'' rather than implicit ``some''.}
empirical evidence.

For the uniform distribution assumption, let us define
\begin{eqnarray}
\Delta\phi &\equiv &
\min \left( \left|\phi - \frac{\pi}{5}\right|,
   \left|\phi - \frac{9\pi}{5}\right| \right)
\label{e-defnDphi}
\end{eqnarray}
for $\phi \in \left[0,2\pi\right]$.
The chance of the observational optimal phase $\phiWMAP$ being close to 
$\pi/5$ or $9\pi/5$ is then
\begin{eqnarray}
P\left( \Delta\phi < \Delta\phiWMAP  \right) 
&=&
\left\{ 
  \begin{array}{l l}
2 \frac{\Delta\phi}{\pi}
, & \mbox{ if }  \Delta\phi \le \frac{\pi}{5}  \\
\frac{1}{5} +  \frac{\Delta\phi}{\pi}
, & \mbox{ if } \frac{\pi}{5} \le \Delta\phi \le \frac{4\pi}{5} .  \\
  \end{array}
  \right.
\label{e-probphi}
\end{eqnarray}
The piecewise nature of this function is because; e.g. there are four ways
in which $\phi$ can differ from $\pm \pi/5$ by a small angle such as
$10\ddeg$, but only two ways in which it can differ from $\pm \pi/5$ by 
a large angle such as $100\ddeg$.

To search for an optimal solution in the parameter space, 
a Metropolis-Hastings version of a 
Markov-chain Monte Carlo (MCMC) method is used 
\nocite{Neal93}(e.g., Sect. 4.2,  {Neal} 1993,  and references therein) over the
five-dimensional parameter space 
\begin{equation}
\{ (l,b,\theta,\alpha,\phi) \},
\label{e-parameterspace}
\end{equation}
where the parameters represent dodecahedron orientation ($l$, $b$, $\theta$), 
circle size $\alpha$ and twist phase $\phi$. 
All parameters are initialised as arbitrary angles,
except that the circle size is constrained to $5\ddeg \le \alpha \le 60\ddeg$
both initially and during the whole chain. 
See \SSS\ref{s-mcmc-method} for more details. 

We also ran some MCMC chains starting
with the parameters of the \nocite{RLCMB04}{Roukema} {et~al.} (2004) ``hint'' of a PDS solution with
circles of size $\sim 11 \ddeg$. If
this solution is correct, then we would expect the chains to remain localised
around that same solution. If the solution is wrong and if a
correct solution
exists, then the chains should move towards that correct solution, even 
though starting at a wrong point. If the solution is wrong and if no correct
solution exists, then we would expect the chains to move randomly and fail
to find a strong maximum.


The correlation function definitions are described in
\SSS\ref{s-corr-method}, the probability estimator used to compare 
multiple-SLS cross-correlation and single-SLS auto-correlation functions 
is presented in \SSS\ref{s-prob-method}, and the 
MCMC method is described in \SSS\ref{s-mcmc-method}.
Results are presented in  
\SSS\ref{s-results}. Discussion follows in 
\SSS\ref{s-disc}
and conclusions are made in
\SSS\ref{s-conclu}.

For references on cosmic topology in general, 
please see the first known article on the subject
\nocite{Schw00,Schw98}({Schwarzschild} 1900, 1998), 
a short beginner's review of cosmic topology \nocite{Rouk00BASI}({Roukema} 2000),
more in-depth reviews 
\nocite{LaLu95,Lum98,Stark98,LR99}({Lachi\`eze-Rey} \& {Luminet} 1995; {Luminet} 1998; {Starkman} 1998; {Luminet} \& {Roukema} 1999),  
workshop proceedings \nocite{Stark98,BR99}({Starkman} 1998; {Blanl{\oe}il} \& {Roukema} 2000), 
and lists of two-dimensional and 
three-dimensional methods (Table~2, \nocite{LR99}{Luminet} \& {Roukema} 1999; 
\nocite{ULL99b,Rouk02topclass,RG04}{Uzan} {et~al.} 1999; {Roukema} 2002; {Rebou\c{c}as} \& {Gomero} 2004). 
For background on spherical, multiply connected spaces, 
see \nocite{Weeks2001}{Weeks} (2001),
\nocite{GausSph01}{Gausmann} {et~al.} (2001), 
\nocite{LehSph02}{Lehoucq} {et~al.} (2002),
\nocite{RiazSph03}{Riazuelo} {et~al.} (2004), and
\nocite{LumNat03}{Luminet} {et~al.} (2003).
For the identified circles principle, of which the present
method can be thought of as an extension, see \nocite{Corn96,Corn98b}{Cornish} {et~al.} (1996, 1998).

One correction to a point made in much of the above literature 
concerns the
independence of the metric
and global topology. For example, 
\nocite{LR99}{Luminet} \& {Roukema} (1999) wrote, ``However, general relativity
deals only with local geometrical properties of the universe,
such as its curvature, not with its global characteristics, namely
its topology''. The error here is that global topology {\em can},
at least in certain cases,
generate a local effect, and thus affect local geometrical properties. See the
$T^3$, Newtonian weak field limit, heuristic calculation in 
\nocite{RBBSJ06}{Roukema} {et~al.} (2007): if the Universe contains inhomogeneities and 
is not expanding perfectly isotropically,
then at least in the $T^3$ case, 
\postrefereechanges{
additional accelerations and decelerations exist in the different fundamental 
directions, in such a way that they tend to equalise
the slightly different expansion rates.}

Comoving coordinates are used when discussing distances 
(i.e. ``proper distances'', \nocite{Wein72}{Weinberg} 1972, equivalent to ``conformal time''
if $c=1$). We write the Hubble constant as $H_0 \equiv 100 h$\kms/Mpc.


\section{Observations} \label{s-wmap}

The analysis presented here uses the Internal Linear Combination 
(ILC)\footnote{\url{http://lambda.gsfc.nasa.gov/data/map/} 
\url{dr2/dfp/wmap_ilc_3yr_v2.fits}}
all-sky map of the 
three-year WMAP data \nocite{WMAPSpergel06}({Spergel} {et~al.} 2007)
and the \nocite{WMAPTegmarkFor}{Tegmark} {et~al.} (2003) foreground-cleaned, Wiener-filtered 
(TOH)\footnote{\url{http://space.mit.edu/home/tegmark/} 
\url{wmap/wiener3yr_map.fits}}
version of the same data. (Unless otherwise stated,
we refer to the three-year data, not the one-year data.)

For our main analyses, we used either 
the ``kp2'' mask to eliminate the Galactic Plane (GP) 
\langed{``Galactic Plane'' is used here as a proper noun, 
like ``Solar System''.}
and associated regions
from the analysis\footnote{Data file:
\url{http://lambda.gsfc.nasa.gov/data/map/}
\url{dr2/ancillary/wmap_kp2_r9_mask_3yr_v2.fits}, 
map projection: 
\url{http://lambda.gsfc.nasa.gov/product/map/}
\url{current/map_images/f02_int_mask_b.png}.} or no mask at all.
We also discuss analyses with the ``kp0'' mask.
The kp2 mask covers about 15\% of the sky, and the kp0 mask covers about
25\% of the sky.

We did not carry out any explicit smoothing of these maps, though the
MCMC method itself could be thought of as a method that (when
analysed statistically) implicitly smooths the information in the
data.


\fcauto

\fccross


\section{Method} \label{s-method}

A GPL (GNU General Public Licence) program 
{\sc circles}\footnote{Version {\sc circles-0.2.4} is the one used
for the main calculations in this paper, along with 
{\sc circles-0.2.4.f13.3}, which differs from the former 
only in the formatted
printing of galactic longitude to standard output and
to a file. Various versions of the packet are downloadable from 
\url{http://adjani.astro.umk.pl/GPLdownload/dodec/}. The software 
presently requires medium to advanced {\sc GNU/Linux}, 
{\sc Fortran77} and {\sc C} experience for a scientific user.}
is available for reproducing the analysis as described below. A typical
12,000 step MCMC chain as described below should take about 3--4
days to run on an x86 type processor running at about 1.5-2 GHz. 

\subsection{Spatial versus angular correlation functions}

Since the na\"{\i}ve Sachs-Wolfe effect is due to locally
isotropic emission, which is what makes the identified circles test
useful for the purposes of testing cosmic topology hypotheses, 
we will assume the same approximation as that required for the circles test 
to reveal a signal, i.e.  that the integrated Sachs-Wolfe effect and the Doppler
contribution are relatively small on the observational angular scales 
corresponding to the length scales of interest. At worst, this 
assumption should weaken any signal found; it is difficult to see
how it could create a false signal.
This assumption is written below in Eq.~(\ref{e-temp-covspace}).


\subsection{Auto- and cross-correlation functions: $\xisc$ vs $\ximc$} 
\label{s-corr-method}

Due to observational constraints, the spatial auto-correlation
function that we are interested in for this test is usually estimated
indirectly, e.g. in angle, and calculated in terms of spherical
harmonic components of the temperature fluctuations considered as a
function on the sphere.

However, the angular correlation function on the SLS is sometimes
calculated. Figure~16 of \nocite{WMAPSpergel}{Spergel} {et~al.} (2003)  
shows the main feature
useful for this test: to first order, the correlation is high at low
separations and low at high separations.  As mentioned above, 
the figure also shows both the
disagreement between the simply connected flat model and the observations
and the surprising flatness of the correlation function over most
angular scales greater than about 40--50$\ddeg$.

Figures~\ref{f-c_auto} and \ref{f-c_cross} show what is described algebraically
in Eqs.~(\ref{e-ximc-pdsbad}) and (\ref{e-ximc-pdsgood}): in a simply
connected model, 
$\ximc(r \ltapprox \rmax \hGpc) \ll \xisc(r \ltapprox \rmax\hGpc)$,
while in a multiply connected model at the correct orientation and 
identified circle size, 
$\ximc(r \ltapprox \rmax \hGpc) \sim \xisc(r \ltapprox \rmax\hGpc)$.
For simplicity, these two diagrams show only one holonomy transformation $g_j$. 
A holonomy transformation is an isometry of the covering space
used to define the PDS, 
i.e. $g_j \in \Gamma$ where $\Gamma = I^*$, the binary icosahedral group,
is the group of
isometries dividing the covering space $S^3$ into 120 multiple copies
of the fundamental domain.

The auto-correlation, $\xisc$, is calculated using Eq.~(\ref{e-xisc-defn}).
To calculate $\ximc$ 
for a given orientation and circle size, we used Eq.~(\ref{e-ximc-two}), with
all 12 of the holonomy transformations 
(those giving the first ``layer'' of spherical dodecahedra neighbouring the
``first'' copy of the fundamental domain).

In principle,  
Eq.~(\ref{e-ximc-two}) could be evaluated over
a ``large'' subset of all members of the subgroup generated by
the holonomy transformations, e.g. $g_2 g_1,
g_4 g_5 g_{12}, \ldots$. However, this would be highly redundant. 
The minimum number of calculations that should be done to 
test a reasonably large sample of different pairs would be
those within one copy of the fundamental domain, plus pairs crossing
the boundaries of that copy of the fundamental domain.

This leads to an alternative way of visualising $\ximc$. It could be
thought of as the auto-correlation function of the full set of all
temperature ``fluctuations'' (or pixels for a given pixelisation of the
sphere) mapped into one copy of the fundamental domain, plus those crossing
``boundaries''.

In practice, we find it simpler to work in the covering space and to 
use just the set of 
\postrefereechanges{
holonomy transformations that give the neighbours 
of a single copy of the fundamental domain, rather than the full
group of 120 holonomy transformations.
}


\subsection{Comoving lengths and angular scales} \label{s-scales}

\fkeyangles

Physically, the separations of interest are lengths of spatial geodesics in the
$S^3$ covering space --- the hypersphere. This is used for both the simply
connected and PDS assumptions, since there is no problem having just a 
single SLS embedded in an $S^3$ covering space.
These separations can be thought of as
arclengths or the angles that subtend them 
on that hypersphere, of radius $R_C$
(e.g. see eq.~(2) of \nocite{RLCMB04}{Roukema} {et~al.} 2004),
embedded in Euclidean 4-space 
\nocite{Rouk01-4D}(e.g.  {Roukema} 2001),
but these
angles are different from angles on the SLS (a 2-sphere of radius $\rSLS$)
as seen from an observer at the centre of that 2-sphere. 

Fig.~\ref{f-keyangles} shows some key relations within the SLS.
From the left-hand triangle, use of the spherical sine and cosine formulae
leads to 
\begin{eqnarray}
\tan \frac{\rSLS}{R_C} = \frac{\tan (\pi/10)}{\cos \alpha},
\label{e-alpha-tri}
\end{eqnarray}
for the ratio of the SLS ($S^2$) 
radius, $\rSLS$, to the radius of the full hypersphere ($S^3$), $R_C$, 
where $\alpha$ is the radius of a matched circle ($S^1$) on the SLS.
For the right-hand triangle 
a single application of the spherical sine formula
is sufficient to obtain an expression for relating the geodesic 
(3-sphere) distance $d(x_i,x_j)$
between a pair of points $(x_i,x_j)$ to an angle $\theta_d(x_i,x_j)$ more familiar
to observers of the cosmic microwave background:
\begin{eqnarray}
\sin \frac{d/2}{R_C} = \sin \frac{\rSLS}{R_C} {\sin \frac{\theta_d}{2}}.
\label{e-d-theta-tri}
\end{eqnarray}

However, what is of interest in the present work is
to estimate the spatial correlation functions
in comoving space rather than angular correlation functions on the SLS.
While it is possible to define a meaningful relation between $d$ and
$\theta_d$ when both $x_i$  and $x_j$ lie on the SLS, a more general
distance $d$ betweeen $x_i$ and $g_k(x_j)$ for one of the 12 holonomy transformations
$g_k$ cannot easily be interpreted as an angle $\theta_d$, since, in 
general, $g_k(x_j)$ does not lie on the SLS.

For the auto-correlation function, there is only one copy of the SLS,
so the maximum separation $d$ at which $\xisc$ could, in principle, be
calculated is the diameter of the SLS, i.e. $d \le 2 \rSLS$.

For the cross-correlation, since copies of the SLS fill (redundantly, with
overlaps) the covering space (the hypersphere), the maximum 
separation $d$ that could, in principle, be used for estimating $\ximc$ 
is $\pi R_C$.\footnote{If the covering space were infinite, then there 
would be no upper limit to the distances at which the cross-correlation could
be calculated, independently 
\protect\langed{``be no upper limit'' ... ``independently''}
of whether or not $\ximc$ at these separations
were meaningful.}
This is several times larger than $2 \rSLS$. However, correlations across
more than half the size (e.g. injectivity radius) of the fundamental domain
contain information redundant with those at smaller separations.




Moreover, as mentioned above, we focus on a range of length scales that are
small enough that the auto-correlation is (certainly)  high and
the cross-correlation (in the case of a correctly oriented PDS, if it
is physically correct) is also high. Above we suggested 
$d \ltapprox \rmax${\hGpc}.
Figure~\ref{f-auto} shows the auto-correlation function estimated
from the ILC WMAP map using the kp2 galactic contamination mask.
The correlations are high 
for $ d \ltapprox 2${\hGpc} and remain positive 
for $ d \ltapprox 4${\hGpc}. They are approximately zero for
separations greater than one SLS radius:
$ 9.5 \hGpc \ltapprox d \ltapprox 19 \hGpc.$


Since we want to use a range of small scales, not just a single
scale, 
\postrefereechanges{ the specific range used in calculations here was 
\begin{equation}
5/90 < d/\rSLS  < 40/90,
\label{e-range-dimless}
\end{equation}
i.e. 
\begin{equation}
0.5 {\hGpc} \ltapprox  d  \ltapprox  4.0 {\hGpc}
\label{e-range-Gpc}
\end{equation}
for $\Omtot \approx 1.0$, $\Omm \approx 0.3$. The fraction $1/90$ is by analogy
with degrees, but as shown in 
Fig.~\ref{f-keyangles} and Eq.~(\ref{e-d-theta-tri}), this does not relate
linearly to an angle on the SLS and is better considered as an arbitrary
unit. The approximate correspondence in angles on the SLS, as given in
Eq.~(\ref{e-d-theta-tri}), is approximately 
$3 \ddeg \ltapprox \theta_d  \ltapprox 25\ddeg$. }


\postrefereechanges{
Since a relatively small number of points, $\Npoint$, is used in 
estimating the correlation, and since we bin into separation intervals,
the inclusion of pairs with smaller separations, $d \ltapprox 0.5${\hGpc},
would be likely to provide only a small, noisy 
contribution to the total correlation. Using these closer pairs without
introducing too much noise would 
require increasing $\Npoint$, thereby slowing down the calculations. 
Nevertheless, this could be examined in future work.}
See \SSS\ref{s-prob-method}
for the values of $\Npoint$ and adopted number of bins.

\subsection{Calculating the holonomy transformation from a pair of matched circles}

In \nocite{RLCMB04}{Roukema} {et~al.} (2004), circle pairs were examined without calculating holonomy transformations
of the 3-manifold, since that was not needed. 

Although it is possible to explore the parameter space
of different possible 3-manifold sizes and orientations directly 
from sets of holonomy transformations without any initial reference
to identified circles, since the latter are derived from the former,
we find it easier to start from the latter.

We represent the parameter space of these different possibilities by
the coordinates of
the face centres of a fundamental domain (spherical dodecahedron), as 
stated above. In this case, it is possible to 
reconstruct the holonomy transformation for a circle pair for the PDS corresponding 
to that circle pair, given the positions of the two matching circles.
Consider the centres of the two corresponding circles as 
euclidean vectors ${\bf s}, M {\bf s}$ in ${\mathbb R}^4$, where $M$ 
is a four-dimensional matrix representation of the holonomy transformation as a left
multiplier whose values numerical values are not yet known. 
Using the term ``eigenplane'' as introduced 
in \nocite{Eigenplanes05}{Roukema} (2005), these two vectors determine one of the two
eigenplanes of the
isometry $M$, as described in Eq.~(15) of \nocite{GausSph01}{Gausmann} {et~al.} (2001).

As in Section~2.3.3 of \nocite{Eigenplanes05}{Roukema} (2005), using ${\bf s}$ and $M {\bf s}$
in Eq.~(19) of \nocite{Eigenplanes05}{Roukema} (2005) gives a second vector,
${\bf t}$, in this same eigenplane, orthonormal to ${\bf s}$:
\begin{eqnarray}
{\bf t} &\equiv& 
  { M {\bf s} - ({\bf s} \cdot M {\bf s}) \; {\bf s}  \over
 \sqrt{ 1 - ({\bf s} \cdot M {\bf s})^2 } },
\label{e-planeone-trivial}
\end{eqnarray}
where $\cdot$ is the inner product on ${\mathbb R}^4$.

Since $M {\bf t}$ lies in the same eigenplane and resolves to the same 
components in the plane as $M {\bf s}$, apart from a rotation by $\pi/2$, 
it follows that
\begin{eqnarray}
M {\bf t} &=& 
  { - ({\bf t} \cdot M {\bf s}) \; {\bf s}  +
    ({\bf s} \cdot M {\bf s}) \; {\bf t}  } .
\label{e-planeone-Mt}
\end{eqnarray}
We find a third vector ${\bf u}$ starting with one of the four Cartesian
axis unit vectors ${\bf u_0}$ 
(that which is furthest in angle from ${\bf s}, {\bf t}$)
\begin{eqnarray}
  {\bf u} \equiv 
  {\bf u_0} 
  - ({\bf u_0} \cdot {\bf s}) {\bf s}
  -  ({\bf u_0} \cdot {\bf t}) {\bf t}  .
\end{eqnarray}

A fourth vector ${\bf v}$ is defined as the cross product 
\langed{You seem to suggest ``crossproduct''. 
Common usage is ``cross~product''\footnote{
\url{http://en.wikipedia.org/wiki/Cross_product} \\
\url{http://mathworld.wolfram.com/CrossProduct.html}.}}
of ${\bf s}$, 
${\bf t}$, and ${\bf u}$, which can be calculated by writing these three vectors
as row vectors 
${\bf s}'$, 
${\bf t}'$, and ${\bf u}'$, respectively
and estimating the determinant of the appropriate 
$4 \times 4$ matrix using the four othonormal basis vectors ${\bf e}_i$ 
in a given orthonormal basis:
\begin{eqnarray} 
 {\bf v} &\equiv & 
    \begin{array}{| c |} 
      {\bf s}' \\
      {\bf t}' \\ 
      {\bf u}' \\
      {\bf e}_1 \; {\bf e}_2 \; {\bf e}_3 \; {\bf e}_4 .
    \end{array}
\end{eqnarray}

Since the zero point of rotation within the ${\bf u, v}$ plane is arbitrary,
we then define the vectors $M{\bf u}$ and $ M{\bf v}$:
\begin{eqnarray}
M{\bf u} &\equiv & (\cos \phi) {\bf u} +  (\sin \phi)  {\bf v}  \nonumber \\
M{\bf v} &\equiv & -(\sin \phi) {\bf u} +  (\cos \phi) {\bf v}
\end{eqnarray}
where $\phi$ is the generalised twist phase mentioned in \SSS~\ref{s-intro}.
We then calculate $M$ from
\begin{eqnarray}
M &=& [ {\bf s}\; {\bf t}\; {\bf u}\; {\bf v} ]^{-1} \;\;
          [ M{\bf s}\; M{\bf t}\; M{\bf u}\; M{\bf v} ].
\end{eqnarray}

\subsection{``Probability'' estimator} \label{s-prob-method}

Our primary aim is to see if we can reject the simply connected model
by looking for the signal expected from a PDS model. This does not require
a true probability for use in an MCMC search of 
parameter space. Instead, it is sufficient to have a function 
similar enough to a probability function that the chains will explore the full
parameter space and converge after
a reasonably short amount of computing time and that they will show
where the maximal correlation lies.

For a given set of parameters $(l,b,\theta,\alpha,\phi)$ 
[see Eq.~(\ref{e-parameterspace})] defining a sixtuplet of circle pairs, 
$\Npoint = 2000$ points are selected from a uniform distribution on the 2-sphere.
The auto- and cross-correlations $\xisc$ and $\ximc$ are calculated for 
all $\Npair = \Npoint (\Npoint -1)/2$ and 
$\Npair = 12 \Npoint^2$ pairs,  respectively, applying 
Eqs~(\ref{e-ximc-two}) and 
(\ref{e-xisc-defn}), respectively.  
Many of these pairs, especially 
in the latter case, fall at separations too large to be of interest for 
our test. We bin the pairs into $n=7$ bins in the comoving separation 
range given by Eq.~(\ref{e-range-Gpc}).

We define the ``probability'' that the function $\ximc$ at 
a given point in parameter space is sampled from 
the known, ``true'' correlation, i.e. the auto-correlation $\xisc$,
by assuming a Gaussian distribution of errors of width $\sigma_i$ 
in each $i$-th bin (unless $\ximc(i) > \xisc(i)$) and assuming
that the bins are independent:
\begin{eqnarray}
P(l,b,\theta,\alpha,\phi) &\equiv& \prod_{i=1}^n 
\left\{
\begin{array}{l }
e^{- \frac{[\ximctiny(i)-\xisctiny(i)]^2}{2\sigma_i^2}} 
   \\
    \quad \quad \quad \quad  \mbox{\rm if } \ximctiny(i) \le \xisctiny(i) \\
{\ } \\
1 + 0.01 \frac{\ximctiny(i)-\xisctiny(i)}{\xisctiny(i)} 
  \\
    \quad \quad \quad \quad \mbox{\rm if } \ximctiny(i) \ge \xisctiny(i) .
\end{array} 
\right. 
\label{e-prob-defn}
\end{eqnarray}
Values $P < 0.01$ are replaced by $P=0.01$ so that the MCMC
chains are ergodic \nocite{Neal93}(e.g., Sect. 4.2,  {Neal} 1993).

The definition of the distribution width, $\sigma_i$, is based on 
numerical experimentation in order to obtain an optimal fit with the 
MCMC method in a reasonable amount of time:
\begin{equation}
\sigma_i = \frac{1}{2} \xisc(i) \sqrt{\frac{N_n}{N_i}},
\label{e-sigma-defn}
\end{equation}
where $N_i$ is the number of pairs contributing to the $i$-th bin 
in the estimate of $\xisc$.  For the largest bin, $i=n$, which gives width
$\sigma_n = \xisc(n)/2$, so a cross-correlation in this bin needs to be 
as high as $ \ximc(n) \ge \xisc(n)/2$  to contribute a probability per
bin of $\exp(-0.5) \approx 0.61$ or higher to the full product.  The weighting 
$\sqrt{\frac{N_n}{N_i}}$ increases $\sigma_i$ in bins
where there are fewer pairs (normally those with $i < n$), thereby
moderately decreasing the strength of those bins' contributions 
to the product in Eq.~(\ref{e-prob-defn}),
so that bins with relatively high Poisson errors 
do not penalise the probability too much. In practice,
the maximum of $\sqrt{\frac{N_n}{N_i}}$ reaches $\sim \sqrt{6}$,
for $\Npoint = 2000$ and $n=7$ bins in 
the range given in Eq.~(\ref{e-range-Gpc}).

For bins in which $\ximc(i) > \xisc(i)$, use of the 
upper expression in Eq.~(\ref{e-prob-defn}) would lead 
to penalising these bins for giving unexpectedly strong cross-correlations.
Would this be reasonable?
When cross-correlating points that
are physically close in comoving space but observed at widely separated 
angles, it is conceivable that the avoidance of foreground effects could
lead to $\ximc(i) \gtapprox \xisc(i)$ in some bins, in which case we would
want to favour these contributions.
To know whether any
significant effect should be expected in practice would require full modelling of
the Doppler and integrated Sachs-Wolfe effects.

However, while it is unclear how significant any such effect would be,
it seems clear that we should not penalise the bins in which
$\ximc(i) > \xisc(i)$.
Here, a very slight excess probability per bin is applied in these bins, as 
defined in the lower expression in Eq.~(\ref{e-prob-defn}). 
A bin in which the cross-correlation is as high as $\ximc(i) = 2 \xisc(i)$,
which is much higher than is likely to occur in practice,
would yield a probability per bin of $1.01$ to the full product, so the 
contribution in practice is likely to be small; i.e., there should be
a slight preference for excess cross-correlations with respect to auto-correlations.

\subsection{Markov Chain Monte Carlo method: Metropolis-Hastings algorithm}
 \label{s-mcmc-method}

The function $P(l,b,\theta,\alpha,\phi)$ defined
in Eq.~(\ref{e-prob-defn}) is used to explore the parameter space 
defined in Eq.~(\ref{e-parameterspace}), maximising $P$ using an
MCMC method with a Metropolis-Hastings algorithm. 

An informal description of this method follows. Starting from a
randomly selected point in parameter space, a ``chain'' of successive
points in parameter space is calculated, such that a new step in the
chain is taken with certainty when the new probability at a newly
chosen test step is higher than the present step, but may or may not
be taken if the probability at the new step is lower. The probability
of advancing in the latter case is the ratio of the new probability
to the probability at the present step. This enables a chain of steps
to seek the region of highest probability, while not getting stuck 
in noisy areas or local hills. For a more formal and complete description
of the method, see, e.g. \nocite{Neal93}{Neal} (1993), and references therein, 
and in particular, Section~4.2 of that paper.

At each step, the $k$-th parameter is chosen randomly 
using a uniform distribution on $\{1,\ldots,5\}$. 
The proposal distribution $S_k(x,x_k^*)$ in the \nocite{Neal93}{Neal} (1993) notation 
is a Gaussian distribution on this $k$-th 
parameter, centred on the
previous state $(l,b,\theta,\alpha,\phi)$, of width 
\begin{equation}
\sigmaMCMC = 10 \ddeg ,
\label{e-sigmaMCMC}
\end{equation}
where $\ddeg$ indicates great circle degrees.  
The Gaussian is truncated (without renormalisation) 
in $\alpha$ at $\alpha \ge 5\ddeg$ and $\alpha \le 60\ddeg$.

The probability
$P$ as defined in Eq.~(\ref{e-prob-defn}) is used 
in Eq.~(4.18) in \nocite{Neal93}{Neal} (1993) for the acceptance function $A(x,x^*)$.
The initial state 
(point in parameter space) is selected randomly from uniform distributions in 
\begin{equation}
  \begin{array}{c}
    0 \le l \le 2\pi \nonumber \\
    0 \le b \le \pi/3 \nonumber \\
    0 \le \theta \le 2\pi/5 \nonumber \\
    5 \ddeg \le \alpha \le 60 \ddeg \\
    0 \le \phi \le 2\pi  .
  \end{array}
  \label{e-parameterspace-limits}
\end{equation}
The region of $(l,b,\theta)$ space used for choosing the initial
values is highly redundant. The range in $(l,b)$ 
defined here
covers a fraction $\sin(\pi/3)/2 \approx 43\%$ of the surface of the sphere; 
while in principle it only needs to cover $1/12$ of the sphere,
i.e., there is a factor of $\approx 5$ redundancy. 

The range in the initial choice of 
``rotation'' angle $\theta$ is due to the nature of 
the dodecahedron.
A single choice of $(l,b)$ defines 
a class of dodecahedrons that share one pair of faces joined by 
the axis joining $(l,b)$ to its antipode $(l+\pi,-b)$. Rotation about 
this axis by $\theta \in [0,2\pi/5)$ gives a class of distinct dodecahedra,
but rotation by $2i\pi/5$ for any integer $i$ gives the same set of face
centres.

\filclbthN

\filclbthS

\ftohlbthN

In the chains, the numerical values of $(l,b,\theta)$ are all allowed to
increase or decrease arbitrarily, beyond the ranges chosen initially.
This is for calculational simplicity, both for speed and for minimising
the chance of introducing errors into the software.
Together, the triple $(l,b,\theta)$ is likely to cover only a relatively small 
fraction of that part of 
the non-periodic Cartesian product space of the three parameters defined
by the least and greatest numerical 
values of $l,b$ and $\theta$ reached in a given chain.

During analysis, probably the simplest way to reduce the redundancy is
by converting each triple $(l,b,\theta)$ to a 12-tuplet
$\{ (l,b)_i, i=1,12\}$, where the first face centre $(l,b)_1$ is 
numerically equivalent to $(l,b)$, converted to a conventional numerical
range (e.g. $l \in [0,2\pi), b \in [-\pi/2,\pi/2]$). 
Although the set of face centres should, in principle, consist of 6
antipodal pairs rather than 12 independent sky positions, the freedom
given to $(l,b,\theta)$ and the stochastic nature of the MCMC chains
and the sampling used to estimate correlation functions imply that the
two elements of any antipodal pair will not in practice 
be perfectly antipodal.

The longest dimension in this parameter space is that of the phase $\phi$,
so this is the dimension that constrains 
the ``burn-in'' time. 
The burn-in time is the number of steps in the chain needed 
to find the region of maximal probability, i.e. of the strongest 
cross-correlations relative to auto-correlations, 
independently \langed{``needed \ldots independently''} of where the chain
started from. Since the longest possible ``distance'' from the optimal region
(if that exists) is $180\ddeg$, 
the number of steps to cross this by a random walk can be estimated as
\begin{equation}
\Nburnin \sim 5 \left( \frac{180\ddeg}{\sigmaMCMC}\right)^2  = 1620
\label{e-burnin}
\end{equation}
steps, where the factor of 5 is due to changing only one parameter at any
step.

Thus, for our primary estimate of the optimal point in parameter space, we
use $\Nchain$ chains starting with different random seeds, each run for 12,000
steps. We ignore the first 2000 steps and use the last 10,000 steps for
further analysis.



\tdodec

\section{Results} \label{s-results}

For the main calculation,
$\Nchain=10$ MCMC chains were run, starting with different random seeds 
for both maps (ILC and TOH), with either no mask or the kp2 mask.
Each run had 12,000 steps, except that for the ILC map and the kp2 mask, 
the total number of chains was greater ($\Nchain=24$).  As expected 
(\SSS\ref{s-method}), each chain took
about 3--4 days to run on a $\sim 2$~GHz x86 type processor on a 
system running a variety of the GNU/Linux operating system.

The MCMC chains used in this paper can be downloaded for 
independent analysis from the file
\url{http://adjani.astro.umk.pl}
\url{/GPLdownload/MCMC/mcmc_RBSG08.tbz}.

\subsection{Optimal dodecahedron orientation: $(l,b,\theta)$ space}
\label{s-res-lbtheta}

Figures~\ref{f-ilc_lbth_N}, \ref{f-ilc_lbth_S}, and \ref{f-toh_lbth_N} 
show the sky positions $(l,b)_{i=1,12}$ implied
by the $(l,b,\theta)$ triples in the MCMC chains 
for which $P > 0.5$ (see Eq.~(\protect\ref{e-prob-defn})). For consistency
between the plots, only 10 of the 24 available chains are 
\langed{``only ... can be'' would be wrong, since we {\bf could} use
all 24 chains, but then we would have to find some way of compensating
for that in order to obtain consistency, and it would just complicate
things. In other words, it would be a bit messy, but possible. Hence, better
to state what we did do rather than what we had to do, since it's not strictly
true that we ``had'' to do it.}
used to create
the plot for the ILC, kp2 case, even though for calculations, we used
all 24. The information in the SGP projections (Fig.~\ref{f-ilc_lbth_S}) 
is equivalent to that in the NGP projections (Fig.~\ref{f-ilc_lbth_N}), 
so we do not show further SGP projections.

It is clear that a consistent, preferred dodecahedron orientation
exists for both the ILC and TOH maps, both with and without 
masking for contamination by residual foregrounds with the kp2 galactic
contamination mask.

We have estimated the preferred values of $(l,b)_i$ for the ILC map using
the kp2 mask, for which we have 24 chains.
Since any individual MCMC chain may fail to find the global maximum due
to starting at a distant point in parameter space and random walking in 
the wrong directions, if we search for preferred values of $(l,b)_i$
separately in each chain and then average them and find the dispersion
in the estimates, we risk having a large dispersion due to individual
chains. On the other hand, if we search for preferred values of $(l,b)_i$
in a concatenation of all 24 chains together, we will not have a 
straightforward way of estimating the uncertainty in the estimates of
$(l,b)_i$. 
Hence, we compromise by grouping the 24 chains together into four
groups of six chains\footnote{The ordering is random, defined by
the numerical values of the random seeds used for the different chains.}.

\falphaphi

For a given group, steps 2001 to 12,000 from each of the six chains are
concatenated. An approximate (to within $\sim 5$--$10 \ddeg$) estimate of the optimal $(l,b)_i$ 
values evident in 
Figs~\ref{f-ilc_lbth_N}--\ref{f-toh_lbth_N} is 
\begin{eqnarray}
\{(l,b)_i\}_0 & = & \{ (5\ddeg,5\ddeg), (50\ddeg,45\ddeg), 
                                (120\ddeg,25\ddeg), \nonumber \\
&& (180\ddeg,60\ddeg), (245\ddeg,20\ddeg), (315\ddeg,50\ddeg) \}
\label{e-initiallb}
\end{eqnarray}
and their six antipodes.\footnote{For example, these values can be read off a Mercator
projection of the same data.}
This initial approximation is iterated
to yield a more precise estimate.
Each iteration uses the mean values of points lying within 
an angle $\beta_j$ of the ($j-1$)-th estimate, starting from $j=1$,
where $j=0$ represents the intial approximation.
The angular radii start at $\beta_1 = 30\ddeg$ 
(so that most of the sphere is covered),
decrease by $1\ddeg$ for the next 10 iterations, and then remain 
constant at $\beta_{j \ge 11} = 20\ddeg$, covering roughly half of the
sphere, until the iteration for a given face number converges.
For a given threshold $\Pmin$, only points for which $P > \Pmin$ 
are used.  The means 
are calculated as means of Cartesian $(x,y,z)$ vectors formed from the 
individual sky positions $(l,b)_i$.

\talphaphi

\tsecondary

\fdiscs

These four groups of six chains are then treated as four independent
estimates in order to estimate the uncertainties due to our
MCMC estimation method.  These uncertainties are
standard errors in the mean calculated in Cartesian space, converted
to a one-dimensional uncertainty in great-circle degrees 
assuming (conservatively) that most
of the uncertainty approximately lies in a single dimension:
\begin{equation}
  \sigmalbth  =  
  \frac{ \sqrt{\sigma_x^2 + \sigma_y^2 + \sigma_z^2} }{
    \sqrt{N-1} } 
\label{e-sigmalth-defn}
\end{equation}
over $N=4$ groups of chains.

The resulting numerical estimates are listed in Table~\ref{t-dodec}.
This table lists sky positions of the best estimates of 
the six dodecahedral face centres  for the ILC map with the kp2 
mask, as shown in Fig.~\protect\ref{f-ilc_lbth_N}.
The columns show
minimum probability $\Pmin$,
face number $i$, 
number $n$ of MCMC steps contributing to the estimate obtained
from the final iteration,
galactic longitude $\lII$ and latitude $\bII$, 
and
the standard error in the mean between these four estimates  
in great circle degrees, $\sigmalbth$ 
[see Eq.~(\protect\ref{e-sigmalth-defn})].
The other 6 faces are directly opposite with identical errors.
As in Fig.~\protect\ref{f-ilc_lbth_N}, 
these face centres were derived
from the MCMC chains 
without any constraint on the twist phase $\phi$.

The final iterated solutions listed in this table are
very insensitive to the initial approximation: varying the latter 
by up to $\sim 10\ddeg$ leads to identical final iterations.
The table shows that 
differing thresholds $\Pmin$ give slightly different solutions,
especially in the coordinates of the 5-th face, but these are
consistent with each other within the estimated uncertainties.


\subsection{Circle size $\alpha$}
\label{s-res-alpha}

The optimal dodecahedral face solution shown in
Figs~\ref{f-ilc_lbth_N}--\ref{f-toh_lbth_N} and Table~\ref{t-dodec} 
is inferred from the data 
without regard to whether the MCMC chains have converged to 
a preferred circle size $\alpha$ and/or twist phase $\phi$.

We estimated the latter using the same points in the MCMC chains used
for the estimate in Table~\ref{t-dodec}. 
For the circle size $\alpha$, since $\alpha$ is restricted to the
range $5\ddeg \le \alpha \le 60\ddeg,$ we take the mean and standard
error in the mean of $\alpha$ as a scalar value. 

However, there are some reasons to be prudent about the estimate of
$\alpha$ obtained here.
Firstly, relatively large changes in $\alpha$ for a fixed dodecahedral
face set $(l,b,\theta)$ and a fixed twist $\phi$ can lead to relatively 
small changes in the distances between a given pair of ``close'' points
on two distinct copies of the SLS, especially when $\alpha$ is ``small''.
This is related to the reason why we expect the cross-correlation method
presented in this paper to work: there should be not only many 
``perfectly matched'' pairs of points, in the sense that they are separated
by a comoving distance of zero on a single copy of the fundamental domain,
but there should also be many other close pairs, extending the matched
circle to a ``matched annulus'', or in the case of a small circle, a
``matched disc''.

Figure~\ref{f-discs} illustrates this. For a fixed twist $\phi$, 
each pair of points on the two parts
of the sky inside of two matched circles of radius $\alpha$ has a separation
of at most the separation $d_c(\alpha)$ between the two centres 
of the circles P and P', as projected onto
the copies of the SLS (not the circle centres in $ \mathbb{R}^4$). 
From Fig.~\ref{f-discs} and
Eq.~(\protect\ref{e-alpha-tri}) we have
\begin{eqnarray}
  d_c(\alpha) &=& 2 [ \rSLS - (\pi/10)\; R_C ] \nonumber \\
  &=& 2 R_C \left[ 
    \mbox{ atan } \left( \frac{\tan (\pi/10)}{\cos \alpha} \right) - \frac{\pi}{10}  \right]
\label{e-disc-sepn}
\end{eqnarray}
For $\Omtot > 1.01$, a circle radius as large as $\alpha = 35\ddeg$ gives
$d_c < 4${\hGpc}.
Hence, given that we correlate pairs with $d \ltapprox 4.0${\hGpc}, 
the main cluster of points in Fig.~\ref{f-alpha-phi} could be
interpreted as a ``true'' matched circle size of $\sim$30--40{\ddeg}.

Secondly, since the width of 
the step size in the MCMC, $\sigmaMCMC = 10 \ddeg$ [Eq.~(\ref{e-sigmaMCMC})],
is a large fraction of the
range in $\alpha$, it is possible that the chains did not have 
enough freedom to favour a value close to either of these limits,
so the statistical nature of the precision of any estimate in 
$\alpha$ is less well-established than for the other parameters, 
where the chains are totally unrestricted.

Inspection of Fig.~\ref{f-alpha-phi} suggests that the upper limit,
$\alpha < 60\ddeg$, did not constrain the chains.
However, there is a sharp cluster
of points at $\alpha=5\ddeg$. This is partly an artefact due to not
renormalising the Gaussian used in deciding whether to move to the
new step in an MCMC chain at the lower cutoff of $\alpha = 5\ddeg$, as mentioned
in \SSS\ref{s-mcmc-method}. Lack of renormalisation artificially increases
the probability (by up to a factor of two) of staying near this boundary.
This clustering may also be caused by the increased noise at small circle radii.
In any case, if the true value of $\alpha$ were close to the $\alpha = 5\ddeg$
limit, then the distribution of $\alpha$ values around this true value 
as estimated by this MCMC method could not be a two-sided Gaussian
distribution, and even a one-sided Gaussian distribution would not necessarily
occur.

Both of these caveats should be kept in mind when using
the value $\alpha \approx 21 \pm 1\ddeg$ from Table~\ref{t-alpha-phi}: 
the systematic error in this estimate is likely to be much larger 
than the random error.
In this table, as in 
Table~\protect\ref{t-dodec}, the standard errors in the mean
are obtained by treating the four groups of six MCMC chains
as independent tests.
Listed are 
minimum probability $\Pmin$,
number $n$ of MCMC steps contributing to the estimate,
$\alpha$ and its standard error in the mean $\sigmaalpha$,
$\phi$ and its standard error in the mean $\sigmaphi$, all four
in degrees. The number $n$ can be a non-integer since for a given
MCMC step, it is possible that some of the face centres 
fall within the convergence radius of the final iteration 
as described in \SSS\protect\ref{s-res-lbtheta}, but other face centres do not.

\subsection{Twist phase $\phi$}
\label{s-res-phi}

As for $\alpha$, we estimated $\phi$ using 
the points in the MCMC chains used
for the optimal dodecahedral face solution presented in 
Table~\ref{t-dodec}. 
Since $\phi$ is free to decrease or increase arbitrarily in the MCMC
chains, we convert $\phi$ values for these individual points to 
Cartesian $(x,y)$ coordinates before averaging and reconvert after 
averaging. The standard error in the mean is conservative, 
assuming that all the error occurs in one dimension.

Table~\ref{t-alpha-phi} shows the estimate of $\phi$ for slightly 
differing thresholds $\Pmin$. Clearly, the value is close to $+36\ddeg$
within about one standard error in the mean, 
with only a slight dependence on the choice of $\Pmin$.
Figure \ref{f-alpha-phi} shows the distribution of $\alpha$ and $\phi$
values corresponding to the favoured face centre orientation.

\frlcmb

Only one point is displayed for any given state, independently 
\langed{``displayed ... independently''}
of how
long a chain spent at that point, so the visual density of points does
not fully match the statistical distribution. Hence, the mean values
listed in Table~\ref{t-alpha-phi} can be offset from the visual
centroids in the figure.

What is clear in the figure is that, in addition to the twist phase
$\phi \approx +36\ddeg$, a weak secondary feature and a very weak
tertiary feature are present, of which the first of these is 
consistent with 
the interesting angle of $\phi \approx -36\ddeg$,
to within the estimated uncertainty. 

\fcross

We estimate the parameters of these two weak features by 
using a subset of solution points arbitrarily cut 
at $8\pi /5 \le \phi \le 2\pi$ for the secondary feature
and
$6\pi /5 \le \phi \le 8\pi/5$ for the tertiary feature.
The results are listed in Table~\ref{t-secondary}.
Comparison of the numbers of steps $n$ in the MCMC chains contributing
to the different signals, as listed in Tables~\ref{t-alpha-phi} and
\ref{t-secondary}, shows that the secondary and tertiary signals are 
represented by about 10 and 30 times less steps (respectively) than
the primary signal.

Clearly, the secondary feature is within 1--2$\sigma$ (standard error 
in the mean) of $-\pi/5$, with approximately the same estimated
mean circle radius $\alpha \approx 21-22\ddeg$ as the primary signal.  
Only one of the 
two twist radii can be valid for the spatial matching of 
opposite faces of the fundamental domain of a PDS model, and they 
cannot both be valid simultaneously. On the other hand, the 
density fluctuations in a PDS model must resolve into the eigenmodes
of the PDS, not of infinite flat space. It is conceivable that
some sort of harmonics could occur, so that secondary and 
tertiary weak features exist in addition to the main 
cross-correlation.


\subsection{Testing the Roukema et al. (2004) hypothesis} 
\label{s-rlcmb04}

If the PDS solution found here is correct, then the solution suggested
in \nocite{RLCMB04}{Roukema} {et~al.} (2004) cannot be correct (independently 
\langed{``be correct independently''}
of any other arguments),
since the parameters are very different.

To test this explicitly, we ran 10 MCMC chains on the ILC map with the kp2
mask, starting at the \nocite{RLCMB04}{Roukema} {et~al.} (2004) solution. If the latter solution were correct
(and if this method is correct), then these chains should remain around that
solution. In fact, as \nocite{KeyCSS06}{Key} {et~al.} (2007) appear to agree, the \nocite{RLCMB04}{Roukema} {et~al.} (2004) solution
is at least a local maximum for correlations along matched circles (though they
argue that it is not statistically significant, as do \nocite{LewRouk2008}{Lew} \& {Roukema} (2008) using a different
method), so the MCMC
chains could conceivably remain ``stuck'' at this local maximum even if a global
maximum exists elsewhere in the parameter space. 
However, Fig.~\ref{f-rlcmb} shows clearly that the MCMC chains failed to
remain near the \nocite{RLCMB04}{Roukema} {et~al.} (2004) solution and moved instead 
to the one found in this paper.  


\subsection{Probability in the case of a simply connected model}

What is the chance that a twist close to $+\pi/5$ could 
have occurred in a simply connected model? 
Conservatively taking the 2$\sigma$ upper limits of the estimates of the twist phase
of the global maximum found by the MCMCs
listed in  Table~\ref{t-alpha-phi}, i.e. 
$|\phi - \pi/5| \ltapprox $5.6-7.8\ddeg,
and using
the probability distribution for a simply connected
model represented by Eqs~(\ref{e-defnDphi}) and (\ref{e-probphi}),
the chance of finding $\phi$ to be this close to one of the two PDS values
in the case of a simply connected model is about {\probifnotPDS}.

\fcircles

\subsection{Cross-correlation $\ximc$ of the preferred solution}

Figure~\ref{f-cross} shows the spatial cross-correlation function
$\ximc$ estimated by assuming that the solution found here is correct (thick
curve), together with the auto-correlation function calculated
assuming that space is simply connected. 
The plot is for our solution corrected to an exact dodecahedral solution
as indicated in the figure caption.

It is clear that $\ximc$ is well above zero in 
the range 
$ 0.5 {\hGpc} \ltapprox d \ltapprox
4.0 {\hGpc}$ and 
is of about the same order of magnitude as the auto-correlation
$\xisc$; in fact, it is above zero in the range 
$ d \ltapprox 6.0 {\hGpc}$. This is clearly consistent with the PDS
hypothesis.
On the other hand, according to the simply connected hypothesis, these strong 
cross-correlations are correlations of 
temperature fluctuations at
points located at positions on the sky separated
by well above 10{\hGpc}. Given the nearly flat, zero auto-correlation
measured on scales above 10{\hGpc}, these temperature fluctuations
should be very weakly correlated according to the simply connected 
hypothesis.

What is striking in Fig.~\ref{f-cross} is that $\ximc$ finishes ---
at half the distance separating matching faces of the fundamental
domain --- just where $\xisc$ seems to become flat and zero-valued.
This is qualitatively consistent with a PDS interpretation
of the ``lack of power on large scales''.

\subsection{Implied matched circles} \label{s-matched-circles}

The method used in this paper uses random pairs of points of
which each point is selected randomly (uniformly) on the 2-sphere,
focussing especially on pairs of points that are relatively close.

However, since we used a weighting to focus on bins in pair separations
in a way that favours bins with more pairs (\SSS\ref{s-prob-method}), 
and since we used $ d \ltapprox 4${\hGpc}, corresponding to 
$\theta_d  \ltapprox 25\ddeg$ [Eq.~(\ref{e-range-Gpc})], most of 
the correlations used to detect the optimal solution are from
\postrefereechanges{ 
pairs with $ d \gg 0.5${\hGpc}, i.e. $ \theta_d \gg 3 \ddeg$,
well above our short separation cutoff and well above the map resolution.
}

While our match is optimal on the scale of several {\hGpc}, it
does not necessarily imply a good set of matched circles using
individual temperature fluctuations. Moreover, the caveats on 
the circle size, explained in \SSS\ref{s-res-alpha}, suggest
that, if the PDS solution found here is correct, then its ``perfect''
matched circle size may well be rather different from the 
best estimate found here. 
On the other hand, the first of these
caveats suggests that circles {\em should} still be approximately
matched for wrong estimates of $\alpha$, as long as the true $\alpha$
is ``small'' (e.g. $\ltapprox 35\ddeg$) 
and/or the wrong estimate is not too far from the
correct estimate.

In Figs~\ref{f-circle-1}--\ref{f-circle-6}, we show the matched
circles for the six circles for the corrected solution used in
Fig.~\ref{f-cross}. 
The two circles in pair 1 clearly have similar overall shapes to one
another, as do the two circles in pair 6.
Circle pair 2 has some regions that match and some of which do not match.
Circle pairs 3, 4, and 5 are badly affected by the kp2 cut, but several 
of the uncut regions do seem to show some correspondence.

\posteditorchanges{The 
severe cuts due to the kp2 mask in circle pairs 3, 4, and 5 illustrate
a difference between a search for ``pure'' identified circles and the 
cross-correlation method, as presented in this paper.
Since the cross-correlation method is not limited to zero separation
pairs (in fact, it uses very few pairs close to zero separation),
it has the advantage of being able
to cross-correlate ``moderately'' close pairs and obtain some
signal for the holonomy transformation $g$ corresponding to a severely cut
circle pair, despite the severity of the cut.}

\filclbthkpzero

\ftohlbthkpzero

\fwmapQ

\section{Discussion}
\label{s-disc}

The PDS model has now satisfied several predictions: not only is there
a large-scale cutoff in power,
but now we find that (i) a PDS solution maximising 
the cross-correlation on {\em implied}
small scales relative to the auto-correlation on {\em certainly} 
\langed{``certainly'' vs ``implied'', not ``certain''}
small scales 
exists, and the implied small-scale cross-correlation is of a similar order
of magnitude to the auto-correlation on certainly small scales,
and (ii) the favoured twist of the ``generalised'' PDS model agrees
surprisingly well with one of the two possible twists to within the accuracy of the 
method used.
Does this mean
that the simply connected model should be rejected in favour of the
PDS? 


Focussing on the second of these tests --- the {\probifnotPDS} chance
in the case of a simply connected model of
having a twist as close to $\pm \pi/5$ as what 
we find here --- there are numerous astrophysical, mathematical, and
software caveats to be considered.
Among these are:
\begin{list}{(\roman{enumi})}{\usecounter{enumi}}
\item 
Could this be an effect of galactic contamination that happens to
mimic a dodecahedral symmetry with a preferred $+\pi/5$ twist when
matching dodecahedral faces?
\item 
Could modifications to the infinite flat model such as our
location in a local void \nocite{InoueSilk06}(e.g.  {Inoue} \& {Silk} 2006) be sufficient to modify
Eq.~(\ref{e-probphi}) so that $\phi \approx \pi/5$ is more likely?
\item 
Could the nature of the comparison test between the auto-correlation 
and cross-correlation functions itself favour a $\pm \pi/5$ twist? Could
the MCMC algorithm itself be at fault?
\item 
Does a simply connected, infinite flat model really imply a 
uniform distribution for twist phases $\phi$?
\item 
The angle $\pi/5$ as a rotation angle in ${\mathbb R}^4$ from one copy
of an SLS to another copy of the SLS is necessarily built into the
software. Although this rotation is orthogonal to the twist, could it
be possible that, due to a subtle programming error, this angle is also
built in as a favoured twist angle?
\end{list}

\talphaphinomask

Alternatively, numerous other checks can be made to test the PDS
interpretation of this result, including:
\begin{list}{(\roman{enumi})}{\usecounter{enumi}} 
\addtocounter{enumi}{5}
\item Can this solution be confirmed on smaller scales?
\item Are the ``secondary'' and ``tertiary'' solutions listed in 
Tables~\ref{t-secondary} consistent with PDS harmonics (eigenmodes)?
\item Is this solution compatible with other PDS analyses of the WMAP data?
\end{list}

\subsection{Galactic contamination?}

Could the dominant signal be an effect of galactic contamination?
Figures \ref{f-ilc_lbth_N}--\ref{f-toh_lbth_N} show that
whether the kp2 mask, covering about 15\% of the sky, or no mask 
at all is used, the same solution is clearly dominant, both 
for the ILC map and the TOH map. 
If galactic contamination were significant, it should show up
more strongly in the plots with no masking. Indeed, 
Figs \ref{f-ilc_lbth_N}--\ref{f-toh_lbth_N} show that
some weak amount of extra signal, offset from the 
main solution, does seem to appear when no mask is used,
consistent with an interpretation as galactic contamination. 
The main solution remains dominant.

These figures only show the orientation of the dominant solution.
Could the twist itself be due to the shape of the mask, e.g.
due to correlating ``western'' mask edges with ``eastern'' mask edges?
This seems unlikely, but Table~\ref{t-alpha-phi-nomask} confirms
that the twist is present with very similar values when no
mask at all is used. 
Each case in this table is for 10 chains. These 
are divided into two groups, each of five
chains. The estimates for a given map/mask combination 
are thus based on only $N=2$ ``independent experiments'',
each of five chains concatenated, rather than $N=4$ groups
of six chains, so they should be less precise than those
for the ILC kp2 case, listed in
Table~\protect\ref{t-alpha-phi}. 

The estimates in this table are for 
smaller numbers of chains and so less precise than those 
for the ILC map with the kp2 mask, but they are clearly consistent
with the latter. The signal is therefore unlikely to
be an artefact due to the shape of the mask. 

Curiously, the best estimates of $\phi$, together with
their uncertainties $\sigmaphi$ in Table~\ref{t-alpha-phi-nomask} 
for the TOH map with and without the kp2 mask, seem to indicate
that the further the mean is from $36.0\ddeg$, the larger the
uncertainty $\sigmaphi$. We have not attempted to interpret this 
statistically, though it does hint that an alternative way
of finding best estimates could lead to a best estimate of $\phi$
closer to $36.0\ddeg$, with a smaller uncertainty.
As stated at the beginning of \SSS\ref{s-results}, our MCMC 
chains have been made publicly available for independent analysis.

Returning to the question of possible galactic contamination,
could it be possible to be even more conservative and mask more
of the sky while not losing the signal? 
The two-point correlation function depends on the numbers of pairs
of points available for sampling a domain. Reducing the coverage of
sky by a fraction $f$ implies a reduction in the number of pairs
by $(1-f)^2$. 

This can also be thought of in terms of matched or nearly matched circles.
For most choices of the twist $\phi$, the masked points in one
member of a circle pair will correspond to unmasked points in the
other pair. Obviously, such pairs cannot be used, even though one
of the points lies outside of the masked region. Hence, the masking 
of pairs grows faster than the masking of individual pixels.
The severe cuts in about half of the circles shown in 
Figs~\ref{f-circle-1}--\ref{f-circle-6} illustrate this.

Any further reduction in the amount of sky used risks increasing 
the presence of noise. A stronger mask used for some studies is 
the kp0 mask. This covers about 25\% of the sky; i.e., about 
46\% of pairs of points used for estimating correlations are masked.
Figures \ref{f-ilc_lbth_kpzero} and \ref{f-toh_lbth_kpzero} show
the $(l,b)_i$ estimates implied by the optimal $(l,b,\theta)$ values
in the ILC and TOH maps, respectively, for the kp0 mask.
Both figures contain the solution found with no mask or the kp2 mask,
but an additional set of optimal points is present, weakly for the TOH
map and strongly for the ILC map. 

Which of these is more correct?  \nocite{WMAPTegmarkFor}{Tegmark} {et~al.} (2003) used a method
expected to contain less non-cosmological signal 
from foregrounds and detector noise outside the GP
than the 
ILC method. This would imply that the additional signal that is strong
in Fig.~\ref{f-ilc_lbth_kpzero} and weak in Fig.~\ref{f-toh_lbth_kpzero} is more likely
to be noise than signal.

In their own study of the first-year ILC and TOH maps by an identified circles method, 
\nocite{Aurich2005circ}{Aurich} {et~al.} (2006) find in their Fig.~9 that that the TOH 
map gives a sharper signal --- at $\Omtot \approx 1.015$ --- than the ILC map.
While this could be a coincidence, it could also indicate that the TOH map
correctly reduces foreground and detector noise contamination in such a way that
two quite different methodologies --- that of \nocite{Aurich2005circ}{Aurich} {et~al.} (2006) and our own ---
yield sharper signals favouring a PDS model.

\fgplanesig

Given that there is disagreement between the two
different maps regarding the additional signal, but the main signal
remains present, galactic contamination does not seem to be a strong
contender in explaining the main signal found for either no mask or the
kp2 mask.

\subsubsection{The expected galactic signal} \label{s-gal-expected}

Another test is to see what signal is {\em expected} from the
Galaxy. 
We ran four 12,000 step MCMC chains on the Q
(41 GHz) band (smoothed) WMAP 
map\footnote{\url{http://lambda.gsfc.nasa.gov/data/map/}
  \url{dr2/skymaps/3yr/wmap_band_smth_imap_r9_3yr_Q_v2.fits}}.  This
map is clearly dominated by the Galaxy.
Figure~\ref{f-wmapQ} shows the ``preferred'' maximal cross-correlations
for $\Pmin > 0.5$ and the corresponding circle sizes and twists $\phi$.

Since the Galaxy signal is nearly axisymmetric, the preferred 
face centres $(l,b)_i$ vary approximately uniformly across the whole sphere,
with a strong axisymmetry around the north-south axis, 
and some small variation with latitude. This is not at all similar
to the dominant signal found in the ILC and TOH maps, which at best
could be seen as approaching some sort of axisymmetry only by
shifting one of the face centres to the NGP (or SGP).

The lower panel in Fig.~\ref{f-wmapQ} shows a very strong preference 
for $\phi = \pi$. This does not match the dominant signal in the ILC
and TOH maps.  
\posteditorchanges{Moreover, 
a favoured twist phase of $\phi=\pi$ should be {\em expected} from a map 
dominated by a strong signal approximating that of the GP,
for (nearly) arbitrary face centres.

This is shown schematically in Fig.~\ref{f-gplane_sig}. Consider a sky 
map that is zero everywhere except on the GP, where it has
some positive value. A circle AB that is translated with zero twist to
the other side of the sky, i.e. to A'B', fails to match the circle A'B'.
B is close to the GP but B' is {\em not} close to the GP, and
A is far from the GP, while A' is close to the GP. Keeping in mind that
the sky here is $S^2$, it is clear that a twist of $180\ddeg$ is required
so that the two points that intersected with the GP in the first
matched circle again intersect the GP after the translation. In other words,
the favoured twist angle is $180\ddeg$.
This geometry applies for arbitrary circles, except for those nearly parallel
or orthogonal to the GP.
Hence, the signal shown in the lower panel of Fig.~\ref{f-wmapQ} is just
what can be expected from a galactic signal, and does resemble our 
detected signal.}

If the signal we found is caused by the Galaxy, then it is not due to
the dominant part of the Galaxy signal, but rather to a more subtle
aspect of the galactic signal which by chance has a PDS-like 
symmetry.

\subsection{Modifications to the infinite flat model}

Is there any modification to the simplest version of the infinite
flat model, such as modelling our 
location in a local void \nocite{InoueSilk06}(e.g.  {Inoue} \& {Silk} 2006) or other 
methods of attempting to explain the lack of large-scale power or the 
``Axis of Evil'' \nocite{LandMagAoE07}(e.g.  {Land} \& {Magueijo} 2007,  and references therein),
that would explain a dodecahedrally symmetric pattern of twists by
$+\pi/5$? 
This could be possible, in principle, although in that case, the question
of which model is preferred by Occam's Razor would arise. This topic
goes beyond the scope of this paper.

\subsection{Is $\phi = \pm \pi/5$ built into our method or software?}
\label{s-toys}

Could the results found here be somehow built into the geometrical,
correlational, or other algorithmical aspects of our method, or 
be caused by one or more bugs in the software?
Extensive cross-checking of different parts of the {\sc circles} package
indicates no sign of an error significant enough to create an 
artificial signal similar to the one found here. However, in principle, any
moderately complex piece of software can always contain bugs.

\fcalib

As discussed in \SSS\ref{s-gal-expected}, analysis of the WMAP Q
map gave a clear signal consistent with what is expected --- an axisymmetric,
nearly uniform distribution of preferred face centres and a 
strongly preferred phase of $\phi=\pi$.
However, this is clearly a very special situation with 
a very particular resulting 
twist phase $\phi$. 
Here we describe a more general test. The principle is to
create a simple, asymmetric pattern in six faces of a spherical 
dodecahedron, translate them to the opposite faces, and
apply a twist. By construction, the cross-correlation should
be very strong for this twist. In general, it should be 
weak for other twists. The MCMC method is then used to try to detect
the preferred twist.

Some experimentation has indicated that the simplest test patterns are not
necessarily sufficient to yield a convergent solution in 5-parameter
space, since it is easy to induce false matches due to various
periodicities between ``adjacent'' copies of the pattern.  The
function should be, in general, circularly asymmetric around each face
centre so that only one favoured twist can be found. In the real 
Universe, it could happen that it is more difficult to find a single
solution than in this artificial case. However, the result found
in the WMAP data
was that a single solution does indeed dominate. What is important
to test is whether an artificial signal can lead to a strong, unique
false detection.

The test pattern we adopted is
\begin{equation}
\left| \cos^5\left(\frac{\pi b}{45}\right) \right| \frac{bl(100-|b|) }{36000|b|},
\label{e-test-pattern}
\end{equation}
where $l$ and $b$ are both in degrees. This 
is a smooth, moderately sinusoidal, asymmetric
function of $l$ and $b$, used for all pixels within 31.7{\ddeg} of the
centre of a dodecahedral face, i.e. within a disc approximately 
touching the edges of the pentagonal border. These pixels are then copied by a 
translation to the opposite
face and rotated by a twist of $\phii$. 


There is only a trivial algorithmic difference between
stepping in all five parameters or stepping in a single parameter. 
In this way, we can test most aspects of our method and software, 
as indicated in Sects~\ref{s-prob-method} and
\ref{s-mcmc-method}, in particular, the estimates of the
correlation functions, the calculation of the 
probability estimator defined in Eq.~(\ref{e-prob-defn}), and
the basic MCMC functionality, 
by starting our simulated maps at the correct input values
$(l,b,\theta,\alpha)$ for the simulated map and holding 
these constant, while starting at an arbitrary phase $\phi$
unrelated to the input value $\phii$ of the simulated map.

If something is wrong with the correlation function estimation,
the probability estimator defined in
Eq.~(\ref{e-prob-defn}) or the basic MCMC functionality, or 
if a phase $\phi = \pm \pi/5$ is somehow accidentally 
built into the software, 
then this should be revealed by toy simulations of this type.
Figure~\ref{f-calib} compares input and output values of
$\phi$ based on 20 MCMC chains, each of 6000 
steps. Each input toy map is generated for a random point 
$(l,b,\theta,\alpha,\phi)$, and the chain is started at
the same $(l,b,\theta,\alpha)$ position but a random twist
$\phi \in [0,2\pi]$. 
Since the artificial maps contain matching 31.7{\ddeg} discs,
we restrict the 
circle size in the MCMC chains to $\alpha \le 30\ddeg$.

Estimates of $\phio$ are based on steps 1001 to 6000 (i.e. removing
a burn-in of 1000 steps) of a simulation, using a 
minimum probability requirement of $P > \Pmin = 0.5$ as for the 
analysis on real data.
The mean of the differences $\delta \equiv \phio-\phii$ for the 20 
chains is $+8.3\ddeg$ 
with a standard error in the mean of  
$4.6\ddeg$.   
Clearly, the input and output twists are statistically close to equal.
Moreover, there is no sign that either of $\pm \pi/5$ are favoured due to
either an intrinsic error in our method or to a software error. 
If the correlation function estimates, ``probability'' estimator, 
and general MCMC functionality contain any hidden errors, then 
these do not explain our estimate of $\phi \approx +\pi/5$.

\subsection{Is the expected distribution of $\phi$ uniform?}

If the simply connected model is assumed, then the spatial auto-correlation
on large scales could conceivably be used to model the expected 
cross-correlations, though this would probably neither be easy nor 
free of model-dependent assumptions.

For a fixed circle radius, as  $|\phi|$ increases from zero to $\pi$,
the separation between identified points increases.
Thus, if the autocorrelation were monotonically decreasing as separation 
increased, then higher values of $|\phi|$ would imply lower correlations.

However, for circle radii $\alpha \ltapprox \pi/4$, which constitute 
almost all the points shown in Fig.~\ref{f-alpha-phi}, the separation
(assuming simple connectedness) between opposite circles is a minimum
of $2\cos (\pi/4) \approx 1.4$ times an SLS radius, i.e. a minimum of 
13.4{\hGpc}. If the auto-correlation function $\xisc$ 
is reasonably isotropic, 
then over the range $|\phi| \in (0,\pi)$, the cross-correlation $\ximc$ 
will sample $\xisc(d \gtapprox 13.4{\hGpc}).$  
Figure~\ref{f-auto} shows that this
function is approximately flat and close to zero, at least
relative to the correlations at $d \ltapprox 4{\hGpc}.$ 

The largest circle radii we considered here are those with
$\alpha = 60\ddeg$. This gives a minimum of one SLS radius 
(9.5{\hGpc}) separating a pair of points under the simply
connected assumption, which is approximately the separation
at which $\xisc$ becomes flat and approximately zero.
Hence, if the auto-correlation estimated under the simply connected 
assumption and shown in Fig.~\ref{f-auto} is approximately correct,
then the expected distribution of $\phi$ for the range of circle
sizes considered here should be close to uniform.

Independently of trying to explain our result in terms of 
a simply connected model, what other evidence could support
the PDS interpretation?

\subsection{Can this solution be confirmed on smaller scales?}

Probably the biggest open question raised from the present work
is the nature of the signal at smaller length/angular scales: 
$r \ltapprox 0.5$ {\hGpc} or $\theta_d \ltapprox 3 \ddeg$.
Investigating this would require much smaller Markov chain steps
than we have used here.

To have our Markov chains converge in a reasonably short time
but also have them start at arbitrary positions in parameter space, we
defined a 10$\ddeg$ width Gaussian for stepping between points. Now
that the globally optimal point in parameter space has been found, chains with
a narrower width, and/or a stricter ``probability'' function for
deciding how probable it is that a point in parameter space
corresponds to a correct solution, could be used to find an
optimal point with higher angular resolution, 
by starting from our present solution as an approximate solution,
rather than starting from arbitrary points in parameter space.

However, if our solution is physically correct, then while there is no
reason why the intrinsic density-density {\em cross}-correlation at
sub-gigaparsec scales should differ from the density-density {\em
  auto}-correlation, we can expect some differences in
temperature-temperature correlations due to various projection
effects: both the Doppler effect and the integrated Sachs-Wolfe effect
(ISW) should cause some differences between the two functions when
correlating temperature fluctuations.  Another complication on small
scales, as \nocite{Aurich2005circ}{Aurich} {et~al.} (2006) have shown, is that the WMAP ILC map
has a lack of signal on these scales.


Furthermore, could the {\em stronger} cross-correlation for our PDS solution at around
3--8{\hGpc} relative to the auto-correlation be explained by ISW
dampening of the auto-correlation on these scales? Since the cross-correlation
temperatures corresponding to these cross-correlations 
are observed, in general, at widely different angles, 
they are less likely to be foregrounded by the same ISW fluctuation, 
so this seems possible at least qualitatively. 

A practical problem in extending the present method to smaller scales 
is the need to use enough pairs of 
points $(x_{i_1}, x_{i_2})$ on the SLS for which 
$ d\left( x_{i_1}, [g_j(x_{i_2})] \right)$ is ``small''.
For a given, finite number of points $\Npoint$, there
are fewer and fewer pairs of points separated by smaller and smaller
separations.

As explained in \SSS\ref{s-prob-method}, we used $\Npoint = 2000$ 
points distributed uniformly on the sphere. In fact, a larger number
of points are generated when a galactic contamination mask is used.
To avoid any chance of bias, the points are generated uniformly, 
excluding those falling within the mask, until the required number of
points have been obtained. Excluding distantly separated pairs of points,
where the distance is the minimum of 
$ \{ d\left( x_{i_1}, [g_j(x_{i_2})] \right) \}$ for the 12 holonomy transformations
$g_j$, is a heavy computational task equivalent to making the main calculation
loop needed for correlation function estimation. 

Testing of smaller scales therefore requires considerably more
pairs of points, and this would seem to scale as
$\Npoint^2$, increasing the calculation time proportionally.
Improvements to this scaling using geometrical arguments
would likely be complex and would risk inducing an ambiguity as to
whether the new results are due to the non-uniform distribution of the
points or to the intrinsic properties of the data.



\subsection{Are the ``secondary'' and ``tertiary'' 
solutions explainable as PDS harmonics (eigenmodes)?}

If the eigenmodes of the PDS imply features such as those shown 
in Fig.~\ref{f-alpha-phi} and 
Table~\ref{t-secondary}, then this would support the PDS
interpretation of the signal found here.

\subsection{Is this solution compatible with other PDS analyses of the WMAP data?}

The \nocite{RLCMB04}{Roukema} {et~al.} (2004) ``hint'' of a solution was based on looking for
identified circles, so, as noted above in Eq.~(\ref{e-ximc-zero}),
was based on using a much smaller part of the available signal to search for a
candidate favoured PDS orientation. \nocite{KeyCSS06}{Key} {et~al.} (2007) and
\nocite{LewRouk2008}{Lew} \& {Roukema} (2008) agree that the solution is a local maximum in the
parameter space, but that it is not statistically significant. Using
the greater amount of information available from the cross-correlation
at separations going up to $r \ltapprox 4${\hGpc} rather than just
$r=0$, the results presented in \SSS\ref{s-rlcmb04} agree that the
\nocite{RLCMB04}{Roukema} {et~al.} (2004) PDS solution is not globally
favoured by the data. Moreover, MCMC chains starting at that
solution move towards and converge towards the solution found
with the present method.

One way of comparing with other PDS analyses, which 
use quite different methodologies to analyse the
WMAP data, is to consider estimates 
of the total density parameter $\Omtot$.
\nocite{LumNat03}{Luminet} {et~al.} (2003) initially favoured an estimate of $\Omtot \approx
1.013$, but in more recent work using many more eigenmodes of the PDS,
they
favour $\Omtot \approx 1.018$ (for $\Omm \approx 0.27$) by requiring maximal suppression of the
quadrupole \nocite{Caillerie07}({Caillerie} {et~al.} 2007).
In analyses based on WMAP first-year $C_l$ estimates, 
\nocite{Aurich2005b}{Aurich} {et~al.} (2005b) 
found $1.016 < \Omtot < 1.020$, and using a combined match circles method,
\nocite{Aurich2005circ}{Aurich} {et~al.} (2006) favoured 
$\Omtot \approx 1.015$ (for $\Omm \approx 0.28$).

Are these consistent with our solution? The total density parameter $\Omtot$
relates tightly to the circle radius $\alpha$, with some dependence on $\Omm$,
e.g. Fig.~2 of \nocite{RLCMB04}{Roukema} {et~al.} (2004) or Fig.~10 of \nocite{Aurich2005b}{Aurich} {et~al.} (2005b). 
The estimates $\Omtot = 1.015$ (for $\Omm = 0.28$) and 
$\Omtot = 1.018$ (for $\Omm = 0.27$) correspond to 
$\alpha = 40.5\ddeg$ and $\alpha = 47.6\ddeg$ respectively.
\posteditorchanges{\nocite{Gundermann2005}{Gundermann} (2005) 
also found 
\langed{Jesper Gundermann died in June 2006, so as in the case 
of ``Newton found or ``Einstein found'' or ``Lema\^{\i}tre found'', the
past tense seems right here.}
$\Omtot = 1.017$ (for $\Omm = 0.26$) as one
of his two solutions; this corresponds to $\alpha = 46.8\ddeg$.}

As discussed in \SSS\ref{s-res-alpha}, the ``matched discs'' interpretation (Fig.~\ref{f-discs}) 
of Fig.~\ref{f-alpha-phi} would imply a true matched circle radius of $\sim$30--40{\ddeg}.
This is clearly compatible with the \nocite{Aurich2005circ}{Aurich} {et~al.} (2006) estimate, 
and possibly compatible with the \nocite{Caillerie07}{Caillerie} {et~al.} (2007)
\posteditorchanges{and \nocite{Gundermann2005}{Gundermann} (2005)
estimates}.
Clearly, a variety of different methods, either requiring 
suppression of large-scale power (other work) or maximising large angular scale
correlations in a PDS-like symmetry (our work), and either 
with the in-built constraint that the twist is $\pm \pi/5$ (other work) or 
without that constraint (our work), seem to converge on similar results.

\section{Conclusion} \label{s-conclu}

It seems hard to avoid the conclusion that cross-correlations of
temperature fluctuations on would-be adjacent copies of the SLS, 
which are distant from one another and are (on average)
very weakly correlated according to the WMAP 3-year observations,
(i) imply a highly cross-correlated ``generalised'' Poincar\'e
dodecahedral space symmetry (Figs~\ref{f-ilc_lbth_N}--\ref{f-toh_lbth_N},
Table~\ref{t-dodec})  at which these on-average uncorrelated
fluctuations happen to be well-correlated with one another 
(Fig.~\ref{f-cross}), and,
moreover, (ii) this favoured solution is dominated by a signal whose
twist phase $\phi$ lies within a few degrees of one of the two twist phases
necessary for a valid PDS model (Table~\ref{t-alpha-phi}).

These two successful predictions of the PDS model follow the 
WMAP confirmation of the generic
prediction of small universe models, a power cutoff in structure 
statistics on large scales, and the solution appears to be 
consistent with quite different PDS analyses of the WMAP data.

Do we really live in a Poincar\'e dodecahedral space? Further
constraints either for or against the model are certainly still
needed, but the evidence in favour of a PDS-like signal in the
WMAP data does seem to be accumulating.

\section*{Acknowledgments}

Helpful comments from Bartosz Lew were greatly appreciated.
Use of the Nicolaus Copernicus Astronomical Center 
\langed{This institute was created with a lot of US funding
and for this reason wishes to stick to US spelling.}
(Toru\'n) computer
cluster is gratefully acknowledged.
Use was made of the WMAP data
\url{http://lambda.gsfc.nasa.gov/product/} and of the
Centre de Donn\'ees astronomiques de Strasbourg 
\url{http://cdsads.u-strasbg.fr}.

\subm{ \clearpage }

\nice{
%

}



\begin{thebibliography}{}

\bibitem[{Aurich}, {Lustig}, \&  {Steiner} 2005a]{Aurich2005a}
{Aurich}, R., {Lustig}, S., \& {Steiner}, F. 2005a, \cqg, 22,  3443, \eprint{astro-ph/0504656}

\bibitem[{Aurich}, {Lustig}, \&  {Steiner} 2005b]{Aurich2005b}
{Aurich}, R., {Lustig}, S., \& {Steiner}, F. 2005b, \cqg, 22,  2061, \eprint{astro-ph/0412569}

\bibitem[{Aurich}, {Lustig}, \&  {Steiner} 2006]{Aurich2005circ}
{Aurich}, R., {Lustig}, S., \& {Steiner}, F. 2006, \mnras, 369, 240,  \eprint{astro-ph/0510847}

\bibitem[{Blanl{\oe}il} \& {Roukema} 2000]{BR99}
{Blanl{\oe}il}, V., \& {Roukema}, B.~F., eds. 2000, ``Cosmological Topology in  Paris 1998'' (Paris: Blanl{\oe}il \& Roukema), \eprint{astro-ph/0010170}

\bibitem[{Caillerie}, {Lachi{\`e}ze-Rey}, {Luminet},  {Lehoucq}, {Riazuelo}, \& {Weeks} 2007]{Caillerie07}
{Caillerie}, S., {Lachi{\`e}ze-Rey}, M., {Luminet}, J.~., {et al.} 2007, \aap, 476, 691, \eprint{0705.0217v2}

\bibitem[{Cornish}, {Spergel}, \& {Starkman} 1996]{Corn96}
{Cornish}, N.~J., {Spergel}, D.~N., \& {Starkman}, G.~D. 1996, ArXiv Gen.Rel.  \& Quant.Cosm. e-prints, \eprint{gr-qc/9602039}

\bibitem[{Cornish}, {Spergel}, \&  {Starkman} 1998]{Corn98b}
{Cornish}, N.~J., {Spergel}, D.~N., \& {Starkman}, G.~D. 1998, \cqg, 15, 2657

\bibitem[{Gausmann}, {Lehoucq}, {Luminet}, {Uzan}, \&  {Weeks} 2001]{GausSph01}
{Gausmann}, E., {Lehoucq}, R., {Luminet}, J.-P., {Uzan}, J.-P., \& {Weeks}, J.  2001, \cqg, 18, 5155, \eprint{gr-qc/0106033}

\bibitem[{Gundermann} 2005]{Gundermann2005}
{Gundermann}, J. 2005, arXiv preprints, \eprint{astro-ph/0503014}

\bibitem[{Inoue} \& {Silk} 2006]{InoueSilk06}
{Inoue}, K.~T., \& {Silk}, J. 2006, \apj, 648, 23, \eprint{astro-ph/0602478}

\bibitem[{Key}, {Cornish}, {Spergel}, \&  {Starkman} 2007]{KeyCSS06}
{Key}, J.~S., {Cornish}, N.~J., {Spergel}, D.~N., \& {Starkman}, G.~D. 2007,  \prd, 75, 084034, \eprint{astro-ph/0604616}

\bibitem[{Lachi\`eze-Rey} \& {Luminet} 1995]{LaLu95}
{Lachi\`eze-Rey}, M., \& {Luminet}, J. 1995, \physrep, 254, 135,  \eprint{gr-qc/9605010}

\bibitem[{Lambert} 1772]{Lambert1772}
{Lambert}, J. 1772, {Anmerkungen und Zus\"atze zur Entwerfung der Land und  Himmelscharten. In {\it Beitr\"age zum Gebrauche der Mathematik und deren  Anwendung}, pt. 3, sec. 6. (English translation: Notes and Comments on the  Composition of Terrestrial and Celestial Maps, Ann Arbor, University of  Michigan 1972)}

\bibitem[{Land} \& {Magueijo} 2007]{LandMagAoE07}
{Land}, K., \& {Magueijo}, J. 2007, \mnras, 378, 153,  \eprint{arXiv:astro-ph/0611518}

\bibitem[{Lehoucq}, {Weeks}, {Uzan}, {Gausmann}, \&  {Luminet} 2002]{LehSph02}
{Lehoucq}, R., {Weeks}, J., {Uzan}, J.-P., {Gausmann}, E., \& {Luminet}, J.-P.  2002, \cqg, 19, 4683, \eprint{gr-qc/0205009}

\bibitem[{Lew} \& {Roukema} 2008]{LewRouk2008}
{Lew}, B., \& {Roukema}, B.~F. 2008, \aap, 482, 747, \eprint{0801.1358}

\bibitem[{Luminet} \& {Roukema} 1999]{LR99}
{Luminet}, J., \& {Roukema}, B.~F. 1999, in NATO ASIC Proc. 541: Theoretical  and Observational Cosmology, 117, \eprint{astro-ph/9901364}

\bibitem[{Luminet}, {Weeks}, {Riazuelo}, {Lehoucq}, \&  {Uzan} 2003]{LumNat03}
{Luminet}, J., {Weeks}, J.~R., {Riazuelo}, A., {Lehoucq}, R., \& {Uzan}, J.  2003, \nat, 425, 593, \eprint{astro-ph/0310253}

\bibitem[{Luminet} 1998]{Lum98}
{Luminet}, J.-P. 1998, Acta Cosmologica, XXIV-1, 105, \eprint{gr-qc/9804006}

\bibitem[{Neal} 1993]{Neal93}
{Neal}, R. 1993, {Probabilistic Inference Using Markov Chain Monte Carlo  Methods (Technical Report CRG-TR-93-1)  [\url{http://omega.albany.edu:8008/neal.pdf}]} (Department of Computer  Science, University of Toronto: Toronto)

\bibitem[{Rebou\c{c}as} \& {Gomero} 2004]{RG04}
{Rebou\c{c}as}, M.~J., \& {Gomero}, G.~I. 2004, Braz. J. Phys., 34, 1358,  \eprint{astro-ph/0402324}

\bibitem[{Riazuelo}, {Weeks}, {Uzan}, {Lehoucq}, \&  {Luminet} 2004]{RiazSph03}
{Riazuelo}, A., {Weeks}, J., {Uzan}, J., {Lehoucq}, R., \& {Luminet}, J. 2004,  \prd, 69, 103518, \eprint{astro-ph/0311314}

\bibitem[{Roukema} 2000]{Rouk00BASI}
{Roukema}, B.~F. 2000, \BASI, 28, 483, \eprint{astro-ph/0010185}

\bibitem[{Roukema} 2001]{Rouk01-4D}
{Roukema}, B.~F. 2001, \mnras, 325, 138, \eprint{astro-ph/0102099}

\bibitem[{Roukema} 2002]{Rouk02topclass}
{Roukema}, B.~F. 2002, in {Marcel Grossmann IX Conference on General  Relativity}, eds V.G. Gurzadyan, R.T. Jantzen and R. Ruffini, World  Scientific, Singapore, p. 1937, \eprint{astro-ph/0010189}

\bibitem[{Roukema} 2005]{Eigenplanes05}
{Roukema}, B.~F. 2005, \aap, 439, 479, \eprint{astro-ph/0409694}

\bibitem[{Roukema}, {Bajtlik}, {Biesiada},  {Szaniewska}, \& {Jurkiewicz} 2007]{RBBSJ06}
{Roukema}, B.~F., {Bajtlik}, S., {Biesiada}, M., {Szaniewska}, A., \&  {Jurkiewicz}, H. 2007, \aap, 463, 861, \eprint{astro-ph/0602159}

\bibitem[{Roukema}, {Lew}, {Cechowska}, {Marecki}, \&  {Bajtlik} 2004]{RLCMB04}
{Roukema}, B.~F., {Lew}, B., {Cechowska}, M., {Marecki}, A., \& {Bajtlik}, S.  2004, \aap, 423, 821, \eprint{astro-ph/0402608}

\bibitem[{Schwarzschild} 1900]{Schw00}
{Schwarzschild}, K. 1900, Vier.d.Astr.Gess, 35, 337

\bibitem[{Schwarzschild} 1998]{Schw98}
{Schwarzschild}, K. 1998, \cqg, 15, 2539

\bibitem[{Spergel}, {Bean}, {Dor\'e}, {Nolta},  {Bennett}, {Hinshaw}, {Jarosik}, {Komatsu}, {Page}, {Peiris}, {Verde},  {Barnes}, {Halpern}, {Hill}, {Kogut}, \& {et al.} 2007]{WMAPSpergel06}
{Spergel}, D.~N., {Bean}, R., {Dor\'e}, O., {et al.} 2007, \apjs, 170, 377, \eprint{astro-ph/0603449}

\bibitem[{Spergel}, {Verde}, {Peiris}, {Komatsu},  {Nolta}, {Bennett}, {Halpern}, {Hinshaw}, {Jarosik}, {Kogut}, {Limon},  {Meyer}, {Page}, {Tucker}, {Weiland}, {Wollack}, \& {Wright} 2003]{WMAPSpergel}
{Spergel}, D.~N., {Verde}, L., {Peiris}, H.~V., {et al.} 2003, \apjs, 148, 175,  \eprint{astro-ph/0302209}

\bibitem[{Starkman} 1998]{Stark98}
{Starkman}, G.~D. 1998, \cqg, 15, 2529

\bibitem[{Starobinsky} 1993]{Star93}
{Starobinsky}, A.~A. 1993, Journal of Experimental and Theoretical Physics  Letters, 57, 622

\bibitem[{Stevens}, {Scott}, \& {Silk} 1993]{Stevens93}
{Stevens}, D., {Scott}, D., \& {Silk}, J. 1993, Physical Review Letters, 71, 20

\bibitem[{Tegmark}, {de Oliveira-Costa}, \&  {Hamilton} 2003]{WMAPTegmarkFor}
{Tegmark}, M., {de Oliveira-Costa}, A., \& {Hamilton}, A. 2003, \prd, 68,  123523, \eprint{astro-ph/0302496}

\bibitem[{Uzan}, {Lehoucq}, \& {Luminet} 1999]{ULL99b}
{Uzan}, J.-P., {Lehoucq}, R., \& {Luminet}, J.-P. 1999, in {Proc. of the  XIX$^{\rm th}$ Texas meeting, Paris 14--18 December 1998, Eds. E. Aubourg, T.  Montmerle, J. Paul and P. Peter, article n$^{\rm o}$ 04/25},  \eprint{gr-qc/0005128}

\bibitem[{Weeks} 2001]{Weeks2001}
{Weeks}, J. 2001, {The Shape of Space (2nd edition)} (Manhattan: Marcel Dekker)

\bibitem[{Weinberg} 1972]{Wein72}
{Weinberg}, S. 1972, {Gravitation and cosmology: Principles and applications of  the general theory of relativity} (New York: Wiley)

\end{thebibliography}
\end{document}